\documentclass[twocol]{ametsocV6.1}

\usepackage{lineno}
\usepackage[greek, english]{babel}
\usepackage[
    colorlinks=true,
    linkcolor=red
]{hyperref}
\IfFileExists{academicons.sty}{\usepackage{academicons}}{}
\usepackage{amsmath}
\usepackage{graphicx}
\newcommand{\ZH}{\ensuremath{Z_{\mathrm{H}}}}
\newcommand{\ZDR}{\ensuremath{Z_{\mathrm{DR}}}}
\newcommand{\KDP}{\ensuremath{K_{\mathrm{DP}}}}
\newcommand{\rhv}{\ensuremath{\rho_{\mathrm{HV}}}}
\newcommand{\VR}{\ensuremath{V_{\mathrm{R}}}}
\newcommand{\VSR}{\ensuremath{V_{\mathrm{SR}}}}
\newcommand{\Vrot}{\ensuremath{V_{\mathrm{rot}}}}
\newcommand{\BRM}{B2K\ensuremath{_{\mathrm{RM}}}}
\newcommand{\ms}{m\,s\ensuremath{^{-1}}}
\newcommand{\mss}{m\ensuremath{^{2}}\,s\ensuremath{^{-2}}}
\newcommand{\Jkg}{J\,kg\ensuremath{^{-1}}}
\newcommand{\si}{s\ensuremath{^{-1}}}
\newcommand{\dg}{\ensuremath{^{\circ}}}
\definecolor{orcidlogocol}{HTML}{A6CE39}

\makeatletter
\DeclareRobustCommand\amsparen[1]{(#1)}
\renewcommand\p@subsection{\thesection}
\renewcommand\p@subsubsection{\thesection\thesubsection\amsparen}
\renewcommand\p@paragraph{\thesection\thesubsection\amsparen{\thesubsubsection}}
\makeatother

\title{Twister After Dark: Diagnosing a Damaging Nocturnal Tornado in Iloilo, Philippines (08 August 2026)}

\authors{Generich H. Capuli\href{https://orcid.org/0000-0003-1253-7043}{\includegraphics[scale=0.5]{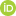}}\aff{a,b}\correspondingauthor{Generich H. Capuli, genecapuli@gmail.com / generich.capuli@pagasa.dost.gov.ph}
}

\affiliation{\aff{a}{Numerical Modelling Section, Research $\&$ Development and Training Division, Department of Science and Technology - Philippine Atmospheric, Geophysical, and Astronomical Services Administration, Brgy. Central, Quezon City 1100, Metro Manila, Philippines}\\
\aff{b}{Project Severe Weather Archive of the Philippines, Quezon City, Philippines}\\
}

\abstract{On $\sim$1430 UTC (2230 LST) 08 August 2026, a significant tornado struck the Mandurriao and Jaro districts of Iloilo City, Philippines. Its parent supercell developed within the southwest monsoon enhanced by Typhoon Dolphin. Synoptic forcing was weak, and in a distinctly nocturnal environment a low-level jet of $\sim$40 kt near 1.5--3 km surged after sunset and advected warm, moist (high-$\theta_e$) marine air onshore. The ERA5 proximity profile depicted MUCAPE of $\sim$2000 \Jkg, negligible inhibition, an LCL below 300 m, and nearly purely streamwise low-level vorticity, whereas deep-layer shear (DLS, 0--6 km bulk wind difference) was below 3 \ms\ because the hodograph folded back above 3 km. Hodographs from the newly operational Iloilo S-band dual-polarization radar (S-ILO), $\sim$5--7 km from the storm, revealed low-level shear (LLS, 0--1 km bulk wind difference) of $\sim$16 \ms\ and 0--1 km storm-relative helicity of 124--211 \mss, exceeding ERA5. The storm exhibited a hook echo with a weak echo hole and a \ZDR\ arc and ring. Three-dimensional analysis revealed an elevated 55--60 dBZ rear-flank core, consistent with a descending reflectivity core, above the weak echo hole and a $\sim$50 dBZ ring interpreted as a debris ball. A survey of 31 damage indicators documented a $\sim$2.9 km long and up to $\sim$340 m wide path rated IF2 on the International Fujita Scale. Processor smoothing of radial velocity and a \rhv\ threshold for rainfall estimation prevented direct identification of tornadic vortex and debris signatures, although a radial velocity simulation indicates that a narrow vortex (core radius $\sim$50--100 m) with a rotational velocity of $\sim$35--45 \ms\ could have been concealed. This case shows that significant tornadoes in the Philippines can develop at night during the monsoon season and that operational radar quality control should be revised to retain severe-storm signatures.
}

\begin{document}

\maketitle

\statement
\begin{itemize}
\item Documents a nocturnal IF2 tornado that struck urban Iloilo City within $\sim$7 km of an S-band dual-polarization radar.
\item The parent supercell formed in a typhoon-enhanced southwest monsoon with weak deep-layer shear (DLS) but strong, highly streamwise low-level shear (LLS).
\item A nocturnal low-level jet surge with warm, moist advection strengthened the LLS after sunset.
\item An elevated rear-flank reflectivity core, consistent with a descending reflectivity core, capped the weak echo hole and debris ball.
\end{itemize}

\section{INTRODUCTION}\label{sec1}

Annually, severe convective storms such as tornadic supercell thunderstorms occur in the Philippine Archipelago, particularly along the Greater Metro Manila Region (GMMR), Western Visayas, and Cotabato provinces. These hazards are by-products of the combination of ample vertical wind shear and atmospheric instability, including a lifting mechanism (i.e., forcing) either from plain-to-mountain flow\footnote{Especially for GMMR and Cotabato provinces as they straddle between two mountainous ranges.} or lowering of the low- and mid-level geopotential heights \citep[{e.g., mountains, tropical cyclones [TCs], easterly waves;}][and references therein]{Kirshbaum2018,Capuli2024,Capuli2026}. Although previous climatological analyses using documented severe weather reports have emphasized the spatial and temporal occurrences of these severe weather events \citep[hereafter SWE;][]{Capuli2024}, the propensity of these severe convective storms remains poorly understood in the region. Still, the documentation of tornadoes remains serendipitous, occurring typically when tornadoes strike cities amidst the mostly sparsely populated areas \citep{Anderson2007,SilvaDias2011,Oliveira2022}. There is evidence, though, that tornadic events in rural lands and forests occur regularly, but are rarely reported, especially when the damage is slight or restricted to vegetation with few damage indicators or low population density \citep{Wagner2019,Wurman2021}\footnote{Also known as the Population Density Effect.}.

Many studies have examined how environmental conditions relate to severe thunderstorms that produce tornadoes or tornado outbreaks \citep[e.g.,][]{Thomspon2012,Lee2013,Coniglio2020,BAGAGLINI2022,Pilguj2022}. This interest has lasted for decades because tornadic storms are among the most destructive natural hazards. Their footprint is local, but their social and economic impacts can be severe \citep[e.g.,][]{Antonescu2017,AndersonFrey2021,Pucik2024}. In the Philippines, however, tornadoes have been sparsely documented, so little is known about the environments and storm characteristics behind them. Tornado environments have been described much more thoroughly elsewhere, notably in the United States \citep{Evans2001,Bunkers2006,Boustead2013} and Europe \citep{Groenemeijer2014,Antonescu2017,Antonescu2019}. More recently, Mexico and Indonesia have also drawn attention \citep{LeonCruz2022,LEONCRUZ2025,Firdaus2025}. Their tropical settings and the way their SWEs vary over time make them useful comparisons for the Philippines.

Generally, severe convective storms (SCS) develop where four ingredients occur together, namely (1) sufficient atmospheric instability, (2) strong vertical wind shear, (3) ample moisture in the lower troposphere and boundary layer, and (4) a lifting mechanism that starts deep moist convection. The lift can come from orography, a frontal boundary, a dryline, or a low-level convergence zone \citep{Rasmussen2003,Thompson2004,Mercer2012,Sherburn2014}. Both \citet{markowski2010} and \citet{DAVIESJONES2015} note that tornadic supercells tend to form where there is considerable overlap between substantial convective available potential energy (CAPE) and deep-layer shear (DLS). CAPE sets the buoyancy available to the updraft while DLS, i.e., the magnitude of the vector wind difference between the surface and 6 km AGL, sustains the mid-level rotation that keeps a supercell organized and long-lived \citep{Rasmussen1998,Brooks2003,Coffer2019}. Interestingly, tornadoes are not confined to isolated supercells as both observations \citep{Nielsen2020} and numerical simulations \citep{Nielsen2018} show that mesoscale convective systems (MCSs) can also produce them.

Case studies that used similar approaches are a useful starting point for understanding severe weather events (SWEs). Studies from several regions show how synoptic-scale and mesoscale environments shape tornadic storms. \citet{Thompson2000} analyzed the 3 May 1999 outbreak over the U.S. Great Plains, which produced the Bridge Creek--Moore, Oklahoma, tornado. \citet{Lopes2025} documented the 11--12 June 2018 severe weather outbreak along the Argentina--Brazil border. In Asia, \citet{Choo2018} examined the significant tornado of 6 May 2012 in Japan, while \citet{Meng2018} documented the deadliest tornado on record in China, which struck Funing County, Yancheng, Jiangsu Province. These studies show that tornadic storms can develop under both strong and weak synoptic forcing, for example from frontal systems, upper-level vortices, or jet streaks. They also need a favorable combination of instability and vertical wind shear, with the addition of complex terrain and mesoscale boundaries that can reinforce vertical motion and help tornadogenesis. Examples include orographic lifting, local outflow boundaries, and storm interactions such as `nudgers' and mergers \citep{Kirshbaum2018,Nixon2024}. Recent work has also examined the radar signatures of tornadic supercells \citep{Clark2024,Wakimoto2025,Lopes2025}. Using dual-polarization radar fields such as differential reflectivity (\ZDR) and the copolar correlation coefficient (\rhv), these studies identified tornadic debris signatures (TDSs), i.e., an area of low \rhv\ ($<$ 0.80) that coincides with high reflectivity, and these studies discussed TDSs alongside other radar-derived storm structures such as \ZDR\ columns in the parent storm. In related work, \citet{Yuan2024} found reduced polarimetric signatures within the low-level mesocyclone of the 14 May 2021 Jiangsu tornadic supercell.

Recent efforts to systematically document severe storm phenomena and develop severe storm report databases have begun in the Philippines through Project Severe Weather Archive of the Philippines \citep[SWAP;][]{Capuli2024}. Despite this effort, tornado damage surveys in the country remain scarce. Recent examples of tornado-focused research in the Philippines include \citet{Capuli2026} and \citet{CapuliCN2026}. Among these studies, \citet{Capuli2026} described the synoptic and mesoscale conditions (including radar features such as BWER and hook echo feature) present during the development of a semi-discrete supercell storm that spawned an EF2 tornado in the rural area of Candating in Arayat, Pampanga on 27 May 2024, located in a favorable area for tornadic storms in the GMMR. The authors showed that the tornado developed in a modest synoptic forcing characterized by lowering of 500-hPa mid-level heights as induced by the migratory TC. This synoptic pattern where mid-level height falls are in-place tend to be associated with major tornado outbreaks found in the U. S. Great Plains' Tornado Alley due to the intense and rapid low-level atmospheric response \citep{Mercer2012,jiang2025}. The lower troposphere exhibited a moist boundary layer, leading to low lifting condensation levels (LCLs), and sufficient southwesterly windflow (or Southwest Monsoon) that generated ample near-surface wind shear (LLS) and DLS in both speed and direction manifested as a curvature in the hodograph profile. It has been long established that ``sickle-shaped'' hodographs (or hook shaped; where a layer of strong LLS sits beneath an abrupt clockwise turning of the shear vector with height) have become accepted as favorable for significant ($\geq$ EF2) tornadoes along cyclonic, right-moving supercells \citep{Weisman1986,Thompson2000,Nixon2022}, especially those that feature strong shear near the surface \citep{Coffer2020}. More recently, \citet{CapuliCN2026} documented a localized outbreak of near-simultaneous tornadic supercells in Camarines Norte that developed within an easterly-wave regime, demonstrating that Philippine tornadic environments are not restricted to the southwest monsoon.

On the night of 08 August 2026, a severe thunderstorm developed over the southern coast of Panay Island and produced a significant tornado that tracked through the Mandurriao and Jaro districts of Iloilo City at approximately 2230 LST (1430 UTC). The tornado uprooted and snapped trees, toppled utility poles, and damaged residential and commercial structures, leaving 148 families (or 587 individuals) affected after their homes and properties sustained damage, as confirmed by the City Disaster Risk Reduction and Management Office of Iloilo (CDRRMO-Iloilo) and local media.

The 8 August 2026 Iloilo tornado is a rare documented case of a tornadic storm in the Philippines. Western Visayas is one of the country's recognized tornado hotspots, yet previous Philippine case studies have focused mostly on Luzon, and few tornadoes from this region have been examined in detail. That makes it a valuable opportunity to study the meteorological processes that govern tornado initiation and development in a tropical environment, in a region that has received little research attention. Although the atmospheric ingredients required for tornadic supercell formation are the same in any part of the world, typical conditions that lead to SCSs in this area of the Philippines may exhibit characteristics that differ from their well-documented U. S. Great Plains or European severe storm environments \citep{Taszarek2020a,Taszarek2020b}. Thus, an improved documentation of how these ingredients are brought together, in the moist tropical environment of the Philippines, can add to the overall knowledge of the distinct atmospheric regimes that are capable of producing tornadic supercells.

Early reports and media of the event were scattered across social media platforms. Project SWAP collected, verified, and archived these materials systematically, and they are included in its fourth data release (DR4). The tornado struck at night, so extremely few photographs or videos of it exist. Damage reports, however, were plentiful and have been compiled into a comprehensive dataset (Section~\ref{sec3}) that provides essential context for assessing the tornado's impact. This study examines the meteorological conditions that led to tornadogenesis, covering the synoptic, mesoscale, and kinematic aspects of the parent storm's environment. Similar to \citet{CapuliCN2026}, this tornado occurred within about 7 km of an operational dual-polarization radar, thus dual-polarization variables can be used to examine the storm's structure, characteristics, and potential debris signatures in greater detail. Using extensive observational data and sounding-derived parameters, this study provides a foundation for future research on tropical tornadoes.

The paper is organized as follows. Section~\ref{sec2} outlines the materials and methods, including the reanalysis dataset, atmospheric profile parameters, and the satellite, radar, surface, and high-resolution optical imagery used in the study. Section~\ref{sec3} presents the environmental context (from the synoptic scale to the mesoscale), the storm evolution and radar-derived kinematics, the surface observations, and the damage analysis. Finally, Section~\ref{sec4} provides the summary of findings and an extended discussion of the results.

\section{METHODOLOGY}\label{sec2}

\subsection{Environmental Reanalysis}\label{sec21}

To characterize the environmental conditions associated with the 08 August 2026 tornado event, this study utilized reanalysis data to assess the potential for SCS development. Reanalysis datasets are widely employed to diagnose atmospheric conditions conducive to severe weather such as tornadoes, damaging winds, and hail-producing storms. They are particularly useful for calculating instability indices and other parameters essential to SCS forecasting \citep{Taszarek2020b}.

For this case, environmental reanalysis data were obtained from the fifth-generation European Centre for Medium-Range Weather Forecast (ECMWF) reanalysis (ERA5) accessed through the Climate Data Store \citep[CDS;][]{Hersbach2020}. ERA5 has demonstrated good performance in depicting vertical profiles of convective environments, especially across the United States, Europe, and even other countries along Southeast Asia \citep{Coffer2020,Taszarek2021a,Pilguj2022b,Firdaus2025}. However, some known biases persist, particularly in the boundary layer. These include discrepancies in low-level parcel characteristics and vertical shear parameters, especially near surface boundaries \citep{King2019}. Such biases are influenced by geographic location and surface elevation. Despite these limitations, ERA5 remains among the most reliable and accessible datasets for investigating severe convective environments, globally \citep{Coffer2020,Taszarek2021b,Taszarek2021a}. For large-scale synoptic analysis, the study domain extended from 4\dg--22\dg N latitude and 110\dg--135\dg E longitude. Meanwhile, the mesoscale sector covered 8\dg--14\dg N and 119\dg--126\dg E. Both domains used a spatial resolution of 0.25\dg{} x 0.25\dg{}. An hourly temporal resolution was used to allow detailed analysis of the atmospheric evolution surrounding the event, which occurred around 14 UTC. The time steps selected for in-depth analysis were 13, 14, and 15 UTC, enabling a focused examination of pre-, during, and post-storm conditions. Throughout this paper, times are given in UTC, with Philippine local standard time (LST = UTC + 8 h) in parentheses.

Although ERA5 provides several convective parameters, including CAPE, it has been shown to substantially underestimate the severity of convective environments. \citet{Capuli2025} demonstrated this for a Philippine hailstorm case in which native ERA5 CAPE of $\sim$2000--2500 \Jkg\ rose to nearly 5000 \Jkg\ when recomputed from the model-level profile with a surface-based (SB) parcel. The difference between a marginally severe environment and one capable of the large hail observed. The bias likely stems from the ERA5 formulation, which lifts candidate parcels from a series of model levels within the lowest $\sim$350 hPa rather than from the near surface, so the selected parcel is frequently more dilute than the true boundary-layer parcel. To address this, both the ERA5 single-level and 137 hybrid sigma-pressure model-level fields were combined and re-gridded onto a common grid, and CAPE was recalculated using most-unstable (MU) parcel which is the parcel yielding the largest CAPE among candidates lifted from levels within the lowest 300 hPa, with the lowest candidate taken from the 2 m AGL. Likewise, other kinematic parameters such as Storm-relative Helicity (SRH), which is not available via the standard data field in ERA5, were computed for the 0--1 and 0--3 km layers using the \textit{xcape} codebase \citep{lepore2021}. Kinematic parameters, such as SRH\footnote{Represented by the vertically integrated storm-relative flux of streamwise vorticity into an updraft.}, the \citet{Bunkers2000} internal-dynamics storm motion (B2K), and the resulting storm-relative wind (\VSR), are computed via the right-moving vector (\BRM), consistent with standard practice for the Northern Hemisphere. This assumes that the storm moves to the right of the non-pressure-weighted mean wind in the 0--6 km layer. Finally, all \textit{xcape}-derived thermodynamic quantities (e.g., MUCAPE, SBCIN), including those underlying the ERA5 mesoscale spatial maps, assume liquid-only pseudoadiabatic ascent, with condensate removed on formation and ice processes neglected as chosen for computational efficiency and consistency across the domain.

The interpretation of all reanalysis-derived parameters (including derived through the Skew-T Hodograph in Section~\ref{sec22}) is nonetheless constrained by the resolution of ERA5 relative to the scale of the setting. Spatial averaging across a land--sea boundary and across smoothed terrain tends to dilute local thermodynamic extremes and to remove the small-scale flow deformation that enhances low-level rotation, so that ERA5-derived (and the recalculated from previous section) CAPE and SRH may either overestimate or underestimate the near-storm environment. Thus, the parameters reported here are interpreted as characterizing the mesoscale environment within which the storms developed.

\subsection{Proximity Sounding Profile}\label{sec22}

Because no radiosonde was launched near the time and location of the event, a model-derived proximity profile was extracted from ERA5 at the S-ILO site (10.772\dg N, 122.579\dg E) for 1400 UTC (2200 LST) using a combination of single-level and 137 hybrid sigma-pressure model-level fields. This profile was analyzed together with the observed soundings from the PAGASA Iloilo station (WMO 98637) at 0000 UTC and the Mactan station (WMO 98646), $\sim$161 km east-southeast of S-ILO, at 1200 UTC to assess the thermodynamic and kinematic conditions preceding and at the time of the tornadic event.

While the ERA5 recalculation described in Section~\ref{sec21} utilizes a liquid-only pseudoadiabatic ascent, the parcel profiles shown on the Skew-T diagrams were computed assuming a non-entraining, irreversible mixed-phase adiabatic process following \citet{Peters2022}. Under pseudoadiabatic ascent, all condensate is assumed to fall out of the parcel immediately \citep{Emanel1994}. Meanwhile, the present formulation for the Skew-Ts instead retains condensate and represents the mixed-phase layer between the 0\,\dg C level and the level of homogeneous freezing ($\sim$$-$40\,\dg C), where liquid and ice coexist. Because it enforces conservation of energy rather than of moist entropy, it avoids the biases that arise when entropy-based formulations are applied across the mixed-phase region. Support for retaining condensate in a tropical context comes from \citet{Xu1989}, who found that mean tropical soundings are nearly neutral with respect to reversible moist adiabatic ascent and are therefore only weakly conditionally unstable, indicating that the tropical environmental temperature profile is more closely matched by an adiabatic than by a pseudoadiabatic parcel. We note that their analysis employed a reversible liquid-only parcel including water loading rather than the mixed-phase formulation adopted here. Thus, it is cited as physical motivation for retaining condensate, not as direct validation of the present parcel model.

In addition, an irreversible adiabatic ascent with entrainment was also employed to compute the Entraining CAPE \citep[ECAPE;][readers are referred to the paper for the full derivation]{Peters2023b}. ECAPE accounts for dilution of updraft plumes due to environmental entrainment, particularly under conditions of limited storm-relative inflow and/or mid- to upper-tropospheric dryness. It therefore provides a more realistic estimate of updraft intensity than the conventional, undiluted CAPE, which can overestimate buoyancy in environments where entrainment is significant. All undiluted Skew-T profiles are calculated with virtual temperature correction. Depending on meteorological context, either MU or SB parcels were selected for key thermodynamic parameters. CAPE and the level of free convection (LFC) were computed using MU parcel profiles, while convective inhibition (CIN) and LCL were determined using SB profiles. Although mixed-layer parcels are often preferred for their robustness, SB and MU parcels were preferred in this study due to their better representation of near-surface instability (or stability) and lowest cloud base heights. These factors are critical in distinguishing SB from elevated convection, particularly in tornadic environments.

To ascertain the kinematic conditions within this tornado event, the hodograph was used. In a conventional, ground-relative hodograph, no transformation is made to the modeled $u$ and $v$ wind components \citep{Markowski2003,Nowotarski2018}. However, this frame does not show the wind profile as experienced by the storm, which governs storm inflow and the ingestion of environmental vorticity. To place the kinematic analysis in the frame of the storm rather than the ground, the wind profile is expressed in storm-relative coordinates. The transformation is a rigid translation of the hodograph by the storm motion vector $\mathbf{c} = (c_u, c_v)$,

\begin{equation}
\mathbf{V}_{\mathrm{SR}}(z)=\mathbf{V}(z)-\mathbf{c}
\label{eq_vsr}
\end{equation}
where $\mathbf{V}(z) = [u(z), v(z)]$ is the ground-relative wind. Because $\mathbf{c}$ is independent of height, the vertical shear vector and the horizontal vorticity are unchanged by the transformation as only the origin of the hodograph moves. Here, $\mathbf{c}$ is the \BRM\ computed with the non-pressure-weighted mean wind and shear of the 0--3 km layer rather than the conventional 0--6 km layer, for consistency with the radar-derived hodographs, whose reliable portion extends only to $\sim$3.5 km AGL (Section~\ref{sec2323}).

The kinematic parameters were calculated from the interpolated $u$- and $v$-wind components from each sounding profile. The bulk wind difference (BWD) was calculated across several vertical layers to assess shear magnitude. These included the layers from 0--500 m (BWD$_{500}$), 0--1 km (BWD$_{01}$; LLS), 0--3 km (BWD$_{03}$; MLS), and 0--6 km (BWD$_{06}$; DLS). SRH is of the greatest dynamical importance in this regard \citep{DaviesJones1984}. The SRH was computed for several layers, namely 0--500 m (SRH$_{500}$), 0--1 km (SRH$_{01}$), 0--3 km (SRH$_{03}$), and 1--3 km (SRH$_{13}$). For the near-ground layer (0--1 km), the mean \VSR\ was calculated relative to the \BRM. This analysis also included the mean streamwise vorticity integrated over 0--500 m and 0--1 km layer represented by $\omega_{s500}$, and $\omega_{s01}$, respectively. These quantities were calculated as

\begin{equation}
\omega_s=\frac{(\nabla\times(\mathbf{v}-\mathbf{c}))\cdot(\mathbf{v}-\mathbf{c})}{\|\mathbf{v}-\mathbf{c}\|}
\label{eq_ws}
\end{equation}
where $\mathbf{v}$ is the three-dimensional environmental wind and $\mathbf{c}$ is the storm motion vector under the assumption of the B2K method. Furthermore, the vertical vorticity along the low-level mesocyclone was calculated. Based on the first-order approximation of tilting efficiency proposed by \citet{Coffer2023}, the vertical vorticity generated from the tilting of streamwise vorticity in a purely streamwise environment can be expressed as

\begin{equation}
\zeta_{\mathrm{LLM}}=\omega_s\,\frac{w}{V_{\mathrm{SR}}}
\label{eq_zeta}
\end{equation}
where $\omega_s$ is the streamwise vorticity\footnote{In this case, angle-average and its maximum $\omega_s$ in the lowest 1 km.}, $w$ is the vertical velocity of the updraft\footnote{CAPE$_{03}$ is used as a proxy for the vertical velocity of the low-level rotating updraft via the parcel-theory relation $w = \sqrt{2\,\mathrm{CAPE}_{03}}$.}, and \VSR\ is the storm-relative wind in the lowest 1 km. This formulation highlights how the conversion of environmental streamwise vorticity into vertical vorticity depends on the interaction and balance between vertical and horizontal motions \citep{DaviesJones1984}. It provides a physically intuitive approximation for assessing the potential for mesocyclone development and subsequent tornadogenesis.

The ``streamwiseness'' ($\widetilde{\omega}_s$) of horizontal vorticity was also calculated by dividing the integrated streamwise component of horizontal vorticity by the integrated total magnitude of the horizontal vorticity in a given layer,

\begin{equation}
\widetilde{\omega}_s=\frac{\omega_s}{\omega}
\label{eq_swise}
\end{equation}

The use of $\widetilde{\omega}_s$ is justified for two reasons. First, near-zero shear conditions are rare in tornado environments \citep{Coffer2020}. Second, the parameter offers intuitive interpretation e.g., a value of 0.95 means that 95\% of the horizontal vorticity is streamwise in nature. More fundamentally, the degree to which vorticity is streamwise remains a fundamental aspect of supercell dynamics and tornadogenesis \citep{Coffer2017}.

Finally, the Deviant Tornado Motion (DTM) was calculated based on the method developed by \citet{Nixon2021}. This parameter evaluates the potential deviation of tornado motion from both the low-level advective flow and the parent storm motion, and offers insight into tornado track behavior relative to ambient wind conditions. All proximity soundings associated in this event were analyzed using SounderPy (v3.1.0) by \citet{Gillett2025}. A list of the evaluated kinematic and thermodynamic parameters, along with their definitions and computation methods, is presented in Table~\ref{Table_1}. Storm-relative quantities in Table~\ref{Table_1} \citep[{SRH, \VSR, $\omega_s$, streamwiseness, DTM, the supercell composite parameter [SCP], and the significant tornado parameter [STP];}][]{Thompson2004,Thomspon2012} were computed against the \BRM, and the SCP and STP are the effective-inflow-layer formulations, which use MU parcel quantities rather than the mixed-layer quantities of the fixed-layer STP.

\begin{table*}[!t]
\caption{Environmental sounding parameters evaluated in this case. An asterisk indicates that the computation failed.}
\label{Table_1}
\centering
\resizebox{\textwidth}{!}{%
\begin{tabular}{lllccc}
\hline\hline
Parameter & Definition & Notes & 00 UTC ILO OBS & 12 UTC MAC OBS & 14 UTC ILO ERA5 \\
\hline
CAPE & Convective Available Potential Energy & MU parcel & 1260 \Jkg & 2532 \Jkg & 2014 \Jkg \\
CAPE$_{03}$ & 0--3 km CAPE & MU parcel & 100 \Jkg & 128 \Jkg & 130 \Jkg \\
ECAPE & Entraining CAPE & MU parcel & * & 1736 \Jkg & 1632 \Jkg \\
CIN & Convective Inhibition & SB parcel & 0 \Jkg & $-$21 \Jkg & 0 \Jkg \\
LCL & Lifted Condensation Level & SB parcel & 910 m & 622 m & 295 m \\
LFC & Level of Free Convection & MU parcel & 958 m & 1081 m & 295 m \\
LR$_{03}$ & 0--3 km Lapse Rate & Low-level Lapse Rate & 5.17\,\dg C\,km$^{-1}$ & 6.58\,\dg C\,km$^{-1}$ & 5.81\,\dg C\,km$^{-1}$ \\
LR$_{36}$ & 3--6 km Lapse Rate & Mid-level Lapse Rate & 5.57\,\dg C\,km$^{-1}$ & 6.07\,\dg C\,km$^{-1}$ & 5.94\,\dg C\,km$^{-1}$ \\
RH$_{13}$ & Mean 1--3 km Relative Humidity &  & 97.20\% & 82.96\% & 84.61\% \\
RH$_{16}$ & Mean 1--6 km RH &  & 96.97\% & 81.98\% & 81.91\% \\
BWD$_{01}$ & 0--1 km Bulk Wind Difference & Low-level Shear (LLS) & 5.16 \ms\ / 10.04 kt & 10.03 \ms\ / 19.51 kt & 10.50 \ms\ / 20.42 kt \\
BWD$_{03}$ & 0--3 km BWD & Mid-level Shear (MLS) & 10.02 \ms\ / 19.49 kt & 16.08 \ms\ / 31.26 kt & 17.64 \ms\ / 34.29 kt \\
BWD$_{06}$ & 0--6 km BWD & Deep-layer Shear (DLS) & 5.83 \ms\ / 11.34 kt & 8.55 \ms\ / 16.62 kt & 2.46 \ms\ / 4.79 kt \\
\BRM & Bunkers RM Storm Motion & RM: Right Mover & 8.89 \ms\ / 17.28 kt & 9.09 \ms\ / 17.66 kt & 13.60 \ms\ / 26.43 kt \\
\VSR & Mean 0--1 km Storm-relative Wind & Using \BRM & 8.33 \ms\ / 16.20 kt & 10.27 \ms\ / 19.96 kt & 11.86 \ms\ / 23.06 kt \\
SRH$_{500}$ & 0--500 m Storm-relative Helicity & Using \BRM & 20 \mss & 57 \mss & 79 \mss \\
SRH$_{01}$ & 0--1 km SRH & Using \BRM & 40 \mss & 101 \mss & 127 \mss \\
SRH$_{03}$ & 0--3 km SRH & Using \BRM & 82 \mss & 247 \mss & 259 \mss \\
SRH$_{13}$ & 1--3 km SRH & Using \BRM & 42 \mss & 146 \mss & 131 \mss \\
$\omega_{s500}$ & 0--500 m Streamwise Vorticity & Using \BRM & 0.005 \si & 0.012 \si & 0.014 \si \\
$\widetilde{\omega}_{s500}$ & 0--500 m Streamwiseness & Using \BRM & 91.61\% & 99.15\% & 98.62\% \\
$\omega_{s01}$ & 0--1 km Streamwise Vorticity & Using \BRM & 0.005 \si & 0.010 \si & 0.011 \si \\
$\widetilde{\omega}_{s01}$ & 0--1 km Streamwiseness & Using \BRM & 95.35\% & 97.61\% & 98.54\% \\
$\zeta_{\mathrm{LLM}}$ & 0--1 km Low-level Vertical Vorticity & MU parcel and \BRM & 0.008 \si & 0.016 \si & 0.015 \si \\
DTM & Deviant Tornado Motion & Using \BRM & 4.65 \ms\ / 9.04 kt & 7.06 \ms\ / 13.73 kt & 10.05 \ms\ / 19.53 kt \\
WMAXSHEAR & Undiluted Updraft Velocity $\times$ BWD$_{06}$ & MU parcel & 293.02 \mss & 608.46 \mss & 156.25 \mss \\
SCP & Supercell Composite Parameter & Effective inflow & 1.13 & 3.32 & 7.44 \\
STP & Significant Tornado Parameter & Effective inflow & 0.25 & 0.74 & 2.28 \\
\hline
\end{tabular}}
\end{table*}

\subsection{Remote Sensing and In-Situ Data}\label{sec23}

\subsubsection{Himawari-9}\label{sec231}

Convective-system evolution was monitored using the Advanced Himawari Imager (AHI) on board Himawari-9 \citep{Bessho2016}. AHI observes in 16 spectral bands, namely three visible (central wavelengths 0.47--0.64 $\mu$m), three near-infrared (0.86--2.3 $\mu$m), and ten infrared (3.9--13.3 $\mu$m). The tornado occurred about two hours after local sunset, so the visible and near-infrared bands carried no useful signal. The analysis therefore relies only on the 10.4-$\mu$m longwave infrared window channel (Band 13, B13), which works the same by day and night. B13 is well suited to delineating cold cloud tops of deep convection [brightness temperature (BT) $<$ 233 K ($-$40\,\dg C)] because water vapor absorbs little at this wavelength \citep{Hirose2019}. It is also commonly used for quickly identifying convective features linked to intense, short-lived rainfall \citep{Wang2022}. The nominal resolution of B13 is 2 km at the sub-satellite point. Full-disk B13 imagery on the GEOS141 fixed grid was retrieved every 10 min from 1400 to 1500 UTC (2200--2300 LST) on 08 August 2026 through the National Computational Infrastructure (NCI) Australia THREDDS data server, although the 1440 UTC scan was unavailable in the archive. This window brackets the S-ILO radar volumes analyzed (2200--2230 LST) and the tornado's lifetime.

\subsubsection{Polarimetric Radar}\label{sec232}

\paragraph{Iloilo Radar and Specifics}\label{sec2321}

\begin{figure}[!t]
\centering
\includegraphics[width=\columnwidth,height=0.82\textheight,keepaspectratio]{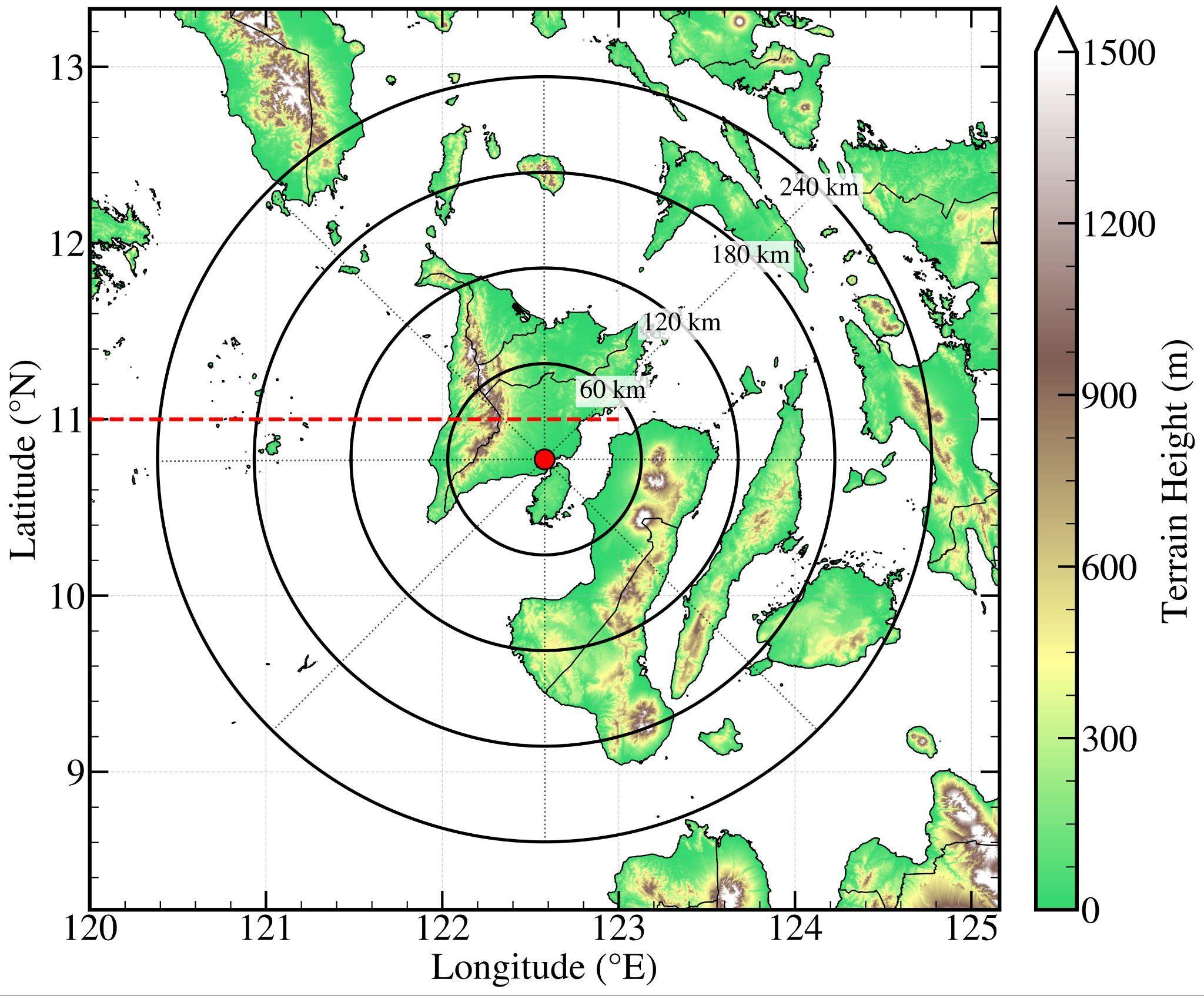}
\caption{Elevation map encompassing S-ILO (red circle) and its 240 km radar coverage with increments of 60 km each. The red dashed line serves as a reference for the dynamical vertical cross section of various atmospheric variables along 11.0\dg N (Fig.~\ref{fig_3}).}
\label{fig_1}
\end{figure}

DOST-PAGASA operates a nationwide network of 17 weather radars, consisting of seven dual-polarization (dual-pol) C-band radars, three dual-pol S-band radars, and seven single-polarization (single-pol) S-band radars. Together, these systems support the agency's capacity for early detection and monitoring of weather systems across the archipelago. The 08 August 2026 tornadic supercell tracked through the domain of the Iloilo S-band radar (hereafter S-ILO), one of the three dual-pol S-band radars, located in Jaro, Iloilo (10.773056\dg N, 122.579444\dg E; 34 m above mean sea level) on Panay Island (Fig.~\ref{fig_1}). S-ILO was inaugurated on 06 March 2026 and began operations on 22 May 2026, less than three months before the event. It provides extensive coverage over the Western Visayas, a region frequently affected by TCs and the southwest monsoon. S-ILO employs a modified Volume Coverage Pattern 11 (VCP11), which optimizes sampling across both low and high elevation angles and is particularly well suited to deep tropical convection. Each volume consists of 14 tilts (0.5--19.5\dg{}) completed every 10 min, with 360 radials at 1\dg{} azimuthal spacing and 250 m range gates out to a Doppler scanning radius of 240 km. The technical specifications of S-ILO are listed in Table~\ref{Table_2}.

\begin{table}[!t]
\caption{Technical specifications of S-ILO.}
\label{Table_2}
\centering
\resizebox{\columnwidth}{!}{%
\begin{tabular}{ll}
\hline\hline
Frequency Band & S-band, 2.868 GHz ($\lambda$ = 10.453 cm) \\
Polarization & Dual-polarization (Dual-pol) \\
Position & 10.773056\dg N, 122.579444\dg E \\
Altitude & 34 m above mean sea level \\
Pulsing Scheme & 4:3 dual-PRF on the lowest 6 tilts (0.5--5.3\dg); \\
 & single PRF on the upper 8 tilts (6.2--19.5\dg) \\
Low / High PRF & 600 / 800 Hz (1.667 / 1.250 ms) dual-PRF; \\
 & 300 Hz (3.333 ms) single PRF \\
Nyquist Velocity & $\pm$15.68 \ms\ (600 Hz), $\pm$20.91 \ms\ (800 Hz); \\
 & extended $\pm$62.72 \ms\ on dual-PRF tilts; \\
 & $\pm$7.84 \ms\ on single-PRF tilts \\
Maximum Range / Doppler Range & 240 km / 240 km \\
Beam Width / Range Resolution (Gate Spacing) & 1.33\dg\ ($h$ and $v$) / 250 m \\
Azimuthal Sampling & 1.0\dg\ (360 rays per sweep) \\
Elevation Angles (14) & 0.5\dg, 1.5\dg, 2.4\dg, 3.4\dg, 4.3\dg, 5.3\dg, 6.2\dg, \\
 & 7.8\dg, 8.7\dg, 10\dg, 12\dg, 14\dg, 16.7\dg, 19.5\dg \\
Volume Cycle Interval & 10 min \\
Start of Operation & 22 May 2026 \\
\hline
\end{tabular}%
}
\end{table}

The volumes were archived in Universal Format (UF), and several of their header entries were found to be inconsistent with the data. These entries were therefore rewritten before analysis, although the data values themselves were not modified. The stored wavelength was unusable and was replaced with $\lambda = 0.10453$ m, and the pulse repetition time (PRT) slot, which carried the pulse repetition frequency (PRF) in Hz, was converted to PRT\,=\,1/PRF. The Nyquist velocity of each sweep was then recomputed from its PRF as $v_N = \lambda\,\mathrm{PRF}/4$. With $\lambda = 0.10453$ m, the PRFs reproduce both scanning modes of S-ILO exactly. The lowest six tilts (0.5--5.3\dg{}) use 4:3 staggered dual-PRF sampling at 600/800 Hz, which yields base Nyquist velocities of $\pm$15.68 and $\pm$20.91 \ms\ and an extended Nyquist velocity of $\pm$62.72 \ms. The remaining eight tilts (6.2--19.5\dg{}) operate at a single PRF of 300 Hz, with a Nyquist velocity of only $\pm$7.84 \ms. The header Nyquist velocity of 30.0 \ms\ on the 1.5--5.3\dg{} tilts was rejected, since \VR\ reaches $\pm$47--61 \ms\ on these tilts and never exceeds the extended limit of $\pm$62.72 \ms.

Beam height above radar level ($h$), and ground range ($s$) were computed with the 4/3-Earth model \citep{Doviak1993},

\begin{equation}
\begin{aligned}
h &= \sqrt{r^{2}+(k_e a)^{2}+2\,r\,k_e a\sin\theta}\;-\;k_e a,\\
s &= k_e a\,\arcsin\!\left(\frac{r\cos\theta}{k_e a+h}\right),
\end{aligned}
\label{eq_beam}
\end{equation}
where $r$ is the slant range, $\theta$ is the elevation angle, $a = 6371$ km, and $k_e = 4/3$.

\paragraph{Radar Quality Control}\label{sec2322}

S-ILO's radar data underwent quality control based on a fuzzy logic approach designed to detect and eliminate interference patterns. This method draws on prior algorithm assessments \citep[e.g.,][and references therein]{Kilambi2018,Overeem2020} and was adapted for both single- and dual-pol radar products following the framework described in \citet{Lin2021}. In addition, these radar volumes had also been subjected to a dual-polarimetric quality control, applied as part of an ongoing project to quality-control radar volumes for operational quantitative precipitation estimation (QPE). Among its criteria, gates with \rhv\ $<$ 0.80 were omitted i.e., no gate in the \rhv\ field falls below this threshold, and masked across all radar fields such as in \ZH, \ZDR, and \KDP\ alike. For QPE, this threshold is a sound choice, since it efficiently removes non-meteorological echoes such as ground clutter, biological targets and clear-air returns that would otherwise contaminate rainfall estimates. For tornadic storms, however, it is detrimental. Lofted debris is characterized precisely by low \rhv, typically $<$ 0.80 \citep{Ryzhkov2005,Bodine2013}, so the same threshold that cleans the volume for rainfall estimation also removes the gates where a TDS would be expected. A TDS can therefore neither be confirmed nor ruled out from the supplied data.

No further correction was applied to the polarimetric fields. Attenuation is negligible at S-band for the short ranges involved, as the storm remained within $\sim$22 km of the radar throughout the analysis period. In addition, S-ILO had only recently been made operational, so no specific attenuation correction, systematic-bias calibration, or wet-radome correction had yet been established for it. Unlike the polarimetric fields, the \VR\ required substantial correction. Two distinct velocity artifacts were present in the S-ILO volumes and were corrected sequentially. The dual-PRF tilts (0.5--5.3\dg{}) are subject to processor errors, in which the unfolding of a gate between the two PRF estimates fails and the retrieved velocity is displaced by an integer multiple of twice one of the base Nyquist velocities. The single-PRF tilts (6.2--19.5\dg{}) are instead subject to conventional aliasing, which is severe at S-ILO, since a $\pm$7.84 \ms\ Nyquist interval is easily exceeded by the 15--25 \ms\ environmental flow. Before either correction, three filters were applied: (1) velocity gates without valid \ZH\ were removed, (2) gates with $|V_{\mathrm{R}}|$ $<$ 1 \ms\ on tilts at or below 1.5\dg{} were removed as residual ground or sea clutter, and (3) isolated velocity patches smaller than 10 contiguous gates were removed with Py-ART's despeckling routine.

The dual-PRF processor errors were corrected first, using the method of \citet{Alford2022}, which builds on the median-reference approach of \citet{Joe2003} and \citet{Holleman2003}. In this method, a gate whose departure from the local reference matches one of the expected error multiples of $2v_L$ or $2v_H$ is shifted back by that multiple. Only the 0.5--5.3\dg{} tilts were treated, using $v_L$ = 15.68 \ms\ and $v_H$ = 20.91 \ms. Once corrected, the dual-PRF tilts cannot fold at the observed speeds, and they therefore served as the unambiguous anchor for dealiasing the single-PRF tilts. Continuity-only dealiasing algorithm such as UNRAVEL \citep{Louf2020} and Py-ART's region-based method lose their reference between the isolated cells of this case, so each folded tilt was instead dealiased against the already-corrected tilts directly below it. This is similar to the vertical-continuity step of four-dimensional dealiasing \citep{James2001}, but anchored to the radar's own dual-PRF tilts rather than to a sounding. Adjacent gates with $|\Delta V|$ $<$ $0.5v_N$ were first joined into regions. Since a fold produces a jump of about $2v_N$, a region never straddles a fold and receives a single fold count $n_k$, so that

\begin{equation}
V_i = V_i^{\mathrm{raw}} + 2\,n_k\,v_N,\qquad n_k\in\{-4,\dots,4\},
\label{eq_fold}
\end{equation}

For every gate, a local reference ($R_i$) was taken from the nearest corrected tilt $b$ below, as a 3 $\times$ 5 ray-by-gate median at the same azimuth and ground range. This reference was projected to the elevation of tilt $s$ and corrected for the environmental shear between the two beam heights,

\begin{equation}
R_i=\left[\frac{\tilde V_b}{\cos\theta_b}+V_{\mathrm{env}}(h_s,\phi)-V_{\mathrm{env}}(h_b,\phi)\right]\cos\theta_s,
\label{eq_ref}
\end{equation}
where $V_{\mathrm{env}}(h,\phi) = u_{\mathrm{env}}(h)\sin\phi + v_{\mathrm{env}}(h)\cos\phi$ is the horizontal radial component of a quadratic fit to the dual-PRF-only wind profile (Section~\ref{sec2323}). The fold counts were then chosen to minimize

\begin{equation}
\begin{aligned}
E(n_k) ={}& \sum_{i\in k}\min\!\big(|V_i-R_i|,\,v_N\big)+\sum_{(i,j)}\min\!\big(|V_i-V_j|,\,v_N\big)\\
&+\lambda\sum_{i\in k}\min\!\big(|V_i-V_{\mathrm{env},i}|,\,v_N\big),
\end{aligned}
\label{eq_cost}
\end{equation}
where the first term anchors the region to the tilt below, the second enforces continuity across region boundaries, and the third is a weak environmental prior ($\lambda$ = 0.2). Each term is capped at $v_N$, so that a few poor references or a genuine sharp shear zone, such as a mesocyclone, cannot dominate the solution. Equation~(\ref{eq_cost}) was minimized by iterated conditional modes \citep{Besag1986}, sweeping the tilts from the bottom up, and a second pass added the reference from the newly corrected tilts above. For the near-storm analysis, a stricter variant was applied as regions were joined only where $|\Delta V|$ $<$ $0.25v_N$ and the reference came from one tilt below only. This was followed by a suspect-region mask that removes regions in which more than half of the gates depart by more than $v_N$ from the 7 $\times$ 21-gate median of their own sweep or by more than $1.5v_N$ from both adjacent tilts. The scheme was verified by folding the dual-PRF tilts artificially at 7.84 \ms\ and recovering them; only 1.6--4.8\% of the gates received a wrong fold, compared with 8--77\% for Py-ART's region-based method and UNRAVEL on the same test.

Despite these corrections, the \VR\ data of S-ILO remain problematic for tornado-scale analysis, for four reasons identified upon checking.

\begin{itemize}
\item \textbf{Velocity sign.} The velocity is stored with the opposite sign to the CF convention (positive toward the radar). With the sign as stored, the retrieved low-level winds blow from the north-northeast and the \BRM\ moves toward the southwest. This contradicts both the east-northeastward storm motion and the southwesterly winds recorded by the automatic weather station (AWS). Therefore, all velocities were multiplied by $-$1.
\item \textbf{Aliasing on the upper tilts.} The upper eight tilts are so heavily aliased that 34--81\% of the raw gates jump by more than $v_N$ relative to the tilt below. Even after dealiasing, 8--29\% of the gates on these tilts still depart from the wind profile by more than $v_N$. Thus, the single-PRF tilts were used only for the wind profile, not for vortex-scale diagnostics.
\item \textbf{Processor smoothing.} The velocity field has been strongly smoothed by the signal processor. Fitting a separable Gaussian to the autocorrelation of the small-scale velocity residual yields an effective filter of $\sigma_{\mathrm{az}}$ = 1.66 radials and $\sigma_r$ = 1.50 gates. This suppresses gate-to-gate shear and can almost entirely hide a vortex narrower than about 150 m.
\item \textbf{Missing data at the debris.} As previously mentioned, the \VR\ gates without reflectivity were discarded, and the \rhv\ $<$ 0.80 gates are missing in \ZH\ and other dual-pol fields. Hence, the \VR\ field is also missing at the gates where debris would be expected.
\end{itemize}

Consequently, no tornadic vortex signature \citep[TVS;][]{Brown1978} could be identified directly in \VR, which motivates the velocity simulation described in subsequent section.

\paragraph{Radar Techniques}\label{sec2323}

Because the echoes in this case were patchy, most single range rings never reached the azimuthal coverage required by a classic velocity--azimuth display \citep[VAD;][]{Browning1968}. The wind profile was therefore retrieved with a volume velocity processing (VVP)-type layer fit \citep{Waldteufel1979}. Layer by layer, all usable gates of all tilts below 45\dg{} within 5--120 km slant range were fitted to a horizontally uniform wind,

\begin{equation}
V_R=u\,\sin\phi\cos\theta+v\,\cos\phi\cos\theta+w_e\sin\theta,
\label{eq_vvp}
\end{equation}
where $\phi$ is the azimuth and $w_e$ combines vertical air motion and hydrometeor fall speed. Layers were centered every 250 m from 125 m to 10 km above ground level (AGL), with a half-thickness of 250 m, and gate heights were computed from Eq.~(\ref{eq_beam}). The fit was solved by least squares with iterative rejection of residuals larger than max(2.5$\sigma$, 2 \ms). A layer was accepted only if it met all of the following criteria, namely (1) at least 100 gates, (2) at least 5 of 12 occupied 30\dg{} azimuth sectors, (3) a condition number of the $(u, v)$ geometry $\leq$ 30, (4) a residual RMS $\leq$ 5 \ms, and (5) at least 60\% of the gates retained.

The retrieval was performed in two passes. The first pass used only the dual-PRF gates, which cannot alias. The layer-by-layer median of this fit over three consecutive volumes ($\pm$10 min) defined a reference profile extending to about 3.4--3.6 km AGL. The second pass added the dealiased single-PRF gates, and a layer was retained only if its wind stayed within 5 \ms\ of the reference. Otherwise, the dual-PRF-only fit was used, and where that layer was also missing, the time-filled reference was adopted (e.g., the 125 m layer at 1410 UTC was taken from 1420 UTC). The 10 m AWS wind nearest to each scan was then added as the lowest level of the profile, and the profile was linearly interpolated to a 10 m grid without vertical smoothing.

Storm motion was computed with the internal-dynamics method of \citet{Bunkers2000} using the non-pressure-weighted mean wind of the 0--3 km layer, because the reliable portion of the radar-derived profile did not extend to 6 km, so the shear vector was taken between the 2.5--3.0 km and 0--0.5 km mean winds, with a deviation of 7.5 \ms. The DTM was computed as the average of the right-moving vector and the 0--500 m mean wind \citep{Nixon2021}. As an independent estimate, the displacement of the column-maximum reflectivity composite (2 km grid, \ZH\ $>$ 10 dBZ) between the 1410 and 1420 UTC volumes was determined by two-dimensional cross-correlation, yielding an observed storm motion of 13.7 \ms\ toward 076\dg{}. The BWD, SRH, critical angle \citep{Esterheld2008}, mean \VSR, and the streamwise vorticity and streamwiseness defined in Section~\ref{sec22} were then computed for each volume relative to the observed storm motion (Table~\ref{Table_3}), and the hodographs were plotted in the same frame.

\begin{table*}[!t]
\caption{Radar-derived kinematics from the volume velocity processing (VVP) wind profiles and hodographs. Storm-relative quantities are computed relative to the observed storm motion (13.7 \ms\ toward 076\dg).}
\label{Table_3}
\centering
\begin{tabular}{lccc}
\hline\hline
Kinematic parameter & 1410 UTC & 1420 UTC & 1430 UTC \\
\hline
BWD$_{01}$ & 16.063 \ms\ / 31.224 kt & 16.250 \ms\ / 31.588 kt & 14.343 \ms\ / 27.881 kt \\
BWD$_{03}$ & 10.641 \ms\ / 20.684 kt & 11.966 \ms\ / 23.261 kt & 16.064 \ms\ / 31.225 kt \\
BWD$_{06}$ & --- & --- & --- \\
\BRM & 13.535 \ms\ / 26.310 kt & 13.827 \ms\ / 26.878 kt & 12.506 \ms\ / 24.310 kt \\
\multicolumn{4}{l}{\textit{Storm-relative (observed storm motion of 13.700 \ms\ / 26.631 kt)}} \\
\VSR & 14.671 \ms\ / 28.518 kt & 14.514 \ms\ / 28.212 kt & 17.707 \ms\ / 34.419 kt \\
SRH$_{500}$ & 114.901 \mss & 114.995 \mss & 100.717 \mss \\
SRH$_{01}$ & 211.451 \mss & 124.379 \mss & 72.898 \mss \\
SRH$_{03}$ & 291.815 \mss & 265.186 \mss & 281.987 \mss \\
$\omega_{s500}$ & 0.0193 \si & 0.0195 \si & 0.0173 \si \\
$\widetilde{\omega}_{s500}$ & 0.7383 / 73.83\% & 0.7589 / 75.89\% & 0.5456 / 54.56\% \\
$\omega_{s01}$ & 0.0150 \si & 0.0103 \si & 0.0074 \si \\
$\widetilde{\omega}_{s01}$ & 0.7224 / 72.24\% & 0.5607 / 56.07\% & 0.3467 / 34.67\% \\
\hline
\end{tabular}
\end{table*}

Since a \rhv-based TDS could not be used, the tornado was tracked through its reflectivity signature instead. On each tilt of the 1420 and 1430 UTC volumes, the weak echo hole (WEH) at the tip of the hook echo was identified in the \ZH\ field and its center was marked manually. Briefly, the WEH is a local \ZH\ minimum, partly as missing data, enclosed by the high-reflectivity ring of the debris ball. Each center was converted to azimuth and range from the radar, to beam height with Eq.~(\ref{eq_beam}), and to latitude and longitude. It was time-stamped from the ray times of its tilt, since successive tilts are scanned about 30 s apart. At 1420 UTC, the WEH could be followed from 2.4\dg{} to 10.0\dg{} (0.48--1.42 km above radar level), and at 1430 UTC from 2.4\dg{} to 5.3\dg{} (0.24--0.43 km). The WEH was enclosed by circles of 0.6 km radius at 1420 UTC and 0.46--0.58 km radius at 1430 UTC.

The vertical structure of the WEH was further examined at 1420 UTC with two vertical cross-sections of \ZH, \ZDR, and \rhv\ through its center, from the surface to 2 km, one north--south (10.628--10.809\dg N along 122.499\dg E) and one east--west (122.442--122.616\dg E along 10.717\dg N). The tracked WEH positions were finally compared with the centerline of the surveyed damage path. The centerline was obtained by projecting the damage points onto their principal axis and smoothing their cross-track offsets with a locally linear Gaussian kernel (0.30 km bandwidth).

To examine the three-dimensional structure of the storm, the reflectivity (\ZH) of the 1420 UTC volume was objectively analyzed onto a Cartesian grid with Py-ART \citep{Helmus2016}, using a Barnes weighting function, a radius of influence that increases with the beam width (minimum of 300 m), and a grid spacing of 0.25 km in the horizontal and 0.2 km in the vertical over a 20 km $\times$ 20 km domain extending to 4 km ARL. Before gridding, every gate was shifted back to the start of the 2.4\dg{} tilt by the storm motion (13.7 \ms\ toward 076\dg{}), because the tilts of one volume are scanned $\sim$31 s apart and a feature moving with the storm otherwise appears displaced downstream on each successive tilt. The gridded field was lightly smoothed (Gaussian filter with $\sigma$ = 0.5 grid points), and closed isosurfaces of \ZH\ at 30, 35, 40, 45, and 50 dBZ were extracted with the marching-cubes algorithm and drawn above the 2.4\dg{} PPI with a vertical exaggeration of 2. The same procedure, applied to \ZH\ of the 1410, 1420, and 1430 UTC volumes, was used to quantify the height and intensity of the reflectivity cores near the hook, and the same advection correction was applied to the WEH centers when assessing the tilt of the WEH (Section~\ref{sec33}).

\paragraph{Radial Velocity Simulation}\label{sec2324}

Because the observed \VR\ is unreliable at the tornado scale, a simulation was performed for two purposes, namely (1) to show what a tornado consistent with the damage survey would have looked like had it been sampled by S-ILO, and (2) to constrain how strong a vortex could have been hidden by the processing. The simulated vortex follows the WEH tracked in the previous subsection. On each tilt of the 1420 UTC volume (particularly 2.4--5.3\dg{}), it was centered on the marked WEH position, so that the simulation tilts with height together with the observed reflectivity signature.

Two tangential wind profiles were used. The first is a Rankine vortex,

\begin{equation}
V_t(\rho)=\begin{cases}
V_{\mathrm{rot}}\,\dfrac{\rho}{R_c}, & \rho\le R_c,\\[6pt]
V_{\mathrm{rot}}\,\dfrac{R_c}{\rho}, & \rho>R_c.
\end{cases}
\label{eq_rankine}
\end{equation}

The second is a Burgers--Rott vortex, which has a smoother core,

\begin{equation}
\begin{aligned}
V_t(\rho) &= V_{\mathrm{rot}}\left(1+\frac{1}{2\alpha}\right)\frac{R_c}{\rho}\left[1-\exp\!\left(-\alpha\frac{\rho^{2}}{R_c^{2}}\right)\right],\\
\alpha &= 1.2564,
\end{aligned}
\label{eq_burgers}
\end{equation}
where $\rho$ is the distance from the vortex center, $R_c$ the core radius (radius of maximum wind) and \Vrot\ is the peak tangential (rotational) speed. A core radius of $R_c$ = 100 m and \Vrot\ = 40 \ms\ were adopted (see below). The cyclonic tangential wind was projected onto the radar beam, $V_R = V_t(\rho)\,(\hat{\mathbf{k}}\times\hat{\boldsymbol{\rho}})\cdot\hat{\mathbf{r}}\,\cos\theta$. It was sampled through the effective radar resolution volume following \citet{Wood1997} and \citet{Brown2002}, using a two-way Gaussian beam with a 1.33\dg{} effective beamwidth, smeared by the 1\dg{} antenna rotation per radial, and a 250 m rectangular range gate, evaluated on 31 $\times$ 9 sub-beam samples per gate. The simulated \VR\ was added to the observed, quality-controlled velocity field without the processor filter, so the simulation represents what the radar would have displayed in unsmoothed data.

The choice of \Vrot\ was constrained statistically with an injection--recovery test. Simulated vortices spanning $R_c$ = 50--400 m and \Vrot\ = 0--80 \ms\ were passed through the beam model and the fitted processor filter, and through a filter 1.5 times wider. They were added to the observed field at 40 random positions within 0.6 km of the WEH. Each injected field was scanned for the maximum cyclonic azimuthal velocity difference ($\Delta V$) within 1 km, using a gate-to-gate detector and a peak detector (radials $\leq$ 2 km apart). Detection thresholds were set at the 95th percentile of the same detectors at 400 random storm-core locations (\ZH\ $\geq$ 35 dBZ, $>$ 3 km from the WEH). The detection probability $P_{\mathrm{det}}(V_{\mathrm{rot}}, R_c)$ obtained from these trials was then combined with the damage rating in a Bayesian framework,

\begin{equation}
\begin{aligned}
&P(V_{\mathrm{rot}}\mid \mathrm{EF2},\ \text{no detection},\ R_c)\\
&\quad\propto P(\mathrm{EF2}\mid V_{\mathrm{rot}})\prod_{\mathrm{tilts}}\big[1-P_{\mathrm{det}}(V_{\mathrm{rot}},R_c)\big]\,P(V_{\mathrm{rot}}),
\end{aligned}
\label{eq_bayes}
\end{equation}
where $P(\mathrm{EF2}\mid V_{\mathrm{rot}})$ is the probability that the peak ground-relative wind, $V_{\mathrm{rot}} + c$, falls within the EF2 range (49.6--60.4 \ms), with the storm translation $c$ drawn from $\mathcal{N}(13.7, 1.5)$ \ms, the non-detection term treats the 2.4\dg{}, 3.4\dg{} and 4.3\dg{} tilts as independent looks, with equal weights on the fitted filter and the 1.5-times filter, and the prior $P(V_{\mathrm{rot}})$ is uniform over 0--80 \ms. For $R_c$ = 100 m, the posterior favors \Vrot\ $\approx$ 35--45 \ms, and \Vrot\ = 40 \ms\ was adopted for the simulation. The EF2 interval is used here because it is defined for a ground-relative gust, which is the quantity that $V_{\mathrm{rot}} + c$ represents; it spans the central speeds of the IF1.5 and IF2 classes (50 and 60 \ms) assigned along the damage path (Section~\ref{sec36}).

\subsubsection{PlanetScope Observations}\label{sec233}

Remote sensing provides innovative technologies for assessing tornado debris \citep{HOQUE2017}. These technologies have shown to be effective in damage assessment following disasters, as demonstrated by numerous research \citep{Xie2016}. Aerial images \citep{schaefer2020}, a form of airborne remote sensing technology, are one important source of data for storm damage assessment.

PlanetScope (PS) is a CubeSat 3U form factor (10 cm $\times$ 10 cm $\times$ 30 cm) satellite constellation operated by Planet Labs, Inc. and was utilized in this study. While in sun-synchronous orbit, these cubesats provide daily 4-band (RGB and NIR) imagery of the Earth at 3 m spatial resolution \citep{PlanetARPS2024}. Currently in orbit are the Dove Classics, the improved Dove-Rs with more spectral and radiometric precision and sharpness, and the hyperspectral SuperDoves. Most images used in the study were daily Dove Classic imagery, or the newer generation Dove-R images (\url{https://www.planet.com/}). These have a Ground Sampling Distance (GSD) of 3--4 m at nadir and a positional accuracy of $<$ 10 m RMSE. We used the level 3B Surface Reflectance (SR; $\rho$) product, which is orthorectified and atmospherically corrected by Planet Labs with their in-house processing. These provide more consistency across time and location localized atmospheric conditions while minimizing uncertainty in the spectral response.

The high-resolution satellite images that were chosen for this analysis are nearly clear images (cloudless, hazeless, and smokeless) within the areas of interest for this analysis of before (05 August 2026) and after the event (11 August 2026), which is $\pm$3 days only from the event (08 August 2026). Furthermore, we apply spatial masking using a sensor-agnostic deep learning model that segments clouds and cloud shadow. This is applied through OmniCloudMask \citep[OCM;][]{Wright2025}, a python library that consists of two ensembled U-Net \citep{Ronnberger2015} models. These constituent models leverage the use of Fastai Dynamic U-Net \citep{Howard2020}, which allows any CNN to be used as a backbone model. The models chosen for this purpose were the ``RegnetY 004'' and the ``ConvNextV2 nano'', supplied with pretrained weights sourced from the timm library \citep{rw2019timm}. With the masked data, the PS satellite imagery can be used to detect the damage from a tornado event using Normalized Difference Vegetation Index (NDVI).

After the clouds were removed from the pre- and post-tornado PS imagery, NDVI was calculated for the pre-tornado (NDVI$_{\mathrm{BEFORE}}$) and post-tornado (NDVI$_{\mathrm{AFTER}}$) scenes. These were then used to compute the percent reduction in NDVI between the two dates. Because it is normalized by the pre-event NDVI, the percent reduction accounts for the initial condition of the vegetation, so a given loss is weighted by how vegetated a pixel was beforehand. Earlier work instead used a simple NDVI difference i.e., \citet{Burow2020} subtracted pre-event NDVI from post-event NDVI to produce an NDVI difference image for three tornadoes from a single outbreak, where vegetation conditions were comparable between images taken one week apart.

NDVI and its percent reduction were calculated as

\begin{equation}
\mathrm{NDVI} = \frac{\rho_{\mathrm{NIR}} - \rho_{\mathrm{Red}}}{\rho_{\mathrm{NIR}} + \rho_{\mathrm{Red}}},
\label{eq_ndvi}
\end{equation}

\begin{equation}
\Delta\mathrm{NDVI}_{\%} = \frac{\mathrm{NDVI}_{\mathrm{pre}} - \mathrm{NDVI}_{\mathrm{post}}}{\max\left(\mathrm{NDVI}_{\mathrm{pre}},\,0.10\right)} \times 100.
\label{eq_dndvi}
\end{equation}

\subsubsection{Automatic Weather Station}\label{sec234}

Surface observations were obtained from the AWS collocated with S-ILO ($\sim$4.8 m above mean sea level), which records temperature, relative humidity, station pressure, wind, and rainfall at 10-min intervals. The local time stamps were converted to UTC. The dewpoint, water vapor mixing ratio, virtual potential temperature ($\theta_v$), equivalent potential temperature ($\theta_e$), and SB LCL were derived with MetPy, with the LCL height obtained from the hypsometric equation over the dry-adiabatic subcloud layer. Perturbations of pressure ($\Delta p'$) and $\theta_v$ ($\Delta\theta_v$) were computed as departures from a centered 3-h running mean, which removes the diurnal cycle and the semidiurnal pressure tide while retaining mesoscale signals such as storm outflow.

The buoyancy of the near-surface air was computed as $B = g\,\Delta\theta_v/\theta_{v,\mathrm{ref}}$, where $\theta_{v,\mathrm{ref}} = \theta_v - \Delta\theta_v$ represents the environmental value. A cold pool was diagnosed where $\Delta\theta_v$ $\leq$ $-$0.5 K and $\Delta p'$ $>$ 0, and its density-current speed was estimated as $C = (2\Delta p'/\rho)^{1/2}$ \citep{Benjamin1968}, where $\rho$ is the air density.

\section{RESULTS}\label{sec3}

\subsection{Overview of Synoptic Environment}\label{sec31}

\begin{figure*}[!t]
\centering
\includegraphics[width=\textwidth,height=0.82\textheight,keepaspectratio]{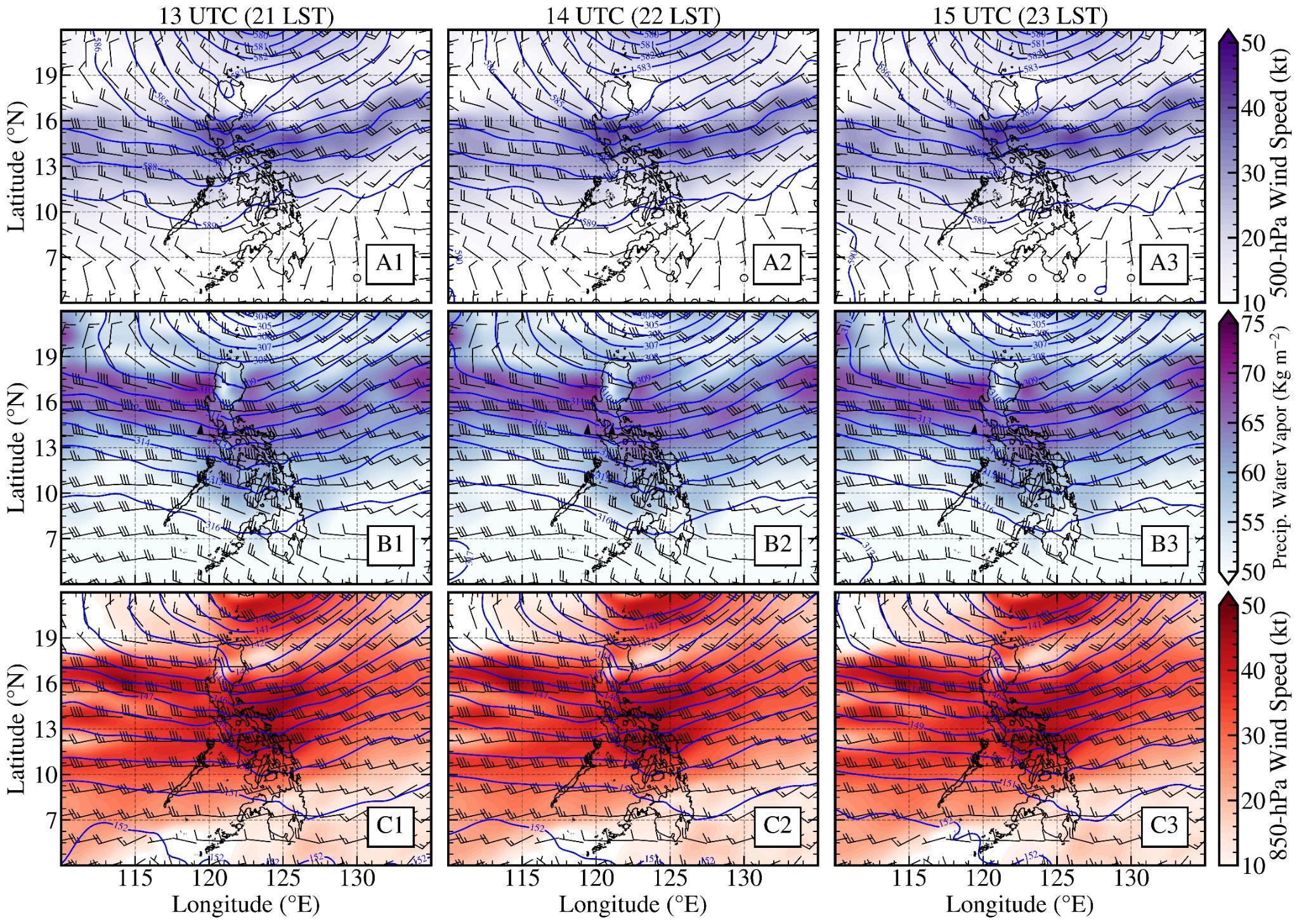}
\caption{Synoptic environment of the Philippine archipelago before (1; 13 UTC/21 LST), during (2; 14 UTC/22 LST), and after (3; 15 UTC/23 LST) the event. (a) 500-hPa geopotential heights (blue; dam) and winds (kt), with wind speeds $>$ 10 kt shaded. (b) 700-hPa geopotential heights (blue; dam), precipitable water (shaded; kg\,m$^{-2}$), and winds (kt). (c) 850-hPa geopotential heights (blue; dam) and winds (kt), with wind speeds $>$ 10 kt shaded.}
\label{fig_2}
\end{figure*}

The synoptic-scale environment from 1300 to 1500 UTC (2100--2300 LST) is shown in Fig.~\ref{fig_2}. As of 0700 UTC (1500 LST), Typhoon Dolphin was located $\sim$825 km north-northeast of extreme northern Luzon and outside the Philippine Area of Responsibility, so that only the southern periphery of its circulation appears along the northern edge of the analysis domain. Although the typhoon did not directly affect the country, it enhanced the southwest monsoon (locally known as \textit{Habagat}), which had brought widespread rainfall over Luzon and the Visayas since early August. At 500 hPa (Fig.~\ref{fig_2}a1--a3), a westerly wind maximum of 25--40 kt extended along 13\dg{}--16\dg N, whereas the flow over the Visayas was comparatively weak (10--20 kt). Geopotential heights over Panay Island remained near 588--589 dam and changed little during the analysis period. Thus, in contrast to the Candating, Arayat, tornado, which occurred under 500-hPa height falls induced by TC Ewiniar \citep{Capuli2026}, synoptic-scale mid-level forcing for ascent was weak over the area of interest, similar to the easterly-wave regime of the Camarines Norte outbreak \citep{CapuliCN2026}.

The monsoon flow was most evident in the lower troposphere. At 700 hPa (Fig.~\ref{fig_2}b1--b3), 30--40 kt westerlies spanned the central Philippines, and precipitable water (PWAT) exceeded 65--75 kg\,m$^{-2}$ along a moisture axis near 14\dg{}--17\dg N, while values of 55--62 kg\,m$^{-2}$ covered the Visayas. At 850 hPa (Fig.~\ref{fig_2}c1--c3), a broad band of 30--45 kt westerly to west-southwesterly flow extended from the South China Sea across Luzon and the Visayas toward the Philippine Sea, with 25--35 kt winds over Panay Island. This band acted as a large-scale low-level jet that sustained the advection of warm, moist marine air from the Sulu Sea and the Panay Gulf onto the southern coast of Panay, similar to the role of the southwest monsoon in the Candating, Arayat, case \citep{Capuli2026}. Because the westerlies at 850 and 700 hPa were considerably stronger than the flow at 500 hPa, the vertical wind shear over the Visayas was concentrated in the lowest $\sim$3 km, a configuration that shaped the hodographs discussed in Section~\ref{sec32}

\begin{figure*}[!t]
\centering
\includegraphics[width=\textwidth,height=0.82\textheight,keepaspectratio]{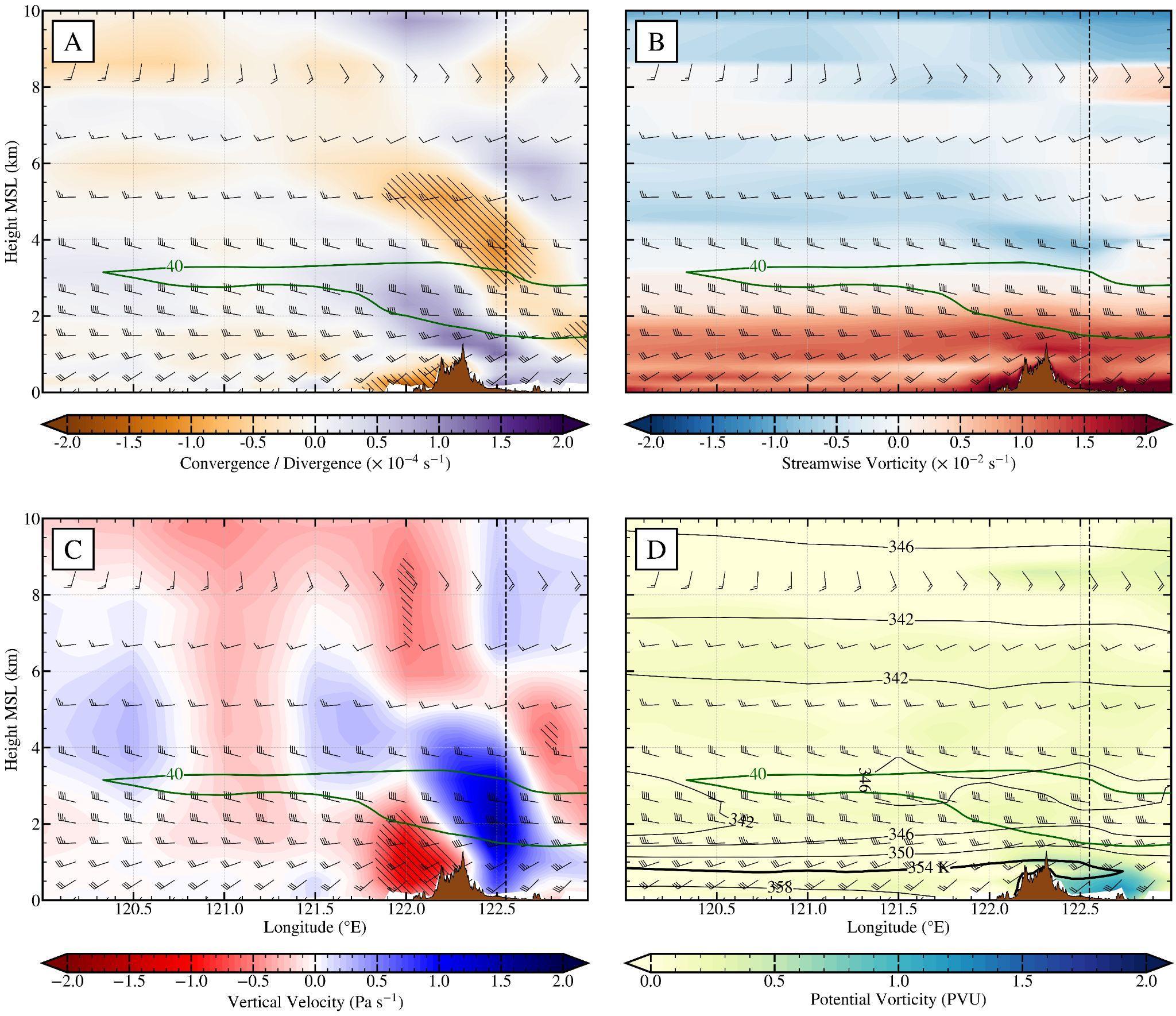}
\caption{ERA5 vertical cross section along 11.0\dg N (red dashed line in Fig.~\ref{fig_1}) at 14 UTC/22 LST of (a) convergence/divergence ($\times$ 10$^{-4}$ \si), with hatching where convergence is stronger than $-$0.50 $\times$ 10$^{-4}$ \si; (b) streamwise vorticity ($\times$ 10$^{-2}$ \si); (c) vertical velocity (Pa \si), with hatching where ascent is stronger than $-$0.50 Pa \si; and (d) potential vorticity (PV; shaded, PVU) with equivalent potential temperature ($\theta_e$; contours every 4 K). All panels show the 40 kt nocturnal wind maximum (green contour), the longitude of the tornado (black dashed line), the terrain (brown), and the tangential and normal wind components (barbs; kt). In (d), the bold 354-K $\theta_e$ contour marks the low-level moisture tongue.}
\label{fig_3}
\end{figure*}

Figure~\ref{fig_3} presents a zonal vertical cross section along 11.0\dg N (red dashed line in Fig.~\ref{fig_1}), $\sim$30 km north of the damage path, at 1400 UTC (2200 LST). The most prominent feature is a nocturnal low-level jet (LLJ), i.e., a 40 kt westerly wind maximum between $\sim$1.5 and 3.3 km MSL over the sea west of Panay that descended eastward to $\sim$1.5--2.8 km over and to the lee of the central Panay mountain range (green contour). The jet coincided with the evening surge of the low-level winds retrieved by S-ILO (Section~\ref{sec34}) and, together with the 25--35 kt flow at 850 hPa over Panay (Fig.~\ref{fig_2}c2), formed the low-level branch of the typhoon-enhanced monsoon. Beneath the jet, the near-surface layer contained positive streamwise vorticity exceeding 1.5 $\times$ 10$^{-2}$ \si\ in the lowest $\sim$0.5 km, with values above 0.5 $\times$ 10$^{-2}$ \si\ extending to $\sim$2 km (Fig.~\ref{fig_3}b), indicating that the horizontal vorticity generated by the strong shear beneath the jet was oriented largely along the storm-relative inflow. Low-level convergence and ascent of 1.0--2.0 Pa \si\ were concentrated on the windward (western) slopes of the mountain range between 121.8\dg{} and 122.2\dg E (Figs.~\ref{fig_3}a and \ref{fig_3}c), consistent with the orographic lifting of the impinging monsoon flow. East of the ridge, near the longitude of the tornado, ERA5 instead depicted descent of 1.0--1.5 Pa \si\ between $\sim$1.5 and 4 km, overlain by a sloping layer of convergence at 3.5--5.5 km. This lee-side descent is likely a terrain-induced response of the coarsely resolved flow; because ERA5 cannot resolve individual convective storms, it should not be interpreted as an absence of ascent in the near-storm environment.

The $\theta_e$ field (Fig.~\ref{fig_3}d) shows that the LLJ was underlain by a tongue of warm, moist air with $\theta_e$ $\geq$ 354 K below $\sim$1 km (bold contour), reaching 358 K near the surface west of Panay, and was overlain by a mid-tropospheric $\theta_e$ minimum below 342 K between $\sim$2 and 7 km. Because the low-level flow was westerly to southwesterly, the jet and the flow beneath it carried this high-$\theta_e$ marine air from the Sulu Sea and the Panay Gulf onshore toward Iloilo City, i.e., the LLJ acted as a conveyor of warm, moist air into the inflow region of the storm. The resulting decrease of $\theta_e$ by $\sim$12--16 K with height indicates a potentially (convectively) unstable column that modest lifting could release. In addition, PV of 0.5--1.2 PVU was concentrated in the lowest $\sim$1 km immediately east of the central Panay range, near and upstream of the event location, whereas PV was generally $<$ 0.3 PVU elsewhere in the lower troposphere. Low-level PV anomalies in the lee of mountains are commonly generated by frictional and diabatic processes as the flow passes over terrain \citep{Aebischer1998}, and they imply locally enhanced low-level vorticity and/or static stability. At the resolution of ERA5, however, this signal is viewed as a qualitative indication that the terrain of Panay modified the low-level flow upstream of Iloilo City rather than as a quantitative estimate of the vorticity ingested by the storm.

These features indicate that the Iloilo tornado should be regarded as a nocturnal event not only in its timing, about 4 h after local sunset, but also in its environment. In the midlatitudes, nocturnal LLJs develop when the frictional coupling between the boundary layer and the flow above weakens after sunset, allowing the residual-layer wind to accelerate through an inertial oscillation \citep{Blackadar1957}, aided by the diurnal cycle of heating over sloping terrain \citep{Holton1967,Shapiro2016}. At $\sim$10.7\dg N, the inertial period is $\sim$64 h, so the inertial acceleration alone is slow. The evening surge over Panay is more plausibly explained by the frictional decoupling of an already strong, typhoon-enhanced monsoon flow, combined with the nocturnal reversal of the land--sea and slope circulations over the island. Whatever its cause, the jet increased the LLS and SRH after sunset while advecting high-$\theta_e$ air beneath it, which is the combination that sustains supercells and favors tornadoes during the nocturnal transition in the midlatitudes \citep{Mead2011,Gropp2018}.

\subsection{Mesoscale Environment}\label{sec32}

\begin{figure*}[!t]
\centering
\includegraphics[width=\textwidth,height=0.82\textheight,keepaspectratio]{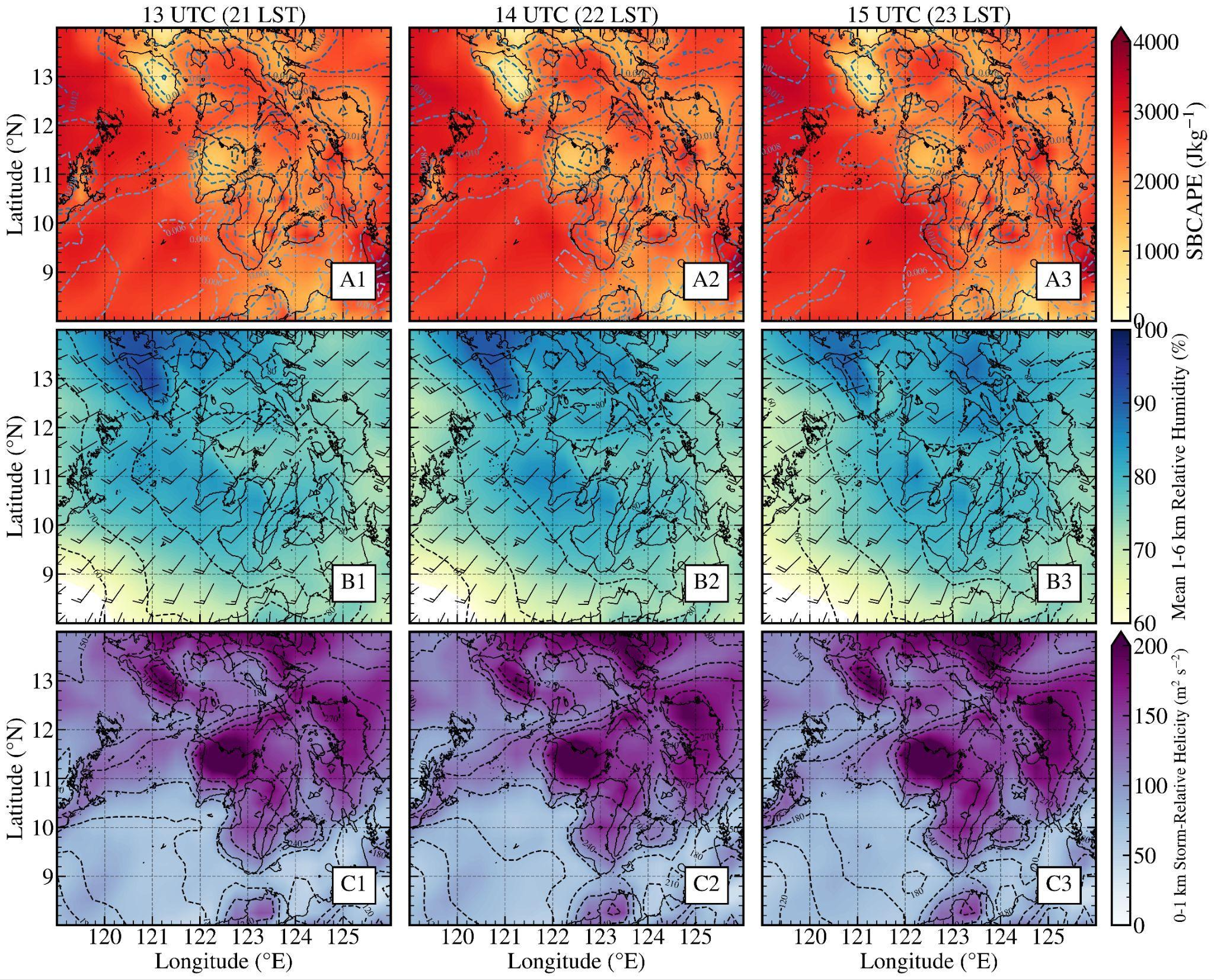}
\caption{Convective and kinematic environment of the Visayas before (1; 13 UTC/21 LST), during (2; 14 UTC/22 LST), and after (3; 15 UTC/23 LST) the event. (a) CAPE (shaded; \Jkg) and 0--1 km streamwise vorticity (blue dashed contours; \si). (b) Mean 1--6 km relative humidity (shaded; \%), mean 1--3 km relative humidity (black dashed contours; \%), and 10 m winds (kt). (c) SRH$_{01}$ (shaded; \mss) and SRH$_{03}$ (black dashed contours; \mss).}
\label{fig_4}
\end{figure*}

The mesoscale environment from 1300 to 1500 UTC (2100--2300 LST) is shown in Fig.~\ref{fig_4}. CAPE of 2500--3500 \Jkg\ covered the Sulu Sea, the Panay Gulf, and the Guimaras Strait (Fig.~\ref{fig_4}a1--a3), whereas values over the interior of Panay decreased to 1000--2000 \Jkg\ as the land surface cooled after sunset. Iloilo City, on the southeastern coast of Panay, lay along the gradient between these two regimes, with CAPE of $\sim$2000--2500 \Jkg. The persistence of large CAPE over the adjacent waters and along the coast reflects the continuous onshore advection of warm, moist marine air by the monsoon flow, which offset the nocturnal stabilization of the boundary layer over land. The $\omega_{s01}$ reached 0.010--0.016 \si\ over Panay and the adjacent straits, overlapping the zone of large CAPE along the southern coast. The mean 1--6 km relative humidity (RH$_{16}$) was 75\%--85\% over the Visayas (Fig.~\ref{fig_4}b1--b3), with RH$_{13}$ exceeding 80\%, and the 10 m winds were westerly to southwesterly at 10--15 kt. Such a moist free troposphere limits the dilution of updrafts by the entrainment of dry environmental air \citep{Peters2023b}, which favors the maintenance of strong updrafts in an environment with only moderate CAPE.

The SRH$_{01}$ (Fig.~\ref{fig_4}c1--c3) exhibited a maximum exceeding 200 \mss\ over northern and central Panay and the Guimaras Strait, with SRH$_{03}$ of 240--270 \mss. Iloilo City lay along the southern edge of this maximum, where the SRH$_{01}$ was $\sim$120--170 \mss. The SRH maximum remained nearly stationary from 1300 to 1500 UTC, indicating that the kinematic environment was sustained by the persistent monsoon flow rather than by a transient mesoscale feature. Collectively, the overlap of moderate CAPE, a moist free troposphere, and enhanced low-level SRH and streamwise vorticity along the southern coast of Panay constituted an environment supportive of supercells capable of producing tornadoes.

\begin{figure*}[!t]
\centering
\includegraphics[width=\textwidth,height=0.82\textheight,keepaspectratio]{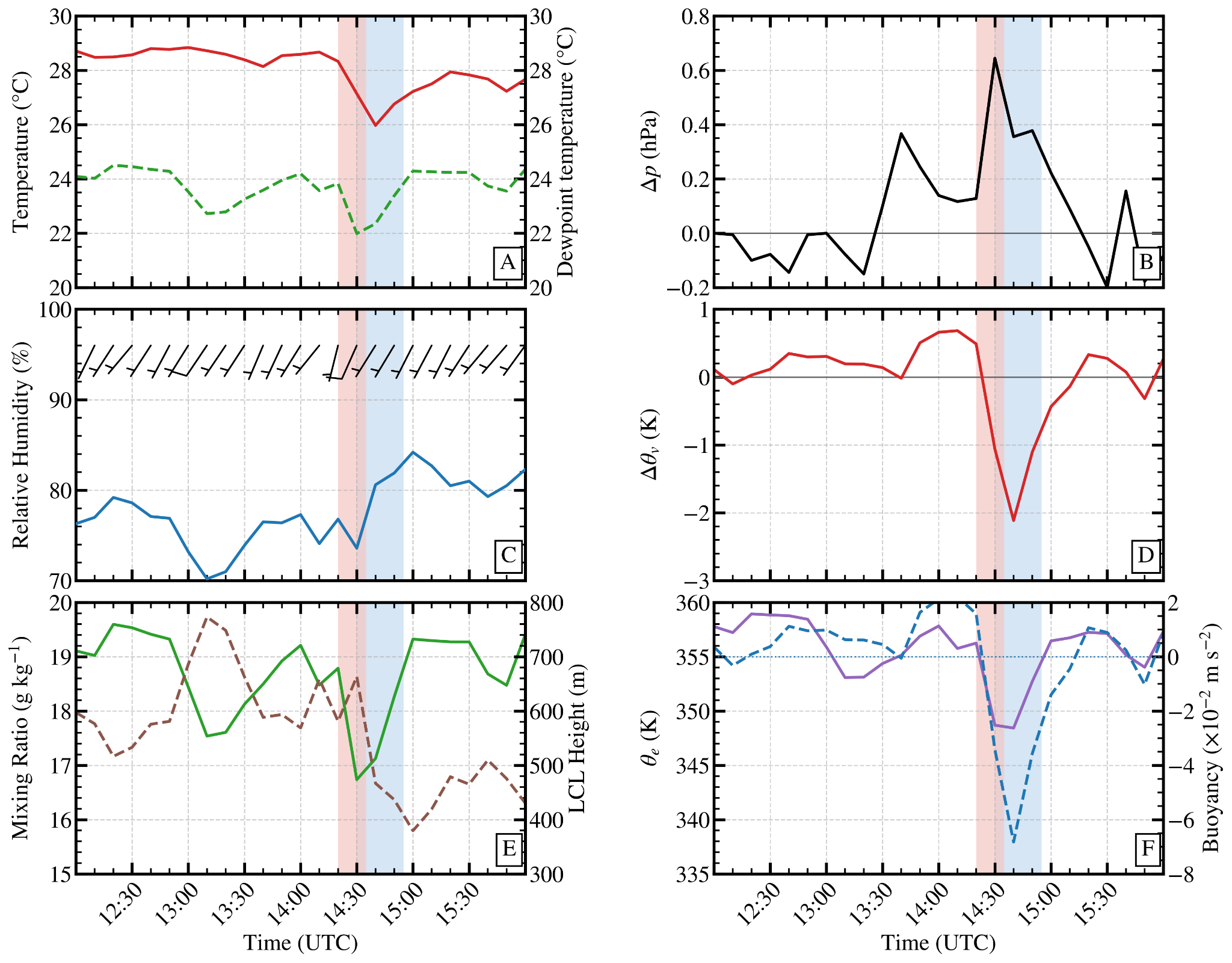}
\caption{Observations from the automatic weather station (AWS) collocated with S-ILO from 12 to 16 UTC at 10-min intervals. (a) Temperature (\dg C; red line) and Dewpoint Temperature (\dg C; green dashed line), (b) Pressure Perturbation (hPa; black line), (c) Relative Humidity (\%; blue line) with surface wind barbs, (d) $\Delta\theta_v$ (K; red line), (e) Mixing Ratio (g\,kg$^{-1}$; green line) and LCL (m; brown dashed line), (f) $\theta_e$ (K; purple) and Buoyancy ($\times$ 10$^{-2}$ m\,s$^{-2}$). Red shading marks the reported tornado period (1420--1435 UTC), and blue shading marks the cold-pool period ($\Delta\theta_v$ $\leq$ $-$0.5 K and $\Delta p'$ $>$ 0). Perturbations are departures from a centered 3-h running mean.}
\label{fig_5}
\end{figure*}

The surface observations at the S-ILO AWS confirm that the coastal boundary layer did not stabilize appreciably after sunset (Fig.~\ref{fig_5}). Between 1200 and 1420 UTC (2000--2220 LST), the temperature remained at 28.1--28.8\,\dg C, the dewpoint at 22.7--24.5\,\dg C, and $\theta_e$ at $\sim$353--359 K under light southwesterly winds. In other words, the near-surface air retained its daytime warmth and moisture rather than cooling radiatively. This persistence reflects the continuous warm, moist advection from the adjacent seas by the onshore monsoon flow beneath the LLJ. Consequently, the storm could continue to ingest SB parcels with negligible CIN (Table~\ref{Table_1}), while the strengthening jet aloft lengthened the low-level hodograph. Supercells that persist into the evening over the Great Plains are similarly characterized by increasing SRH and only small increases in inhibition during the nocturnal transition \citep{Gropp2018}.

\begin{figure*}[!t]
\centering
\includegraphics[width=\textwidth,height=0.82\textheight,keepaspectratio]{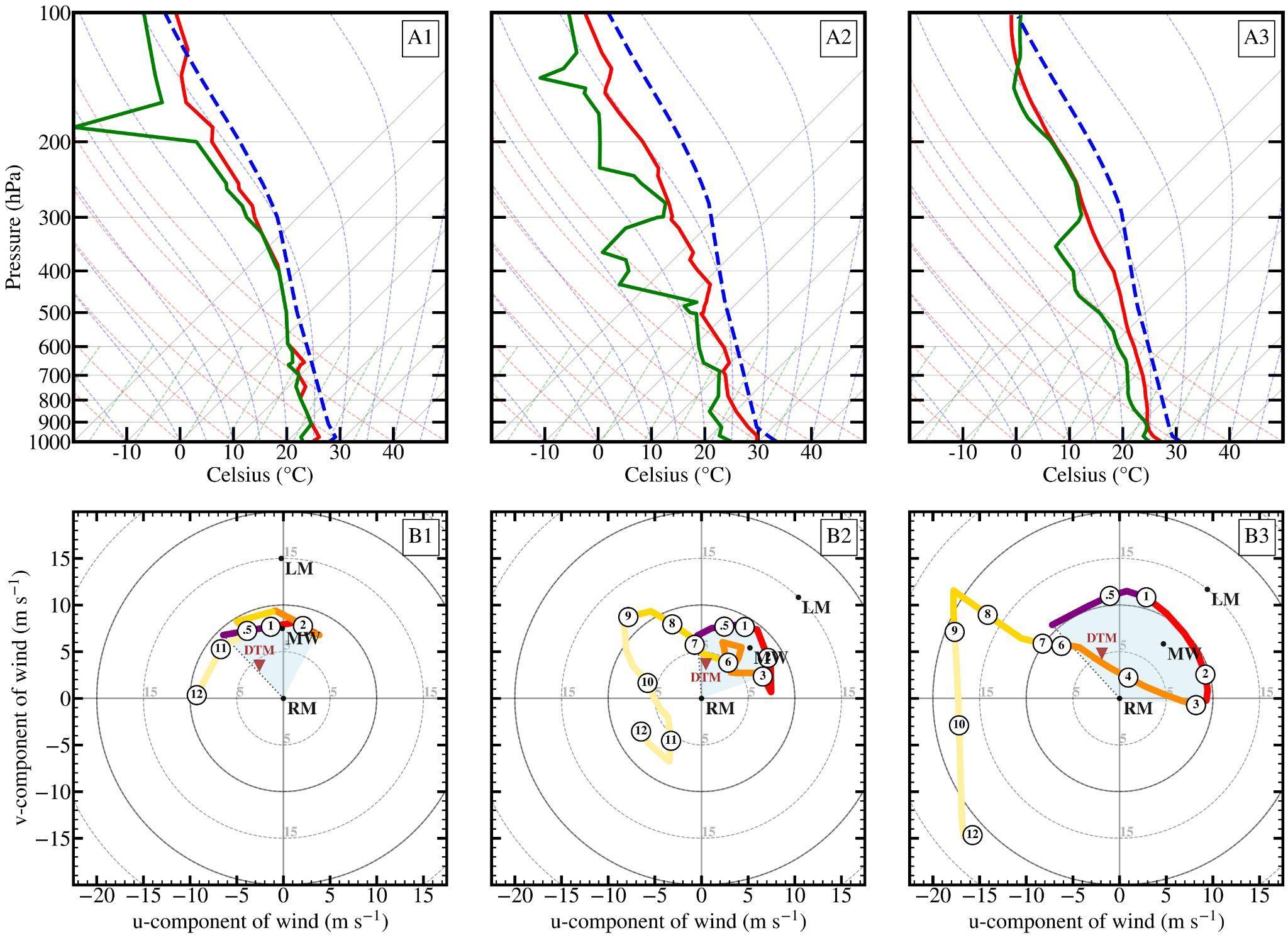}
\caption{Thermodynamic profiles and hodographs representing (1) the pre-convective environment from the PAGASA Iloilo radiosonde (WMO 98637) at 00 UTC/08 LST, (2) the environment $\sim$2 h before the event from the Mactan radiosonde (WMO 98646), $\sim$161 km east-southeast of S-ILO, at 12 UTC/20 LST, and (3) the ERA5 proximity profile at S-ILO at 14 UTC/22 LST. (a) Skew-T log-p diagrams of temperature (red;\,\dg C), dewpoint temperature (green;\,\dg C), and the non-entraining, irreversible mixed-phase MU parcel profile (blue dashed). (b) Storm-relative hodographs (\ms) relative to the 0--3 km Bunkers right-moving storm motion (RM), with height labels in km AGL, the left-moving (LM) and mean-wind (MW) motions, the deviant tornado motion (DTM), and the SRH$_{03}$ (blue shading).}
\label{fig_6}
\end{figure*}

The thermodynamic and kinematic evolution at the point scale is shown in Fig.~\ref{fig_6} and summarized in Table~\ref{Table_1}. The 0000 UTC (0800 LST) Iloilo sounding (Fig.~\ref{fig_6}a1) was nearly saturated from the surface to $\sim$300 hPa (RH$_{13}$ and RH$_{16}$ of $\sim$97\%), with weak lapse rates (LR$_{03}$ of 5.17 \,\dg C\,km$^{-1}$ and LR$_{36}$ of 5.57 \,\dg C\,km$^{-1}$), MUCAPE of 1260 \Jkg, and an LCL of 910 m. This moist-adiabatic profile was typical of the rainy morning conditions within the monsoon, and the accompanying hodograph (Fig.~\ref{fig_6}b1) was small, with an LLS of 5.2 \ms\ and an SRH$_{01}$ of only 40 \mss. By 1200 UTC (2000 LST), the Mactan sounding (Fig.~\ref{fig_6}a2) displayed a markedly more unstable profile, with MUCAPE of 2532 \Jkg, ECAPE of 1736 \Jkg, SBCIN of $-$21 \Jkg, an LCL of 622 m, and a steeper low-level lapse rate (LR$_{03}$ of 6.58 \,\dg C\,km$^{-1}$). A dry layer between $\sim$450 and 350 hPa reduced RH$_{16}$ to $\sim$82\%. Its hodograph (Fig.~\ref{fig_6}b2) had LLS and MLS of 10.0 and 16.1 \ms, respectively, an SRH$_{03}$ of 247 \mss, and near-surface horizontal vorticity that was almost purely streamwise ($\widetilde{\omega}_{s500}$ of 99\%). Although Mactan is $\sim$161 km east-southeast of S-ILO, it sampled the same monsoon airmass a few hours before the event.

The ERA5 proximity profile at 1400 UTC (2200 LST; Fig.~\ref{fig_6}a3), which is closest in time and space to the event, depicted MUCAPE of 2014 \Jkg, ECAPE of 1632 \Jkg, no CIN, and an LCL and LFC of only 295 m. The ratio of ECAPE to MUCAPE ($\sim$0.81) indicates that entrainment would have diluted the updraft only modestly, owing to the moist free troposphere (RH$_{13}$ of 85\% and RH$_{16}$ of 82\%) and the strong storm-relative inflow (mean \VSR\ of 11.9 \ms). The hodograph (Fig.~\ref{fig_6}b3) exhibited pronounced clockwise curvature in the lowest 3 km, with LLS and MLS of 10.5 and 17.6 \ms, respectively. The SRH$_{500}$, SRH$_{01}$, and SRH$_{03}$ were 79, 127, and 259 \mss, respectively, and the streamwiseness was $\sim$98\%--99\% in both the 0--500 m and 0--1 km layers. Above $\sim$3 km, however, the winds weakened and backed with height, so that the hodograph folded back toward its low-level segment. The DLS was therefore only 2.5 \ms, and WMAXSHEAR was only 156 \mss. This DLS is far below the $\sim$20 \ms\ commonly associated with supercells \citep[e.g.,][]{Thompson2003}. Nevertheless, the effective-layer formulations of the SCP (7.44) and STP (2.28), which measure the shear over the lower half of the MU parcel's depth rather than the fixed 0--6 km layer, remained large. The first-order estimate of low-level vertical vorticity from the tilting of streamwise vorticity (Eq.~(\ref{eq_zeta})) yielded $\zeta_{\mathrm{LLM}}$ of 0.015 \si, comparable to that at Mactan (0.016 \si) and nearly twice that at Iloilo at 0000 UTC (0.008 \si).

Overall, the three profiles describe a transition from a saturated, weakly unstable morning environment to an evening environment with moderate instability, very low LCLs, and strong, highly streamwise LLS. The combination of low LCLs ($<$ 300 m in ERA5), negligible CIN, and large low-level SRH and streamwise vorticity is characteristic of environments that support significant tornadoes \citep{Thompson2003,Coffer2019,Coffer2020}, whereas the weak DLS is atypical of such environments in the midlatitudes. As shown in Section~\ref{sec34}, the radar-derived wind profile further indicates that ERA5 underestimated the LLS in the near-storm environment.

\subsection{Storm Evolution and Life Cycle}\label{sec33}

\begin{figure*}[!t]
\centering
\includegraphics[width=\textwidth,height=0.82\textheight,keepaspectratio]{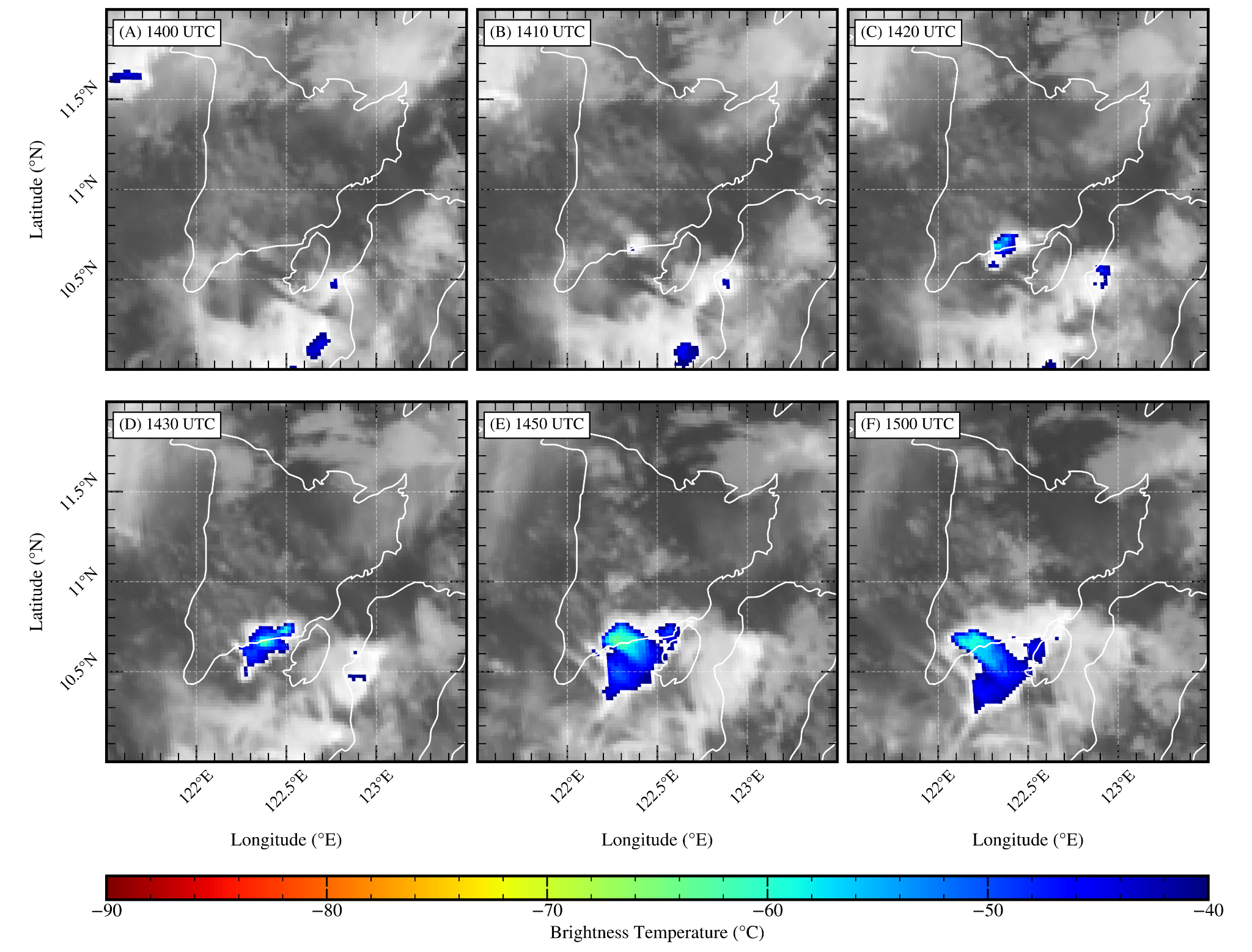}
\caption{Himawari-9 AHI 10.4-$\mu$m (B13) brightness temperature depicting the evolution of the 08 August 2026 tornadic storm at (a) 1400, (b) 1410, (c) 1420, (d) 1430, (e) 1450, and (f) 1500 UTC (2200--2300 LST). Brightness temperatures colder than $-$40\,\dg C are color-shaded. The 1440 UTC scan was unavailable.}
\label{fig_7}
\end{figure*}

Figure~\ref{fig_7} shows the evolution of the tornado-producing storm in 10-min Himawari-9 B13 imagery from 1400 to 1500 UTC (2200--2300 LST). Imagery before 1400 UTC is not shown, but the earlier stages of the storm's life cycle are described below for context.

At 1200 UTC (2000 LST; not shown in Fig.~\ref{fig_7}), the convective environment over Western Visayas was characterized by a warm, moist, and conditionally unstable airmass within the typhoon-enhanced southwest monsoon, as sampled by the Mactan sounding (Section~\ref{sec32}). This is consistent with Heavy Rainfall Warning No. 15, issued at 1200 UTC by the DOST-PAGASA Visayas PAGASA Regional Services Division (PAGASA-VisPRSD), which reported light to moderate rains with occasional heavy rains over Western Visayas, including Iloilo. As convection persisted over the region in the Himawari-9 imagery, PAGASA-VisPRSD issued a thunderstorm watch at 1400 UTC (2200 LST) for possible thunderstorm activity over parts of Western Visayas, including Negros and Iloilo.

By 1410 UTC (2210 LST), the parent cell was already evident in S-ILO reflectivity west-southwest of the radar, having approached from $\sim$21.5 km at 1400 UTC to $\sim$14.2 km from S-ILO at 1410 UTC, yet B13 depicted only a small, relatively warm cloud top along the southern coast of Panay (Fig.~\ref{fig_7}b). The first cloud top colder than $-$40\,\dg C appeared at 1420 UTC (Fig.~\ref{fig_7}c), with BTs of $\sim-$55\,\dg C over the southern coast of Panay near 10.7\dg N, 122.4\dg E, indicating that the updraft had by then penetrated the upper troposphere. At 1425 UTC (2225 LST), PAGASA-VisPRSD issued a thunderstorm advisory reporting that moderate to heavy rain showers with lightning and strong winds were being experienced over Pavia, Iloilo City, and Oton. The advisory added that these conditions could also affect Dumangas, Zarraga, Leganes, and Santa Barbara in Iloilo, and Buenavista and Jordan in Guimaras. This advisory was issued only minutes before the tornado damage began in Mandurriao (Section~\ref{sec36}).

At 1430 UTC (Fig.~\ref{fig_7}d), when the tornado was producing damage in Mandurriao and Jaro, the tornadic storm occupied the eastern tip of a cloud area colder than $-$40\,\dg C that extended $\sim$25--30 km west-southwestward along the southern coast of Panay, with a BT of only $\sim-$45 to $-$55\,\dg C over Iloilo City. By 1450 and 1500 UTC (Figs.~\ref{fig_7}e and \ref{fig_7}f), the tornadic storm appeared as a small, separate cloud top of $\sim-$45 to $-$50\,\dg C over Iloilo City and the northern Guimaras Strait, east of a much larger and colder cloud shield that had developed over southern Panay and expanded toward the south and southwest over the Panay Gulf, reaching a minimum BT of $\sim-$68\,\dg C $\sim$25 km west of Iloilo City at 1450 UTC. In the Mactan sounding, BTs of $-$45 to $-$55\,\dg C correspond to cloud tops at $\sim$12--13 km, whereas $-$68\,\dg C corresponds to $\sim$15 km and the cold-point tropopause lay at $\sim$16.7 km ($-$82.5\,\dg C). The tornadic storm was therefore low-topped relative to the surrounding deep convection, with cloud tops $\sim$2--3 km lower than those of the adjacent cloud shield and $\sim$4 km below the tropopause, and no overshooting top was evident. The southwestward expansion of the larger cloud shield is consistent with the northeasterly to easterly winds of $\sim$6--15 \ms\ between 200 and 150 hPa in the Mactan sounding. Low-topped tornadic supercells are common in tropical environments with moderate CAPE, weak DLS, and strong LLS, such as those of landfalling TCs \citep{McCaul1991}, and the shallow cloud top of the Iloilo storm agrees with the moderate CAPE and the low-level concentration of the shear in its environment. Because the updraft occupied only a few $\sim$2 km pixels, its BT may also have been warmed by sub-pixel effects, so these cloud-top heights are approximate. The tornado was thus produced by a relatively shallow storm whose cloud top did not stand out in the satellite imagery against the colder convection nearby, which underscores the importance of radar for detecting such storms at night.

\begin{figure*}[!t]
\centering
\includegraphics[width=\textwidth,height=0.82\textheight,keepaspectratio]{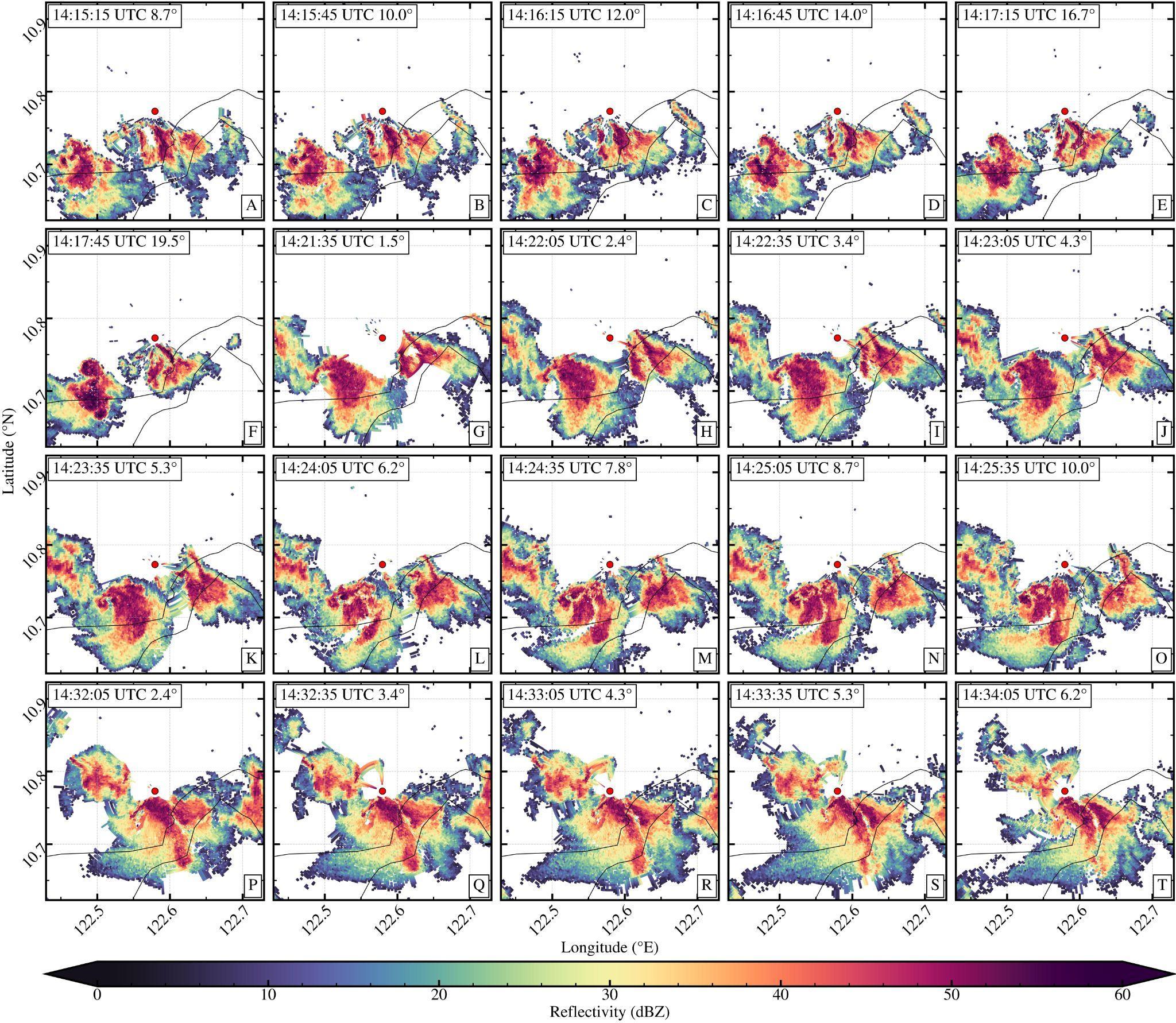}
\caption{Time sequence of Reflectivity (dBZ) PPIs within $\sim$15 km of the S-ILO radar, 14:15:15--14:34:05 UTC (22:15--22:34 LST), compiled from the 14:10, 14:20 and 14:30 UTC volume scans. Panels are in chronological order (A--T), with successive tilts $\sim$30 s apart: (A--F) 8.7--19.5\dg{} from the 14:10 UTC volume, (G--O) 1.5--10.0\dg{} from the 14:20 UTC volume and (P--T) 2.4--6.2\dg{} from the 14:30 UTC volume. Each panel is labeled with its sweep time (UTC) and elevation angle. The red dot marks the location of S-ILO.}
\label{fig_8}
\end{figure*}

\begin{figure*}[!t]
\centering
\includegraphics[width=\textwidth,height=0.82\textheight,keepaspectratio]{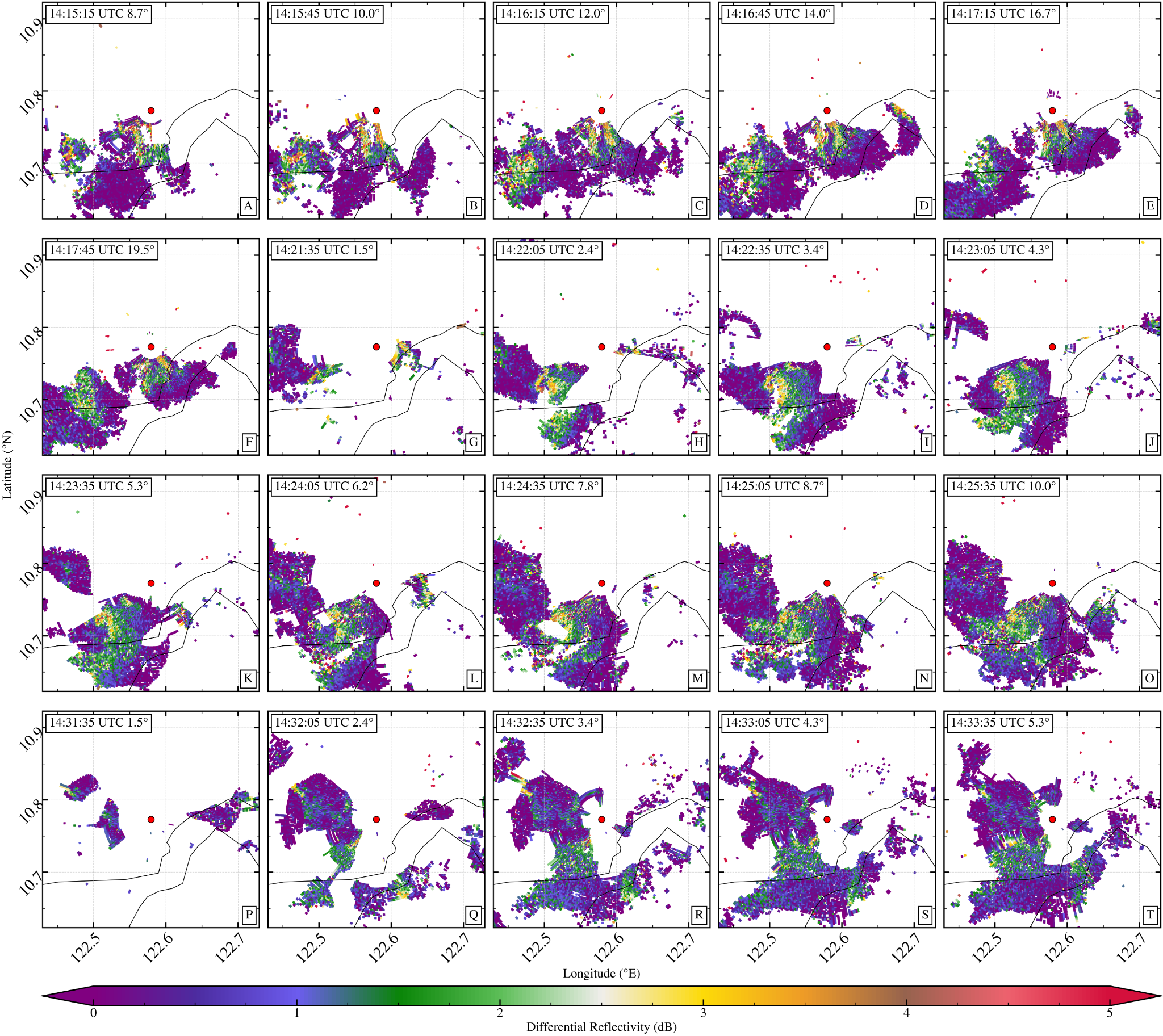}
\caption{Similar to Figure~\ref{fig_8}, but for Differential Reflectivity (dB) PPIs within $\sim$15 km of the S-ILO radar, 14:15:15--14:33:35 UTC (22:15--22:33 LST), compiled from the 14:10, 14:20 and 14:30 UTC volume scans. Panels are in chronological order (A--T), with successive tilts $\sim$30 s apart: (A--F) 8.7--19.5\dg{} from the 14:10 UTC volume, (G--O) 1.5--10.0\dg{} from the 14:20 UTC volume and (P--T) 1.5--5.3\dg{} from the 14:30 UTC volume. Each panel is labeled with its sweep time (UTC) and elevation angle.}
\label{fig_9}
\end{figure*}

The radar evolution of the storm is shown in Figs.~\ref{fig_8} and \ref{fig_9}, which compile the upper tilts of the 1410 UTC volume and the 1420 and 1430 UTC volumes into a time sequence with successive tilts $\sim$30 s apart. Between 1400 and 1430 UTC, the centroid of the $\geq$ 45 dBZ core moved from $\sim$21.5 km west-southwest of S-ILO to within $\sim$4 km of the radar, consistent with the observed storm motion of 13.7 \ms\ toward 076\dg{}. The upper tilts of the 1410 UTC volume (Figs.~\ref{fig_8}a--f), which sampled the tornadic storm between $\sim$2.1 and 4.7 km above radar level (ARL), show a core exceeding 50--55 dBZ near 10.69\dg N, 122.49\dg E at every tilt, indicating that large hydrometeors were sustained well above the cloud base by a strong updraft. At 14.0\dg{} and 16.7\dg{} (Figs.~\ref{fig_8}d and \ref{fig_8}e), this core was curled into a comma shape with a notch of weaker echo along its eastern flank, a configuration consistent with a rotating updraft at midlevels.

\begin{figure*}[!t]
\centering
\includegraphics[width=\textwidth,height=0.82\textheight,keepaspectratio]{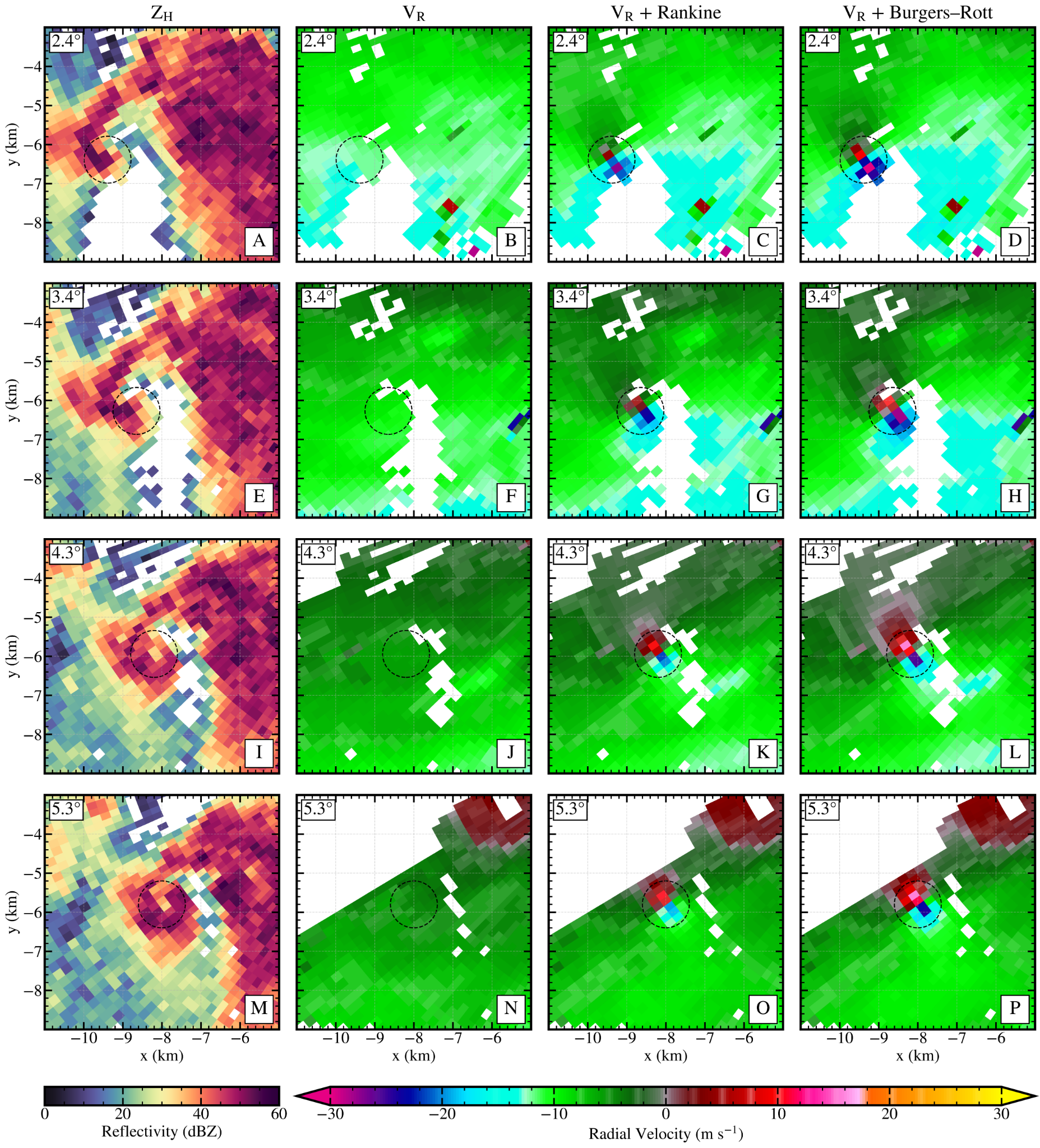}
\caption{1420 UTC S-ILO observations (\ZH\ and \VR) and simulated vortex signatures at the hook echo and weak echo hole of the Jaro tornadic supercell at elevation angles of 2.4\dg{}, 3.4\dg{}, 4.3\dg{} and 5.3\dg{}. Columns: (A, E, I, M) \ZH\ (dBZ), (B, F, J, N) observed \VR\ (\ms), (C, G, K, O) observed \VR\ plus a simulated cyclonic vortex with a Rankine profile, and (D, H, L, P) a Burgers--Rott profile. They are sampled with the radar's beam (1.33\dg{}, 1\dg{} azimuthal sampling) and 250 m gate spacing. The simulated vortices have $R_c$ = 100 m and \Vrot\ = 40 \ms\ and are centered on the WEH (dashed circles); $x$ and $y$ are the distances (km) east and north of S-ILO.}
\label{fig_10}
\end{figure*}

By the 1420 UTC volume (Figs.~\ref{fig_8}g--o), the storm had acquired a classic low-level supercell structure. At the lowest usable tilts (1.5\dg{}--4.3\dg{}, $\sim$0.3--0.8 km ARL), the core exceeded 55 dBZ, and a hook-shaped appendage developed along its southwestern flank, with a WEH at the tip of the hook (Figs.~\ref{fig_10}a, \ref{fig_10}e, \ref{fig_10}i, and \ref{fig_10}m). The WEH could be followed from 2.4\dg{} to 10.0\dg{} (0.48--1.42 km ARL) and was enclosed by a ring of 50--55 dBZ, the reflectivity structure commonly described as a debris ball. Its center shifted $\sim$3.5 km toward the east-northeast between the 2.4\dg{} and 10.0\dg{} scans, which were collected $\sim$3.6 min apart. When each tilt is shifted back to the 2.4\dg{} scan time using the storm motion, however, the WEH centers from 2.4\dg{} to 10.0\dg{} lie within $\sim$0.5 km of one another ($x$ $\approx$ $-$9.4 to $-$8.8 km and $y$ $\approx$ $-$6.4 to $-$6.1 km relative to S-ILO), indicating that the WEH column was nearly vertical in the lowest $\sim$1.4 km and that the apparent tilt was produced by the translation of the storm during the volume scan. At 6.2\dg{}--10.0\dg{} ($\sim$0.8--1.4 km ARL; Figs.~\ref{fig_8}l--o), a weak echo vault bounded on its southern side by a curved band of high reflectivity was evident near 10.71\dg N, 122.52\dg E, indicating a bounded weak echo region (BWER) above the hook where the inflow ascended into the updraft. In the 1430 UTC volume (Figs.~\ref{fig_8}p--t), the storm core had reached the near range of the radar, the hook and WEH were located over Jaro at 0.24--0.43 km ARL, and a band of moderate reflectivity trailed to the south-southwest along the rear flank.

The \ZDR\ fields (Fig.~\ref{fig_9}) reveal several polarimetric features of the supercell. In the 1410 UTC volume (Figs.~\ref{fig_9}a--f), \ZDR\ of 1.5--3 dB extended through the $\sim$2--4.7 km layer within and adjacent to the reflectivity core, indicating large, oblate raindrops lofted by the updraft to just below the environmental 0\,\dg C level ($\sim$5.1 km MSL in the Mactan sounding). Whether this enhancement continued above the melting level as a \ZDR\ column could not be determined because the highest tilt (19.5\dg{}) reached only $\sim$4.7 km at that range. In the 1420 UTC volume, an arc of \ZDR\ of 2.5--3.5 dB lined the western and southwestern reflectivity gradient adjacent to the hook echo at 2.4\dg{}--5.3\dg{} (Figs.~\ref{fig_9}h--k). This \ZDR\ arc is attributed to the size sorting of hydrometeors by strong storm-relative winds and low-level helicity, which concentrates large drops along the inflow side of the forward-flank precipitation \citep{Kumjian2008,Kumjian2009}. At 6.2\dg{}--8.7\dg{} (Figs.~\ref{fig_9}l--n), enhanced \ZDR\ of 2--3 dB also surrounded the weak echo vault identified in the reflectivity field, forming a \ZDR\ ring around the BWER that marks the periphery of the updraft \citep{Kumjian2008}. In the 1430 UTC volume (Figs.~\ref{fig_9}p--t), \ZDR\ near the storm core was mostly $<$ 1.5 dB, and the near-range sampling did not reveal a coherent arc.

\begin{figure*}[!t]
\centering
\includegraphics[width=\textwidth,height=0.82\textheight,keepaspectratio]{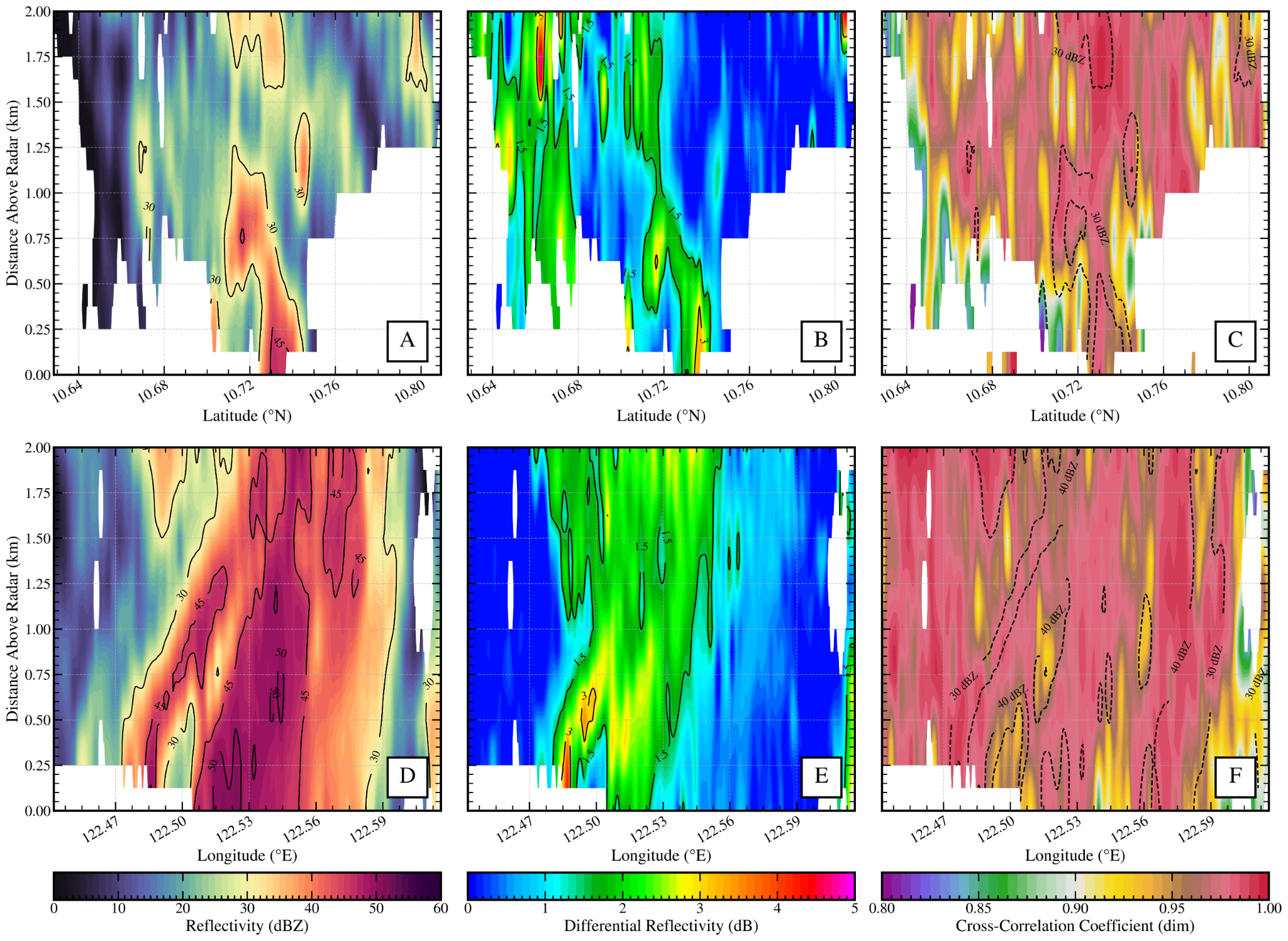}
\caption{Vertical cross-sections of the Jaro tornadic supercell at 2220 LST (1420 UTC), 08 August 2026, along the two transects described in Section~\ref{sec2323}: (a--c) south--north from 10.63\dg N, 122.50\dg E to 10.81\dg N, 122.50\dg E, and (d--f) west--east from 10.71\dg N, 122.44\dg E to 10.71\dg N, 122.61\dg E. (a, d) Reflectivity (dBZ), (b, e) Differential Reflectivity (dB), and (c, f) Cross-Correlation Coefficient (dimensionless). For Panels A and D, solid contours mark reflectivity at 30, 45 and 50 dBZ, while differential reflectivity (b and e) at 1.5 and 3 dB; dashed contours in panels C and F demarcate the reflectivity at 30, 40 and 50 dBZ.}
\label{fig_11}
\end{figure*}

The vertical structure of the WEH at 1420 UTC is shown in the cross sections of Fig.~\ref{fig_11}. In the south--north section along 122.499\dg E (Figs.~\ref{fig_11}a--c), a $\geq$ 45 dBZ column extended from the surface to $\sim$1 km ARL near 10.72\dg{}--10.74\dg N and tilted slightly northward with height, while missing and low-reflectivity gates occupied the area immediately to its south (10.70\dg{}--10.715\dg N) below $\sim$0.6 km, i.e., at the location of the WEH. \ZDR\ exceeded 3 dB at the base of the high-reflectivity column and was 1.5--2 dB through the lowest $\sim$0.75 km. In the west--east section along 10.717\dg N (Figs.~\ref{fig_11}d--f), the $\geq$ 45 dBZ core sloped eastward with height from $\sim$122.50\dg E near the surface to $\sim$122.53\dg E at 1.25 km ARL, and a \ZDR\ maximum exceeding 3 dB occupied the western (inflow) edge of the core between 0.3 and 0.65 km ARL. In both sections, \rhv\ decreased to $\sim$0.85--0.93 in the lowest $\sim$0.6 km around the WEH, compared with $>$ 0.97 in the surrounding rain. Such reductions may arise from a mixture of large drops and hail, from non-uniform beam filling across the strong reflectivity gradients of the hook, or from small amounts of debris mixed with rain. Because all gates with \rhv\ $<$ 0.80 had been removed by the quality control applied for rainfall estimation (Section~\ref{sec2322}), the absence of lower values cannot be used to exclude a TDS. Likewise, the missing gates at the center of the WEH are consistent both with genuinely weak echo in a hydrometeor-depleted vortex core and with debris gates removed by the threshold.

\begin{figure*}[!t]
\centering
\includegraphics[width=\textwidth,height=0.82\textheight,keepaspectratio]{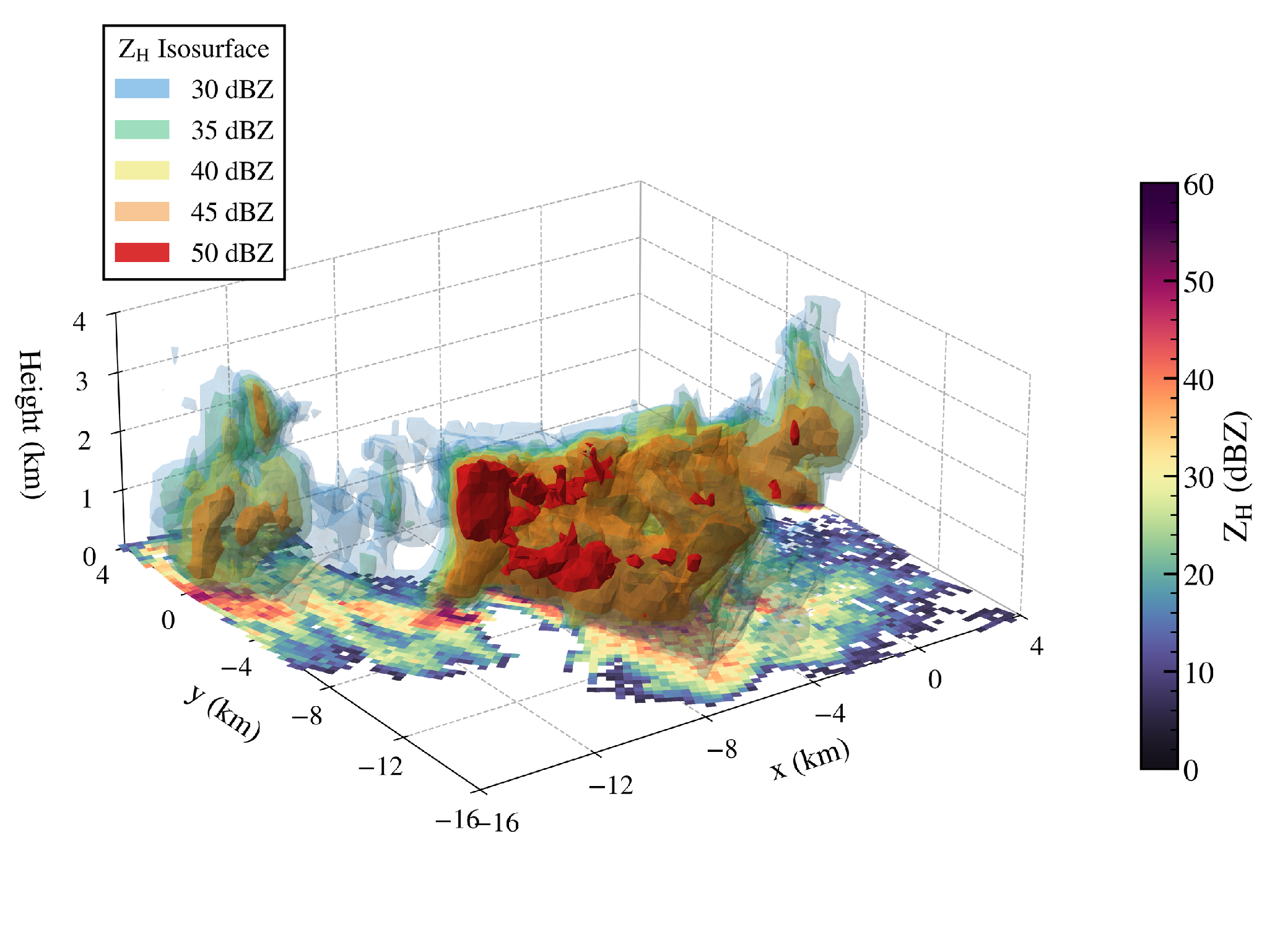}
\caption{Three-dimensional reflectivity structure of the Jaro tornadic supercell in the 1420 UTC (2220 LST) S-ILO volume, viewed from the southwest. Isosurfaces of \ZH\ at 30 (blue), 35 (green), 40 (yellow), 45 (orange), and 50 dBZ (red) are drawn above the 2.4\dg{} PPI (floor; dBZ). Each tilt was shifted to the 2.4\dg{} scan time using the storm motion (13.7 \ms\ toward 076\dg{}), objectively analyzed onto a 0.25 km $\times$ 0.25 km $\times$ 0.2 km grid, and drawn with a vertical exaggeration of 2; $x$ and $y$ are the distances (km) east and north of S-ILO.}
\label{fig_12}
\end{figure*}

The three-dimensional structure of the storm in the 1420 UTC volume is shown in Fig.~\ref{fig_12}, in which the tilts were shifted to the 2.4\dg{} scan time and objectively analyzed onto a Cartesian grid (Section~\ref{sec2323}). The 45 dBZ isosurface encloses the main core of the storm from near the surface to $\sim$2.5--3 km ARL, above which the isosurfaces are truncated because the highest tilt (19.5\dg{}) reached only $\sim$3.9 km at this range. Additional cells to the northwest and east of the tornadic storm are also evident. The most conspicuous feature is a compact core exceeding 55 dBZ (maximum of $\sim$60 dBZ) between $\sim$1.6 and 2.6 km ARL on the western (rear) flank of the storm, centered $\sim$0.4 km north of the WEH column at $x$ $\approx$ $-$9.0 km and $y$ $\approx$ $-$5.8 km, which is enclosed by the large 50 dBZ (red) isosurface on the western flank of the storm in Fig.~\ref{fig_12}. Directly beneath this core, the reflectivity decreased to $\sim$15--35 dBZ within the WEH below $\sim$1.2 km, while a ring of 50--52 dBZ enclosed the WEH on its western and eastern sides between $\sim$0.2 and 1.0 km ARL, as also seen in the vertical cross sections through the WEH (Fig.~\ref{fig_11}). Thus, the WEH and the surrounding debris-ball ring were capped by an intense, elevated core. Because the same reflectivity structure can be produced by hydrometeors, the debris interpretation of the ring requires polarimetric confirmation \citep{Ryzhkov2005,Bodine2013}, which the quality control applied to the data precluded.

The location of the elevated core, on the rear flank of the storm and immediately above the hook and WEH, resembles the descending reflectivity core (DRC) documented by \citet{Rasmussen2006}, a reflectivity protuberance pendant from the rear-flank echo overhang at 3--6 km that descends into the hook appendage, often before tornadogenesis. The preceding volume supports such a descent. At 1410 UTC, in the storm-relative position corresponding to the hook at 1420 UTC, $\geq$ 55 dBZ echo occupied $\sim$2.4--5.4 km ARL, while the reflectivity below $\sim$1.4 km did not exceed 44 dBZ and no hook was present. By 1420 UTC, the core had lowered to $\sim$1.6--2.6 km, and the hook, WEH, and debris-ball ring had formed beneath it. By 1430 UTC, when the tornado was producing damage in Jaro, the hook region below $\sim$2.2 km contained 46--51 dBZ without a distinct elevated maximum, consistent with the core having descended and spread through the hook, although the proximity of the storm to the radar limited the sampled depth to $\sim$2 km. This sequence, a lowering of the rear-flank core between 1410 and 1420 UTC followed by the onset of damage $\sim$10 min later, is consistent with the timing reported by \citet{Kennedy2007}, who found that 59\% of 64 supercells produced DRCs and that 30\% of these DRCs occurred from 10 min before to 5 min after tornadogenesis. The limitations of this inference are discussed in Section~\ref{sec4}

\subsection{Radar-Derived Near-Storm Kinematics and Vortex Simulation}\label{sec34}

\begin{figure*}[!t]
\centering
\includegraphics[width=\textwidth,height=0.82\textheight,keepaspectratio]{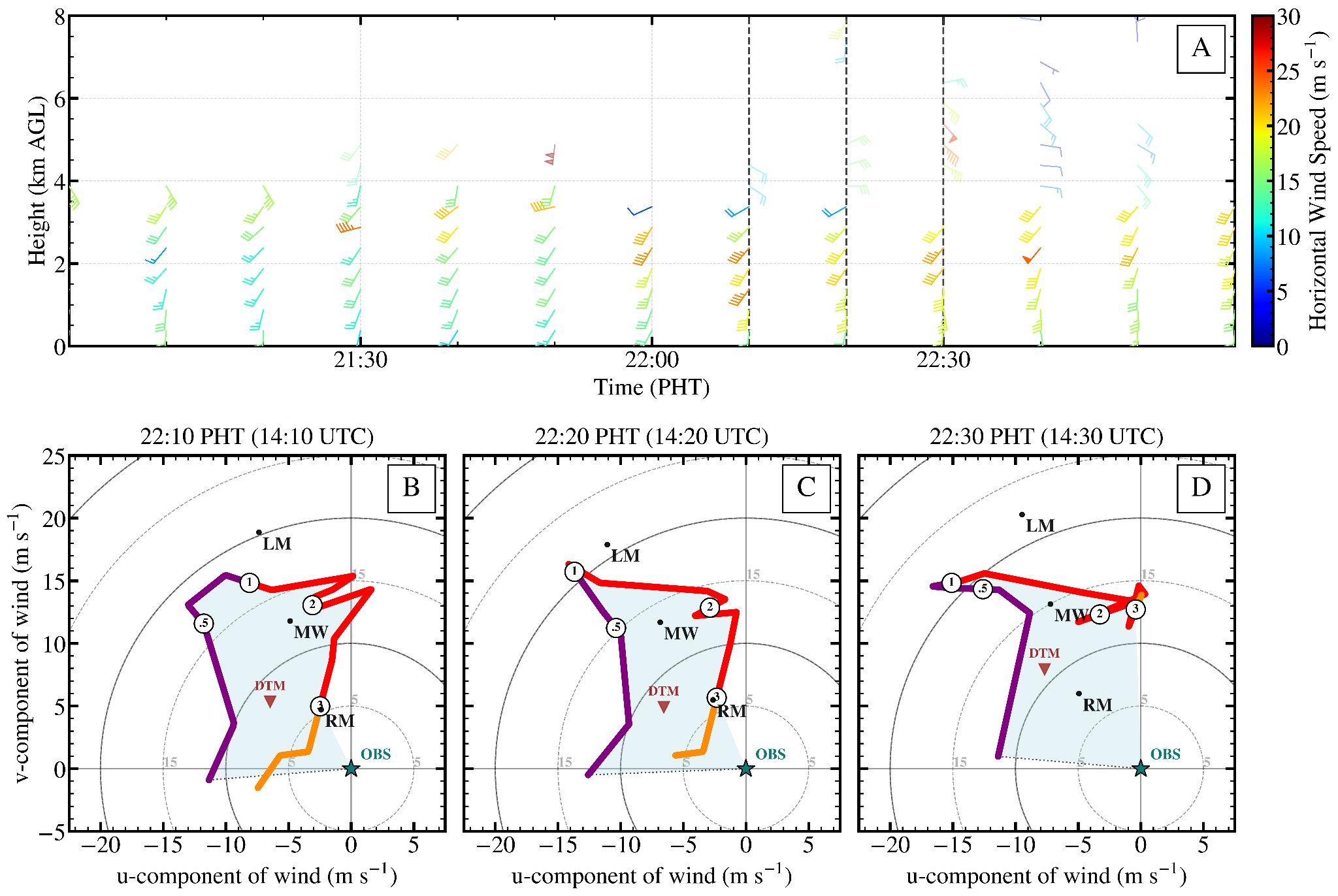}
\caption{(a) VVP-derived wind profile (barbs colored by horizontal wind speed; \ms) from 2100 to 2300 LST (1300--1500 UTC) derived from S-ILO. (b--d) Radar-derived hodographs at 1410, 1420, and 1430 UTC (2210, 2220, and 2230 LST; black dashed lines in panel a), plotted relative to the observed storm motion (OBS; 13.7 \ms\ toward 076\dg{}), with height labels in km AGL. The \BRM, Bunkers left-moving (LM), and mean-wind (MW) motions, the deviant tornado motion (DTM), and the SRH$_{03}$ (blue shading) are also shown.}
\label{fig_13}
\end{figure*}

The VVP-derived wind profile from 2100 to 2300 LST (1300--1500 UTC; Fig.~\ref{fig_13}a) documents the evolution of the low-level flow upstream of the storm. Before $\sim$2130 LST, the winds in the lowest 3 km were southwesterly at $\sim$10--15 \ms. From $\sim$2130 to 2230 LST, the winds between $\sim$1 and 3 km AGL strengthened to 20--25 \ms, whereas the 10 m wind measured at the S-ILO AWS remained only $\sim$2--5 \ms, so that the LLS increased as the storm approached Iloilo City. This intensification is consistent with the nocturnal wind maximum depicted by ERA5 near 1.5--3 km (Fig.~\ref{fig_3}) and resembles the evening strengthening of the LLJ that enhances LLS and SRH in nocturnal tornado environments elsewhere \citep{Kis2010,Mead2011}.

The radar-derived hodographs at 1410 and 1420 UTC (Figs.~\ref{fig_13}b and \ref{fig_13}c; Table~\ref{Table_3}), plotted relative to the observed storm motion, exhibited a long, clockwise-curved 0--1 km segment, with an LLS of 16.1--16.3 \ms\ ($\sim$31 kt) and a mean \VSR\ of 14.5--14.7 \ms. Relative to the observed motion, the SRH$_{500}$, SRH$_{01}$, and SRH$_{03}$ were 115, 211, and 292 \mss\ at 1410 UTC and 115, 124, and 265 \mss\ at 1420 UTC, respectively, with an $\omega_{s500}$ of $\sim$0.019 \si\ and a streamwiseness of 74\%--76\%. Critical angles of 48\dg{}--60\dg{}, computed against the \BRM, indicate a substantial streamwise component of the near-ground vorticity, although less than the $\sim$90\dg{} that corresponds to purely streamwise vorticity \citep{Esterheld2008}. These values exceed their ERA5 counterparts at 1400 UTC (LLS of 10.5 \ms\ and SRH$_{01}$ of 127 \mss), confirming that ERA5 underestimated the LLS near the storm, as is common for reanalyses within the boundary layer and near surface boundaries \citep{King2019}. Near-ground SRH and streamwise vorticity of this magnitude are among the best discriminators between tornadic and nontornadic supercell environments in the United States \citep{Coffer2019,Coffer2020}.

The \BRM\ computed from the 0--3 km layer (13.5--13.8 \ms\ from 230\dg{}--234\dg{}) closely matched the observed speed but was $\sim$20\dg{} to the left of the observed motion (13.7 \ms\ toward 076\dg{}), i.e., the storm moved to the right of the predicted supercell motion by 5--6 \ms. A rightward deviation of this kind increases the storm-relative inflow and SRH available to the storm and has been associated with tornadic supercells \citep{Bunkers2022}. In this case, the SRH$_{01}$ at 1410 UTC increased from 158 \mss\ relative to the \BRM\ to 211 \mss\ relative to the observed motion. By 1430 UTC (Fig.~\ref{fig_13}d), the SRH$_{01}$ decreased to 73 \mss\ and the 0--500 m streamwiseness to 55\%. This hodograph, however, does not represent the inflow environment, because the storm core was then within $\sim$4 km of the radar and the AWS wind used as its lowest level had been modified by the storm's outflow (Section~\ref{sec35}). The retrieved SRH values carry an uncertainty of approximately $\pm$30--50 \mss.

Despite this favorable kinematic environment, the observed radial velocity at the WEH did not resolve a tornado-scale couplet. At 1420 UTC, the sign-corrected radial velocity around the WEH was smooth and predominantly inbound at $\sim$10--15 \ms\ from 2.4\dg{} to 5.3\dg{} (Figs.~\ref{fig_10}b, \ref{fig_10}f, \ref{fig_10}j, and \ref{fig_10}n). The storm-relative velocity showed the correct cyclonic sense, with outbound flow on the northwestern side and inbound flow on the southeastern side of a broad shear zone, but the largest cyclonic azimuthal velocity difference within 1 km of the WEH was only 4.0, 4.5, and 4.8 \ms\ at 2.4\dg{}, 3.4\dg{}, and 4.3\dg{}, respectively, and the gate-to-gate difference was only 1.0--1.5 \ms. These values are well below the detection thresholds derived from the storm core (7.9--12.2 and 4.7--6.3 \ms, respectively) and lie between the 10th and 45th percentiles of the storm-wide distribution. At 1430 UTC, the cyclonic shear strengthened aloft ($\Delta V$ of $\sim$17 \ms\ at 5.3\dg{}), but the 2.4\dg{} circle enclosing the WEH contained no valid velocity gates.

Figures~\ref{fig_10}c, \ref{fig_10}g, \ref{fig_10}k, and \ref{fig_10}o show the same observations with a cyclonic Rankine vortex ($R_c$ = 100 m, \Vrot\ = 40 \ms) superimposed at the WEH, and Figs.~\ref{fig_10}d, \ref{fig_10}h, \ref{fig_10}l, and \ref{fig_10}p show the equivalent Burgers--Rott vortex. Sampled through the S-ILO beam without the processor filter, either vortex would have produced a compact couplet of $\sim$30--40 \ms\ across adjacent gates at every tilt from 2.4\dg{} to 5.3\dg{}, centered within the WEH and readily identifiable as a TVS \citep{Brown1978}. The Burgers--Rott profile, with its smoother core and broader outer flow, yields a couplet of similar magnitude that extends over a few more gates. Hence, a tornado consistent with the surveyed damage would have been conspicuous in unfiltered data, and its absence from the archived field is attributable mainly to the processing rather than to the absence of a vortex.

The injection--recovery test quantifies this attribution. With the fitted processor filter, vortices with $R_c$ $\geq$ 150 m and \Vrot\ $\geq$ 20 \ms\ would have been detected at $\geq$ 90\% of the injected positions and are therefore excluded, whereas narrower vortices could have remained hidden up to \Vrot\ of $\sim$30 \ms\ for $R_c$ = 100 m, $\sim$40 \ms\ for $R_c$ = 75 m, and $\sim$60 \ms\ for $R_c$ = 50 m. In unfiltered data, the corresponding detection limits would have been only 12, 16, and 24 \ms, respectively. Combined with the damage rating (Eq.~(\ref{eq_bayes})), the posterior for $R_c$ = 100 m favors \Vrot\ $\approx$ 35--45 \ms, i.e., the lower end of the range permitted by the damage, because larger values become increasingly difficult to conceal. The width of the damage path (Section~\ref{sec36}) independently constrains $R_c$ to $\sim$30--115 m, so that the radar non-detection and the damage survey are mutually consistent with a narrow tornado of moderate rotational velocity translating at $\sim$14 \ms.

\subsection{Surface Observations of the Storm Outflow}\label{sec35}

The S-ILO AWS, located $\sim$5--7 km north to north-northeast of the damage path, recorded the pre-storm and outflow conditions (Fig.~\ref{fig_5}). From 1200 to 1420 UTC (2000--2220 LST), the surface air was warm and humid, with temperatures of 28.1--28.8\,\dg C, dewpoints of 22.7--24.5\,\dg C, relative humidity of 70\%--79\%, mixing ratios of 17.5--19.6 g\,kg$^{-1}$, and $\theta_e$ of $\sim$353--359 K. The corresponding SB LCL ranged from $\sim$520 to $\sim$760 m, higher than the $\sim$300 m depicted by ERA5 but still low. Light southwesterly winds of 2--3 \ms\ prevailed, consistent with the onshore monsoon flow.

The outflow arrived at 1430 UTC (2230 LST), coincident with the passage of the storm core $\sim$4 km south of the radar. Between 1420 and 1440 UTC, the temperature fell by 2.4 K (from 28.3 to 26.0\,\dg C), the dewpoint decreased from 23.8 to 22.0\,\dg C, and $\theta_e$ decreased by $\sim$7.8 K (from 356.3 to 348.4 K). The pressure perturbation peaked at +0.64 hPa at 1430 UTC, the virtual potential temperature perturbation reached $-$2.1 K at 1440 UTC, and the corresponding buoyancy was $-$0.068 m\,s$^{-2}$. The 10-min mean wind increased to 4.7 \ms\ from 204\dg{}, and no rainfall was recorded by the gauge during the event. The cold pool persisted from 1430 to 1450 UTC, after which the temperature and $\theta_e$ recovered and the LCL lowered to $\sim$380--470 m as the surface air moistened. The density-current speed estimated from the pressure and density perturbations was $\sim$8--11 \ms, slower than the storm motion of 13.7 \ms.

The modest $\theta_v$ deficit ($\sim$1--2 K) and the small negative buoyancy indicate that the outflow sampled by the AWS was relatively warm and only weakly negatively buoyant. \citet{Markowski2002} found that the rear-flank outflow of tornadic supercells, particularly of those producing significant tornadoes, tended to have smaller $\theta_v$ and $\theta_e$ deficits than that of nontornadic supercells, because weakly negatively buoyant air can be more readily lifted and stretched by the low-level updraft. Because the AWS sampled the outflow $\sim$4 km from the storm core and $\sim$5--7 km from the tornado path, it is not possible to determine whether this air originated from the rear-flank or the forward-flank downdraft. Nevertheless, the observations indicate that the near-surface outflow of the Iloilo storm was not strongly negatively buoyant, consistent with a moist, low-LCL environment in which the evaporative cooling of rain is limited.

\subsection{Damage Analysis}\label{sec36}

\begin{figure*}[!t]
\centering
\includegraphics[width=\textwidth,height=0.82\textheight,keepaspectratio]{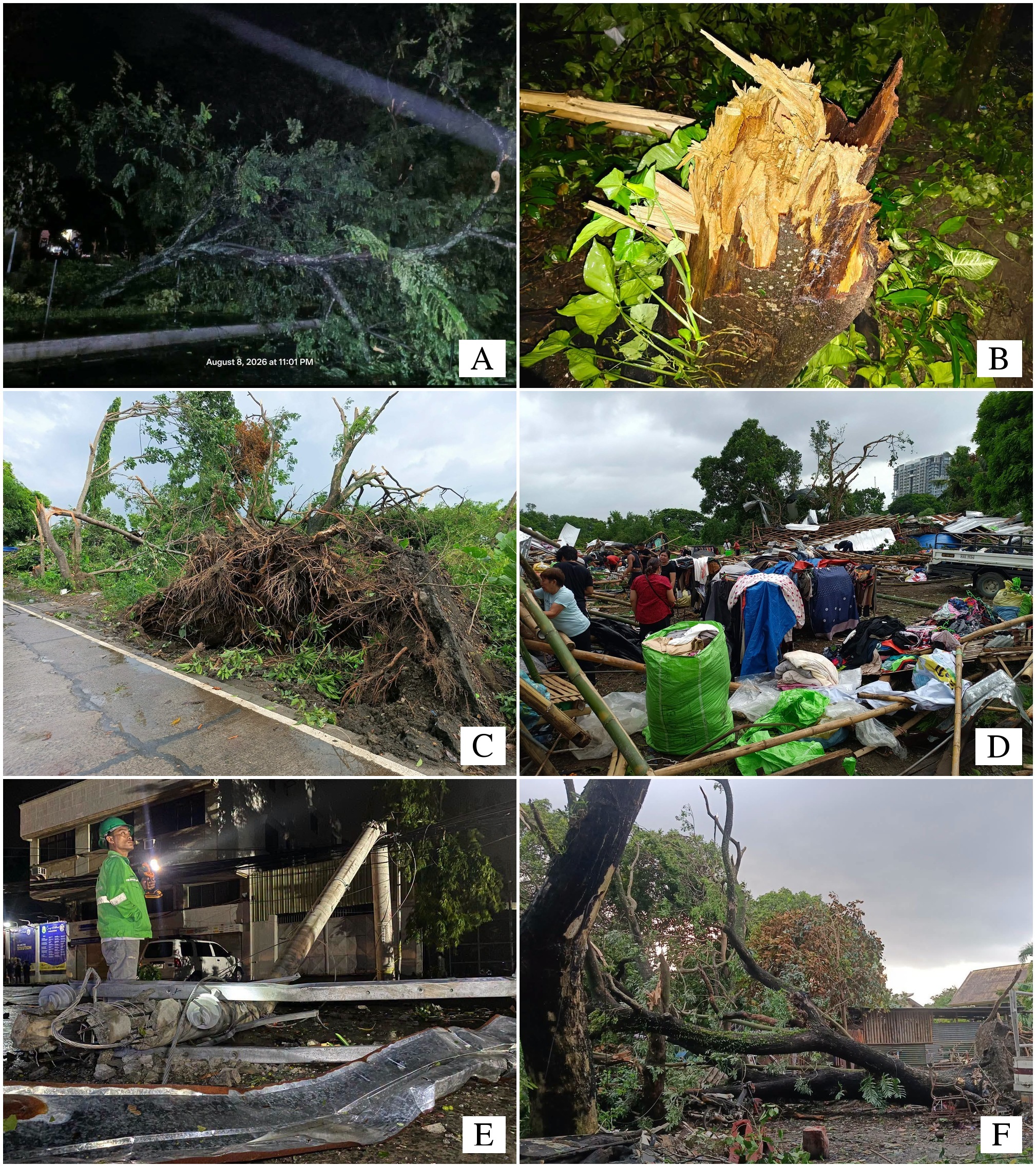}
\caption{Representative damage from the Iloilo tornado, ordered left to right and top to bottom: (a) a large tree toppled across a road; (b) a snapped hardwood trunk; (c) an uprooted tree with its root plate exposed; (d) collapsed bamboo-framed stalls and scattered roofing; (e) a snapped concrete utility pole with downed lines and roofing sheets; and (f) snapped and uprooted large trees adjacent to a building.}
\label{fig_14}
\end{figure*}

\begin{figure*}[!t]
\centering
\includegraphics[width=\textwidth,height=0.82\textheight,keepaspectratio]{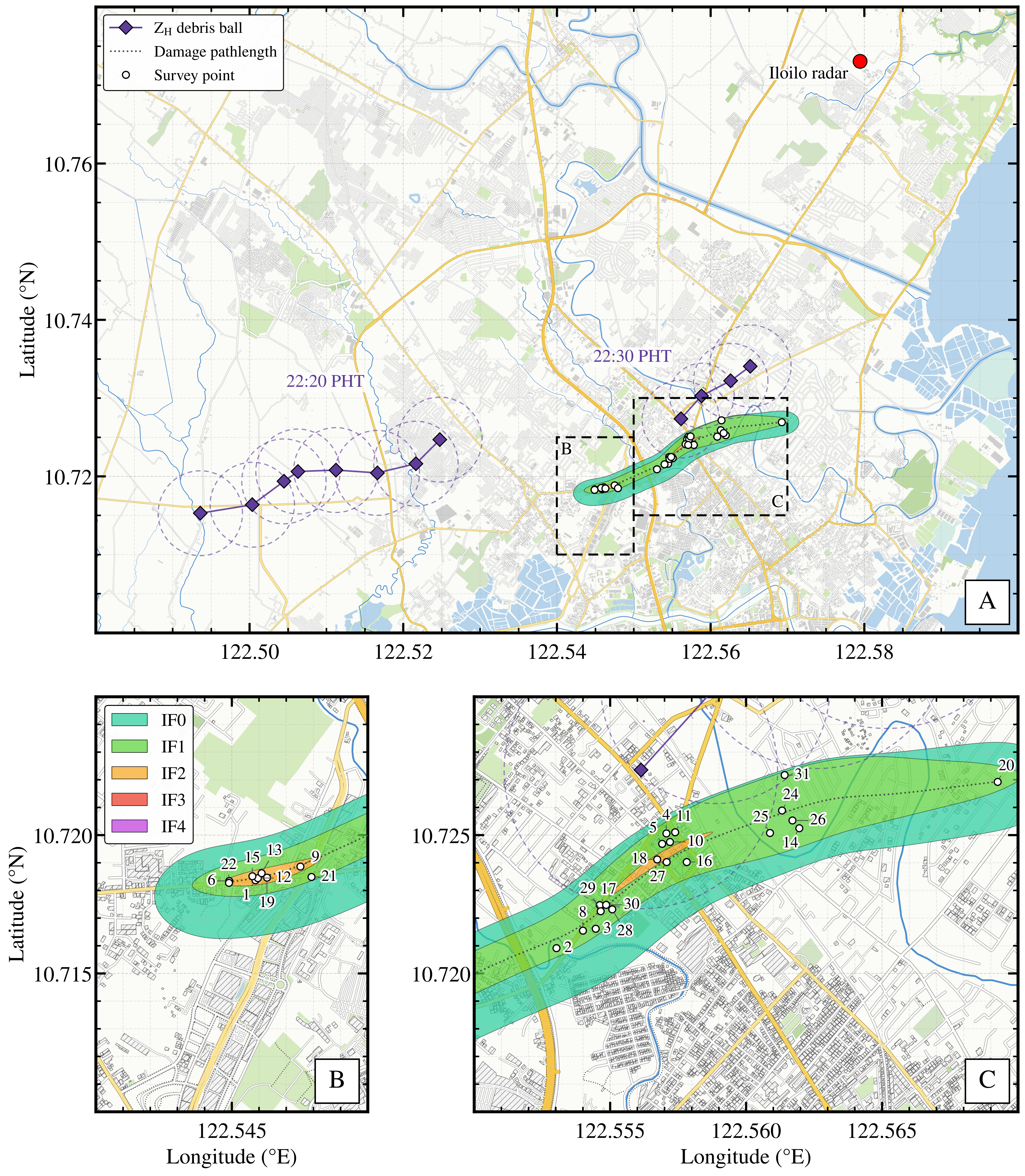}
\caption{(a) Map of the tornado damage path and survey points (white circles, numbered as in the damage inventory). Shading denotes the IF rating of the modeled damage swath, and the dotted line is the kernel-smoothed damage centerline. Purple diamonds and dashed circles mark the centers and enclosing circles of the weak echo hole (\ZH\ debris ball) tracked at 2.4\dg{}--10.0\dg{} in the 2220 LST volume and at 2.4\dg{}--5.3\dg{} in the 2230 LST volume, and the red circle marks S-ILO. (b, c) Enlarged sections of the path near (b) the Iloilo Business Park, Mandurriao, and (c) Jaro Plaza to Cubay, Jaro. Barangay boundaries are shown in gray.}
\label{fig_15}
\end{figure*}

A ground damage survey was conducted along the affected areas of Mandurriao and Jaro and was supplemented by photographs, videos, and CCTV footage archived by Project SWAP. A total of 31 damage indicators (DIs) were documented, of which 29 were geolocated (Figs.~\ref{fig_14} and \ref{fig_15}). The DIs were rated using the International Fujita (IF) Scale \citep{Groenemeijer2023}, which rates intensity using the instantaneous three-dimensional wind at the height of the damage, admits half-step ratings, and contains DIs that are not calibrated exclusively to construction practices in the United States. The documented DIs included trees of the strong, average, and weak strength classes (TR-S, TR-A, and TR-W), tree stands (TS), building roofs and structures of various sturdiness classes (BR and BS), non-structural building elements (BN), poles (PT), fences (FC), signs (SN), and carports (CP). Of the 31 DIs, 4 were rated IF2, 12 were rated IF1.5, 12 were rated IF1, and 3 were rated IF0.5. Several tree ratings in the IF Scale are open lower bounds (e.g., the uprooting of a strong tree, TR-S DoD 3, is rated $\geq$ IF1.5), so that the tabulated value represents the minimum wind consistent with the observed failure.

The four IF2 indicators occurred in two clusters. Near the Iloilo Business Park in Mandurriao, a utility pole (PT-S, DoD 2; IF2, 60 \ms), a strong tree with a snapped trunk (TR-S, DoD 5; $\geq$ IF2), and a building of sturdiness class AB (BS-AB, DoD 2; $\geq$ IF2) reached IF2. Near Jaro Plaza, a sign at the gate of the Jaro Metropolitan Cathedral (SN-M, DoD 2; $\geq$ IF2) also reached IF2. Neighboring indicators in both areas, including uprooted strong trees and damaged roofs, were rated IF1.5, while the eastern portion of the path near Benedicto and Cubay was dominated by IF1 damage to trees, tree stands, fences, and building structures, with an uprooted strong tree (IF1.5) near Benedicto. The tornado is therefore rated IF2 on the IF Scale, which corresponds approximately to EF2 on the Enhanced Fujita scale. This comparison is indicative only, as the two scales differ in reference height, wind-averaging convention, and indicator inventory. Representative damage is shown in Fig.~\ref{fig_14}.

The surveyed damage path (Fig.~\ref{fig_15}) began near the Iloilo Business Park in Mandurriao (10.7183\dg N, 122.5449\dg E) and ended near Barangay Cubay in Jaro (10.7269\dg N, 122.5693\dg E). A kernel-smoothed centerline fitted to the survey points yields a path length of $\sim$2.9 km oriented toward 065\dg{}, with a median width of $\sim$170 m and a maximum width of $\sim$340 m, the latter near Benedicto, where IF1 damage spread over $\sim$220 m. At the storm translation speed of 13.7 \ms, this path would have been traversed in $\sim$3.5 min. A translating Rankine vortex fitted to the ratings along the path ($R_c$ of 50--100 m and a translation of 13.7 \ms) required a median \Vrot\ of $\sim$40 \ms\ (10th--90th percentiles of 29--50 \ms) and placed the vortex center 30--120 m north of the damage centerline, consistent with the stronger winds on the right side of a translating vortex. The resulting IF swath (Fig.~\ref{fig_15}) contains two IF2 patches, at the Iloilo Business Park and near the Jaro Metropolitan Cathedral, connected by a continuous IF1 corridor. Notably, the path orientation (065\dg{}) was $\sim$10\dg{} to the left of the parent-storm motion (076\dg{}), consistent in sense with the DTM estimated from the radar hodographs, although smaller in magnitude \citep{Nixon2021}.

The WEH positions tracked by S-ILO provide an independent constraint on the location and timing of the tornado. In the 2230 LST (1430 UTC) volume, the WEH centers at 2.4\dg{}--5.3\dg{} (0.24--0.43 km ARL) were located 0.39--0.87 km north-northwest of the damage centerline, with the edges of their enclosing circles within 0--0.3 km of the centerline. The 2.4\dg{} center, scanned at 2232:25 LST, lay only 0.27--0.31 km from the DIs at Jaro Plaza. The increasing offset with height indicates a slight north-northwestward tilt of the WEH with height, while the residual offset at the lowest tilt is comparable to the combined uncertainty of the manual WEH marking, the $\sim$0.13 km beam width at this range, and the 0.25 km gate spacing. Interpolating between the 2.4\dg{} WEH positions of the 2220 and 2230 LST volumes, which were 7.0 km and 599 s apart (11.7 \ms\ toward 079\dg{}), places the WEH over the start of the damage path near 2230--2231 LST and over Jaro Plaza near 2232 LST, so that the tornado likely affected the surveyed area between $\sim$2230 and $\sim$2234 LST (1430--1434 UTC), within the reported tornado period of 1420--1435 UTC (Fig.~\ref{fig_5}). This timing is consistent with the arrival of the outflow at the S-ILO AWS at 1430 UTC. In the 2220 LST volume, however, the WEH was already present $\sim$5 km west-southwest of the start of the surveyed path, where no damage was documented. This suggests either that the low-level circulation preceded tornadogenesis by several minutes or that weak damage along this segment went unreported, and the available data cannot distinguish between these possibilities.

\begin{figure*}[!t]
\centering
\includegraphics[width=\textwidth,height=0.82\textheight,keepaspectratio]{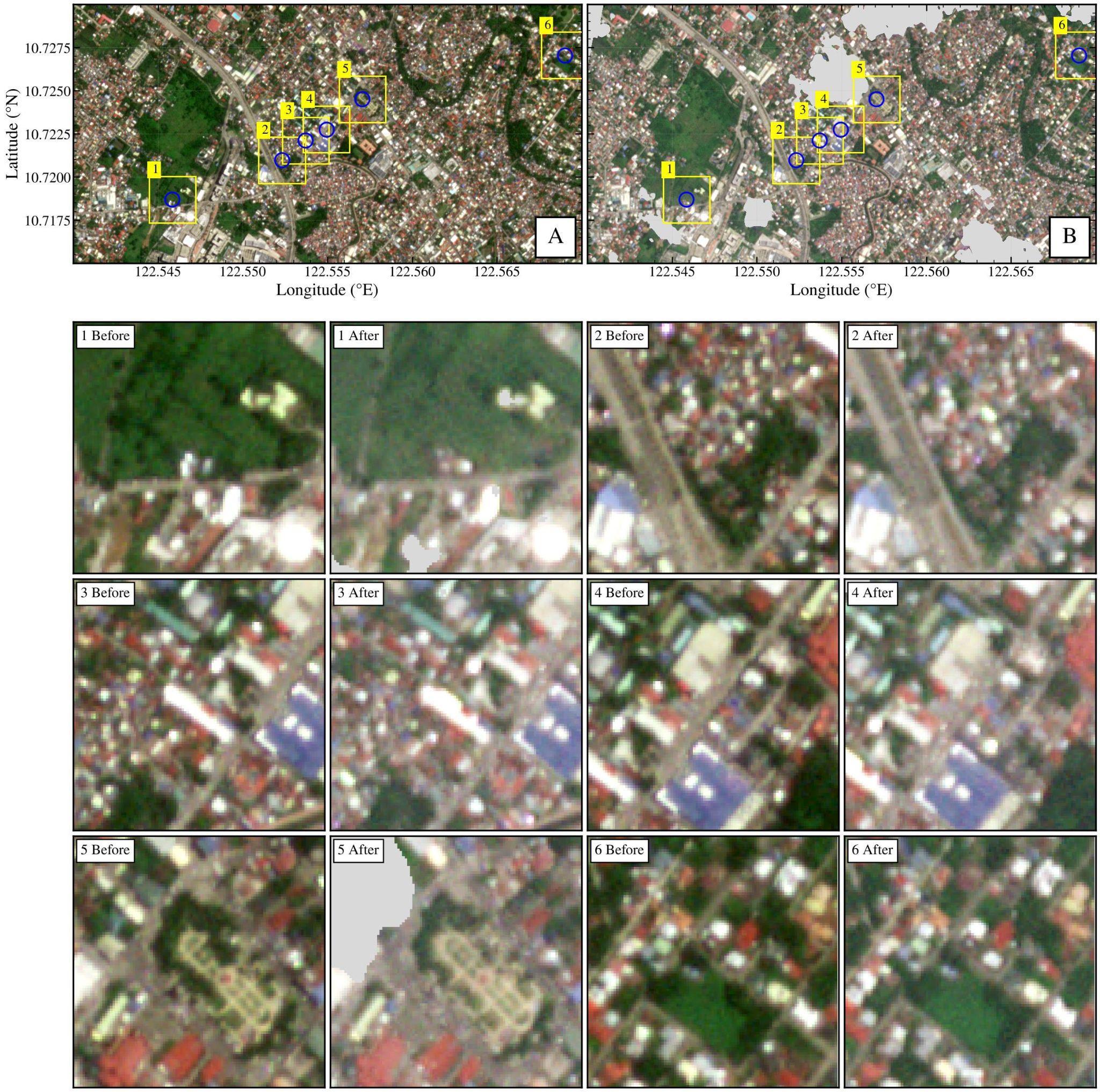}
\caption{PlanetScope true-color composites from (a) 05 August 2026 and (b) 11 August 2026, before and after the Iloilo tornado, with cloud-masked areas in gray. Yellow boxes (1--6) mark the sites shown in the enlarged before-and-after pairs below, and blue circles mark sites of interest along the damage path (Section~\ref{sec36}).}
\label{fig_16}
\end{figure*}

\begin{figure*}[!t]
\centering
\includegraphics[width=\textwidth,height=0.82\textheight,keepaspectratio]{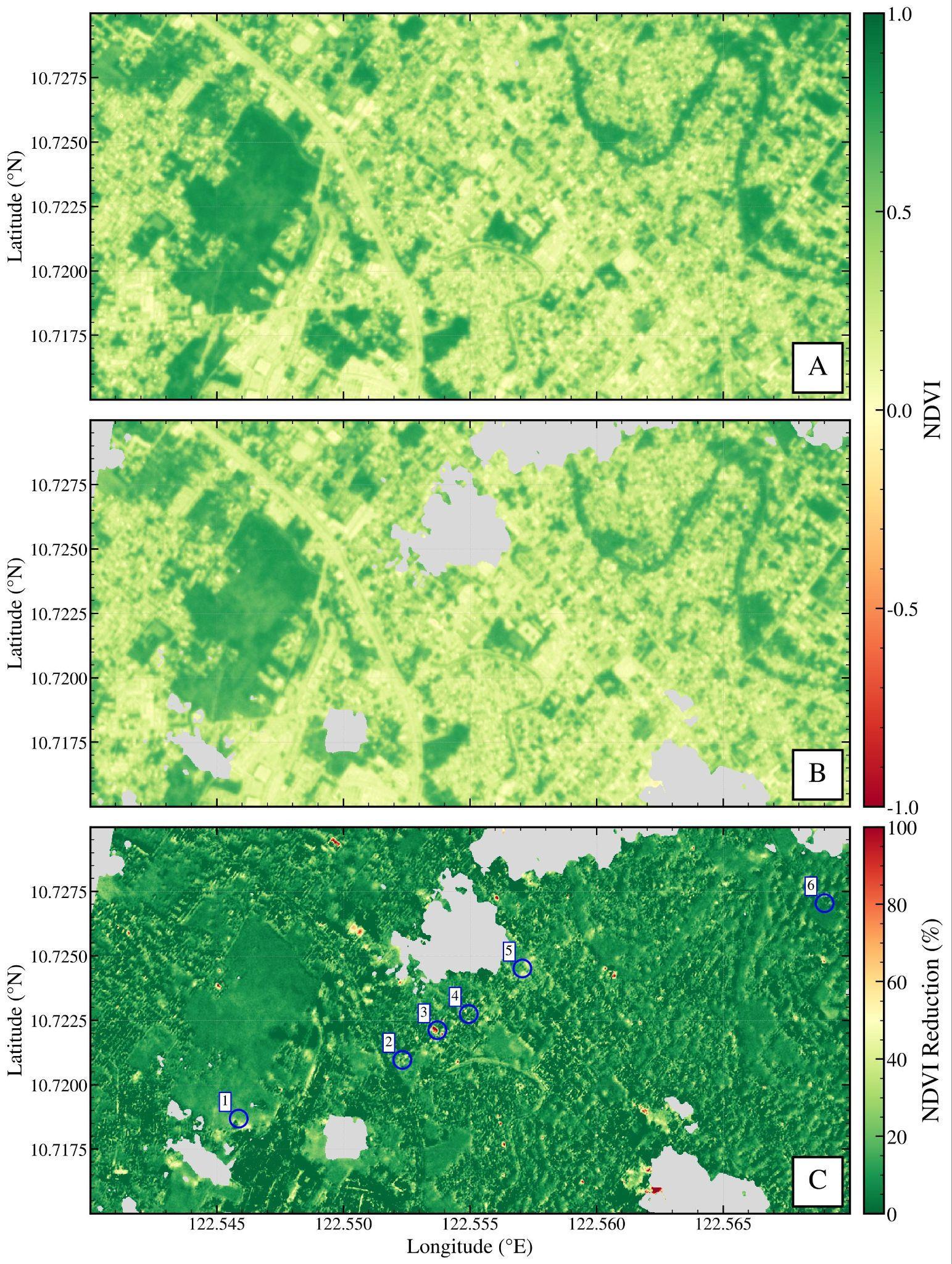}
\caption{(a) NDVI$_{\mathrm{BEFORE}}$ on 05 August 2026, (b) NDVI$_{\mathrm{AFTER}}$ on 11 August 2026, and (c) percent reduction in NDVI between the two dates (Eq.~(\ref{eq_dndvi})). Gray areas are cloud-masked. In (c), the numbered blue circles mark the six sites shown in Fig.~\ref{fig_16}.}
\label{fig_17}
\end{figure*}

The PlanetScope imagery provides supporting evidence that helps delineate the track of the tornado (Figs.~\ref{fig_16} and \ref{fig_17}). Because the pre- and post-event scenes were acquired only 3 days before and after the event (05 and 11 August), changes in land cover unrelated to the tornado are expected to be minor, so that the changes along the surveyed path can be attributed largely to the tornado. Six sites along the track were examined in paired true-color subsets (Fig.~\ref{fig_16}). The most pronounced change occurred at the Iloilo Business Park (site 1), where the damaged shops and commercial establishments were located. There, the dark and textured canopy of the vegetated lot adjacent to the shops became lighter and more uniform on 11 August, and the small structures along its southern edge appeared altered, consistent with the defoliated and uprooted trees and the roof and structural damage documented by the survey near the IF2 indicators. The NDVI reduction (Fig.~\ref{fig_17}c) likewise showed a coherent area of $\sim$20\%--50\% reduction around site 1, which contrasts with the near-zero reduction over the undisturbed vegetation to the north and west. Near Jaro Plaza (site 5), the canopy of the trees surrounding the plaza and the cathedral appeared thinned, consistent with the uprooted and snapped trees documented there, and a cluster of 80\%--100\% reductions occurred near site 3. At the remaining sites, which are dominated by roofs and pavement, the changes were subtler at the $\sim$3 m resolution of the imagery, as expected for a narrow ($\leq$ 340 m) path through a densely built-up area where the dominant DIs (roofs, poles, and signs) produce little change in vegetation indices. Taken together, the sites of change are aligned along the damage centerline from the Iloilo Business Park to Jaro Plaza, so the PlanetScope imagery corroborates the location and orientation of the track derived from the ground survey and the WEH positions, as in previous applications of satellite imagery to tornado tracks \citep{Burow2020,CapuliCN2026}.

\begin{figure*}[!t]
\centering
\includegraphics[width=\textwidth,height=0.82\textheight,keepaspectratio]{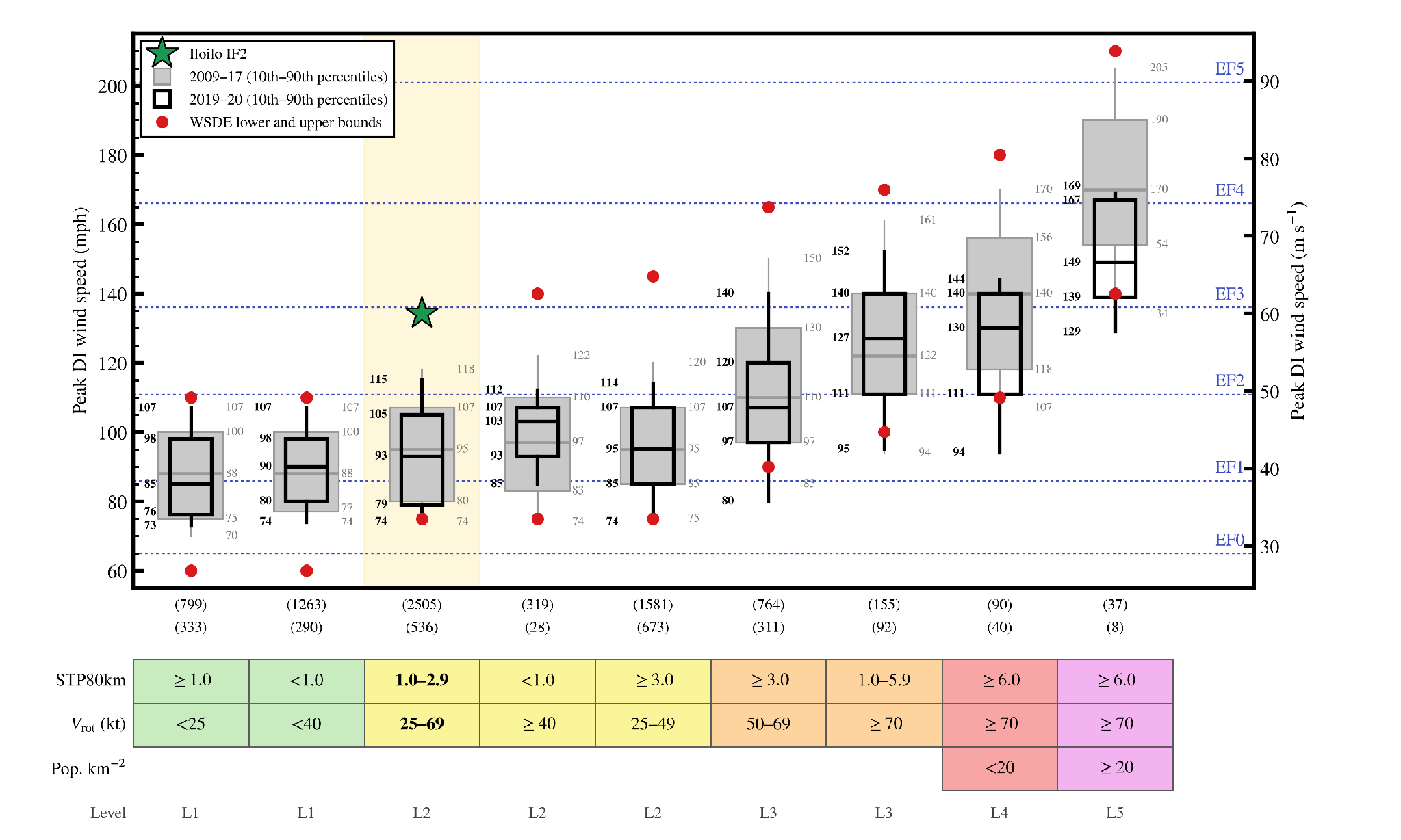}
\caption{Peak damage indicator (DI) wind speed (mph, left; \ms, right) for nine combinations of the significant tornado parameter within 80 km (STP80km), the 0.5\dg{} rotational velocity (\Vrot; kt), and the population density (people\,km$^{-2}$), recreated from Fig.~13 of \citet{Smith2020}. Gray boxes denote the 2009--17 sample and black hollow boxes the independent 2019--20 sample. Boxes span the 25th--75th percentiles, whiskers the 10th--90th percentiles, and the horizontal bars mark the medians, with the percentile values annotated (2019--20 in bold on the left and 2009--17 in gray on the right). Red circles are the lower and upper bounds of the wind speed damage estimate (WSDE), and the blue dotted lines are the EF-scale thresholds. Sample sizes are given in parentheses (top, 2009--17; bottom, 2019--20), and the colored cells denote the damage-intensity levels L1--L5. The green star marks the Iloilo tornado (IF2, $\sim$60 \ms) in the STP80km 1.0--2.9 and \Vrot\ 25--69 kt combination (shaded column).}
\label{fig_18}
\end{figure*}

To place the Iloilo tornado in the context of impact-based forecasting, its environment and damage intensity were compared with the conditional distributions of \citet{Smith2020}, who related the peak DI wind speeds of damage-surveyed tornadoes in the contiguous United States to the 0.5\dg{} \Vrot\ of the nearest WSR-88D, the maximum STP within 80 km of the tornado (STP80km), and the population density, and who grouped these combinations into five calibrated levels of potential damage intensity. Their box-and-whisker distributions are recreated in Fig.~\ref{fig_18}. The ERA5 proximity profile at 1400 UTC yielded an effective-layer STP of 2.28 (Table~\ref{Table_1}). For \Vrot, the simulated velocity couplets of a vortex consistent with the damage ($R_c$ = 100 m and a peak tangential speed of 40 \ms; Fig.~\ref{fig_10}), sampled with the beam and gate of S-ILO, were used. Defined, as in \citet{Smith2020}, as half the peak cyclonic $\Delta V$ across the couplet and sampled over all positions of the vortex relative to the radials and range gates, the beam-sampled \Vrot\ at 2.4\dg{}--5.3\dg{} was $\sim$12--18 \ms\ (median $\sim$13--14 \ms) for the Rankine profile and $\sim$16--24 \ms\ (median $\sim$18--20 \ms) for the Burgers--Rott profile, i.e., $\sim$23--46 kt with medians of $\sim$26--38 kt (Table S6 of the supplementary material). The Iloilo tornado therefore falls in the level-2 combination of STP80km of 1.0--2.9 and \Vrot\ of 25--69 kt (green star in Fig.~\ref{fig_18}). This assignment holds for the median of every profile and tilt, whereas the least favorable sampling of the Rankine vortex ($\sim$12 \ms, or $\sim$23 kt) falls at the lower edge of this combination. Because these \Vrot\ values come from simulated rather than observed velocity couplets, this placement is illustrative. It indicates the signature that S-ILO would likely have displayed for a vortex consistent with the damage, not an independent radar estimate of the tornado intensity. For this combination, which is the most populated in the sample of \citet{Smith2020}, the median peak DI wind speed was 93--95 mph ($\sim$42 \ms), the interquartile range spanned 79--107 mph (35--48 \ms), and the 90th percentile was 115--118 mph (51--53 \ms), whereas the wind speed damage estimate (WSDE) ranged from 75 to 135 mph (34--60 \ms). The peak wind speed of the IF2 rating, $\sim$60 \ms\ ($\sim$134 mph), therefore lies above the 90th percentile and at the upper bound of the WSDE of this combination, which places the Iloilo tornado in the upper tail of the damage intensities expected for its environment and rotation strength.

Two factors may explain this upper-tail outcome. First, the beam-sampled \Vrot\ underestimates the true rotation of a vortex that is narrower than the radar beam \citep{Wood1997}. At $\sim$10--11 km from S-ILO, the effective beam is $\sim$0.25 km wide, so the simulated couplets retained only $\sim$29\%--59\% of the 40 \ms\ peak tangential speed. If the same circulation had been sampled at closer range or with finer resolution, yielding a \Vrot\ of $\geq$ 70 kt ($\geq$ 36 \ms), the tornado would instead fall in the level-3 combination of STP80km of 1.0--5.9 and \Vrot\ $\geq$ 70 kt. In that combination, $\sim$134 mph lies close to the median (122--127 mph) and within the interquartile range (111--140 mph) of the peak DI wind speed (Fig.~\ref{fig_18}). Second, the tornado tracked through densely built-up districts, where the population density exceeds the threshold of 20 people\,km$^{-2}$ used by \citet{Smith2020} by orders of magnitude. \citet{Smith2020} found that, for comparable \Vrot\ and STP80km, tornadoes that affected areas above this threshold produced higher peak DI wind speeds, because more DIs were available to record the strongest winds. The abundance of DIs in Mandurriao and Jaro, from the shops of the Iloilo Business Park to the trees, poles, and signs of Jaro Plaza, likely increased the chance that the strongest winds were captured by the survey. These comparisons are indicative rather than quantitative, because the distributions of \citet{Smith2020} were derived from WSR-88D observations at 0.5\dg{}, from STP80km rather than from a single proximity profile, and from EF-scale DIs in the United States, whose construction practices differ from those in the Philippines.

The same environment also allows the forecastability of the event to be assessed retrospectively. In the convective outlooks of the U.S. Storm Prediction Center (SPC), the categorical risk is derived from the probability of a tornado, damaging wind, or large hail within 25 miles (40 km) of a point, and a 5\% tornado probability on the Day 1 outlook corresponds to a Slight Risk, i.e., level 2 of 5 \citep{Grams2026}. The environment of southern Panay on the evening of 08 August combined MUCAPE of $\sim$2000 \Jkg, negligible CIN, LCLs below 300 m, an STP of 2.28, and an SCP of 7.44 with an LLS of $\sim$16 \ms\ and SRH$_{01}$ of up to 211 \mss\ in the radar-derived profiles, while the weak DLS and the weak synoptic-scale forcing favored only isolated supercells within the monsoon. Had such an outlook been issued, these ingredients would most likely have supported a 5\% tornado probability, i.e., a Slight Risk (2/5) with a tornado threat, over the southern coast of Panay. A higher category, such as an Enhanced Risk driven by a 10\% tornado probability with significant (EF2+) tornado potential, would have been difficult to justify given the expected low coverage of supercells, even though the tornado ultimately reached IF2 intensity. Because the LLJ and the LLS strengthened after sunset, such an outlook would also have needed to emphasize that the threat would peak during the night, when the vulnerability to tornadoes is greatest \citep{Ashley2008}.

\section{SUMMARY AND DISCUSSION}\label{sec4}

This study documented the tornado that struck the Mandurriao and Jaro districts of Iloilo City at approximately 1430 UTC (2230 LST) on 08 August 2026, using ERA5 reanalysis, observed and model proximity soundings, Himawari-9 infrared imagery, dual-polarization observations from S-ILO, surface observations, a ground damage survey, and PlanetScope imagery. The tornado was produced by a supercell that developed near the southern coast of Panay and moved east-northeastward at 13.7 \ms\ across Iloilo City. The damage survey documented a $\sim$2.9 km long and up to $\sim$340 m wide path, and the tornado was rated IF2 on the basis of a utility pole, a snapped strong tree, structural damage to a building, and a sign at the Jaro Metropolitan Cathedral. To our knowledge, this is among the first nocturnal tornadoes in the Philippines documented with both a ground damage survey and dual-polarization radar observations, and it adds a well-documented case from Western Visayas, one of the country's recognized tornado hotspots.

The tornadic storm developed within the southwest monsoon as it was enhanced by Typhoon Dolphin, which was located far to the north-northeast of Luzon. Synoptic-scale forcing for ascent was weak, as 500-hPa heights over the Visayas remained nearly steady, similar to the weakly forced easterly-wave regime of the Camarines Norte outbreak \citep{CapuliCN2026} but unlike the 500-hPa height falls that accompanied the Candating, Arayat, tornado \citep{Capuli2026}. Instead, the monsoon acted as a broad LLJ that supplied deep moisture (PWAT of 55--62 kg\,m$^{-2}$) and concentrated the vertical wind shear in the lowest $\sim$3 km. The resulting environment combined moderate instability (MUCAPE of $\sim$2000 \Jkg\ and ECAPE of $\sim$1600 \Jkg), negligible CIN, very low LCLs, and strong, highly streamwise LLS, while the DLS was weak because the hodograph folded back above $\sim$3 km. Such a hodograph, dominated by LLS rather than DLS, differs from the environments of most tornadic supercells in the U.S. Great Plains. It resembles, however, the tropical-cyclone and high-shear, low-CAPE environments in which tornadic supercells are sustained by strong LLS and storm-relative inflow despite modest DLS \citep{McCaul1991,Sherburn2014}.

The environment of the Iloilo tornado was distinctly nocturnal. Several factors explain how a tornadic supercell was sustained $\sim$4 h after sunset over land, when the stabilization of the boundary layer typically reduces SB instability. First, a nocturnal LLJ of $\sim$20 \ms\ (40 kt) near 1.5--3 km, evident in both ERA5 and the radar-derived wind profiles, surged from $\sim$2130 to 2230 LST and increased the LLS and SRH as the storm approached Iloilo City. A similar evening enhancement of LLS has been documented in nocturnal tornado environments of the United States \citep{Kis2010,Mead2011}. Second, the jet and the onshore flow beneath it advected warm, moist marine air ($\theta_e$ of 353--359 K at the AWS) from the Panay Gulf, so that the coastal boundary layer barely cooled after sunset and CAPE along the coast remained $\sim$2000--2500 \Jkg\ while it decreased inland. The tornado occurred within $\sim$3--4 km of the coast, where this effect was greatest. Third, the moist boundary layer kept the LCL low and limited the negative buoyancy of the outflow, as indicated by the modest $\Delta\theta_v$ of $\sim$2 K at the AWS, which favors the lifting and stretching of near-surface air by the low-level updraft \citep{Markowski2002}. Fourth, the storm moved $\sim$20\dg{} to the right of the \BRM, which increased its storm-relative inflow and effective SRH \citep{Bunkers2022}. Terrain may have further modified the low-level flow, as ERA5 depicted enhanced low-level PV and convergence in the vicinity of the central Panay range. However, the resolution of ERA5 is insufficient to determine whether terrain-induced vorticity contributed to tornadogenesis, as suggested for the Candating, Arayat, and Camarines Norte cases \citep{Capuli2026,CapuliCN2026}.

The S-ILO observations revealed several hallmark features of a tornadic supercell, including a hook echo with a WEH enclosed by a ring of high reflectivity, a \ZDR\ arc along the inflow side of the forward flank, a \ZDR\ ring around a low-level BWER, and enhanced \ZDR\ extending to just below the 0\,\dg C level. The three-dimensional analysis further showed an elevated rear-flank core of 55--60 dBZ directly above the WEH, which had lowered by $\sim$1--2 km since the previous volume and is consistent with a DRC \citep{Rasmussen2006,Kennedy2007}. The WEH tracked across Iloilo City in two consecutive volumes, was nearly vertical in the lowest $\sim$1.4 km once the scan-time displacement was removed, and passed within a few hundred meters of the surveyed damage at the time of the damage, indicating that it marked the tornado or its parent low-level circulation. At the same time, this case illustrates how routine data processing can conceal tornado-scale signatures. The archived velocity field had been smoothed by the signal processor, the \rhv\ $<$ 0.80 threshold applied for rainfall estimation removed the gates where a TDS would be expected, and the velocity sign stored in the UF files was reversed relative to the CF convention. Consequently, neither a TVS nor a TDS could be identified directly, even though the tornado passed only $\sim$5--7 km from the radar. The radial velocity simulation and the injection--recovery analysis showed that a tornado consistent with the damage would have produced a conspicuous couplet in unfiltered data, and that the non-detection is consistent with a narrow vortex ($R_c$ $\approx$ 50--100 m) with \Vrot\ of $\sim$35--45 \ms, in agreement with the width of the damage path and the Rankine swath model. Because radar-based intensity estimates are biased low for vortices comparable to or smaller than the beam \citep{Wood1997}, these values should be regarded as consistent with, rather than independent of, the damage rating.

It is instructive to compare the Iloilo tornado with the two Philippine tornadic supercells documented previously. The Candating, Arayat, tornado \citep[EF2;][]{Capuli2026} occurred around midday within the southwest monsoon, which acted as an LLJ, under subtle synoptic-scale ascent influenced by the approach of TC Ewiniar, in an environment of large undiluted instability from steep low-level lapse rates, weak LLS, and highly streamwise near-surface vorticity. The Camarines Norte outbreak \citep[IF2.5, comparable to the EF3 range;][]{CapuliCN2026} occurred near 0600 LST within an easterly-wave regime, with MUCAPE of $\sim$2200 \Jkg, LCLs below 200 m, and modestly curved but strongly streamwise hodographs. The Iloilo tornado (IF2) occurred at night within the typhoon-enhanced southwest monsoon, with MUCAPE of $\sim$2000 \Jkg, LCLs below 300 m, and strong, streamwise LLS enhanced by a nocturnal LLJ. Despite their different synoptic regimes and times of day, the three cases share low LCLs, moderate rather than extreme instability, and near-ground horizontal vorticity that was almost purely streamwise, whereas the DLS varied widely and was weak in the Iloilo case. These shared characteristics suggest that, in the moist tropical environment of the Philippines, the near-ground kinematic profile and the cloud-base height may discriminate tornadic supercells more effectively than the DLS, a hypothesis that should be tested with a larger sample of events from the SWAP database.

Several limitations should be noted. First, no radiosonde was launched near the time and location of the event, so the near-storm thermodynamic profile relies on ERA5, whose resolution smooths the land--sea and terrain contrasts that likely mattered in this case and which underestimated the LLS relative to the radar-derived profile. Second, the radar-derived hodographs extend only to $\sim$3.5 km AGL, depend on the AWS wind for their lowest level, and were contaminated by outflow at 1430 UTC. Third, the AWS is located $\sim$5--7 km from the damage path, so the surface observations describe the storm's outflow rather than the immediate environment of the tornado. Fourth, the 10-min volume interval of S-ILO is too coarse to resolve the descent of the rear-flank core and the genesis of the low-level vortex, and the three-dimensional analysis relies on \ZH\ alone. The lowering of the core between 1410 and 1420 UTC is therefore inferred from two volumes rather than observed, whereas mobile radars have documented the descent of such cores at much finer temporal resolution \citep{Byko2009}, and the storm-relative position assigned to the 1410 UTC core depends on the assumed storm motion. In addition, DRCs also occur in nontornadic supercells and have limited value as a stand-alone tornado precursor \citep{Kennedy2007}, so the DRC in this case is interpreted as consistent with, rather than diagnostic of, tornadogenesis. In the long term, rapid-scan X-band radars offer the most promising solution to these sampling limitations. Mobile, rapid-scanning, polarimetric X-band radars can complete a 10-elevation volume in $\sim$20 s \citep{Pazmany2013} and have documented tornadogenesis and tornado dissipation in supercells at temporal resolutions of seconds to tens of seconds \citep{Houser2015}. Dense networks of small X-band radars spaced tens of kilometers apart sample the lowest kilometers of the atmosphere that long-range radars miss because of the curvature of the Earth, including damaging winds and heavy rainfall \citep{McLaughlin2009}, while multiparameter phased-array weather radars resolve the three-dimensional structure of storms within $\sim$30 s over a $\sim$60 km radius for the early detection of torrential rainfall and tornadoes \citep{Takahashi2019}. Such systems, deployed to complement S-band radars such as S-ILO over densely populated areas, would resolve the fine structure not only of tornadic storms but also of other thunderstorms capable of producing hazards, such as downbursts, large hail, and flash floods. Fifth, because the tornado occurred at night, no photographs or videos of the tornado itself were available, and the timing of the damage was inferred from the radar rather than from eyewitness reports. Finally, the absence of high-resolution numerical modeling limits the assessment of terrain effects, of the dynamics of the nocturnal LLJ over Panay, and of the storm-scale processes that led to tornadogenesis. Future work would benefit from convection-permitting or large-eddy simulations (e.g., WRF or CM1) of this event.

Based on this case, we offer several recommendations. First, the quality control applied to S-ILO should be relaxed and made storm-aware so that the structures that identify SCSs and their hazards are not removed or heavily masked. The fixed \rhv\ $<$ 0.80 threshold is appropriate for QPE, but in this case it removed exactly the gates in which lofted debris would be expected, together with the corresponding \ZH, \ZDR, and \VR. Rather than deleting such gates, the quality control could flag them while archiving the base moments unaltered, so that the QPE product is computed from the flagged field and the original data remain available for severe-weather analysis. Constraints can then be added to retain low-\rhv\ gates where they are physically consistent with a TDS, i.e., where they coincide with \ZH\ $\geq$ 35--40 dBZ, \ZDR\ near 0 dB, and a hook echo, mesocyclone, or strong azimuthal shear \citep{Ryzhkov2005,Bodine2013}. Alternatively, a hydrometeor classification algorithm with an explicit debris class can separate debris from clutter and biological scatterers objectively \citep{Snyder2015}. Likewise, the near-zero-velocity clutter filter could be restricted to clutter-prone gates identified from a static clutter map, and the spatial smoothing of \VR\ by the signal processor should be reduced or, at minimum, the unfiltered velocity and spectrum width should be archived, since the processing in this case concealed a tornado-scale couplet that would otherwise have been conspicuous at a range of only $\sim$6--12 km. Second, the header metadata of the archived UF volumes (wavelength, PRF, and Nyquist velocity) and the sign convention of the velocity field should be verified for each newly commissioned radar before the data are distributed. Third, a severe-weather scanning mode with more frequent low-level updates than the current 10-min volume would allow rapidly evolving features, such as DRCs and low-level mesocyclones, to be resolved, and a network of rapid-scan X-band radars over the major urban centers would provide this capability in the long term. Fourth, as recommended by \citet{CapuliCN2026}, a rapid-response damage survey capability trained in the IF Scale should be activated whenever a tornado is reported, since damage in urban areas is cleared within hours, as occurred in Iloilo City, where clearing operations began on the night of the event. Fifth, the combination of the near-storm environment (e.g., STP) and radar-derived rotation could support impact-based tornado warnings and probabilistic convective outlooks in the Philippines, following the framework of \citet{Smith2020} and the SPC outlooks, provided that these relations are recalibrated with local cases from Project SWAP. Moreover, because the tornado occurred at night and only minutes after a thunderstorm advisory was issued, this case underscores the elevated threat posed by nocturnal tornadoes \citep{Ashley2008} and the value of radar-based nowcasting of low-level mesocyclones near densely populated coastal cities.

In summary, the Iloilo tornado demonstrates that significant tornadoes in the Philippines can develop at night within the typhoon-enhanced southwest monsoon, in environments characterized by a nocturnal LLJ surge with warm, moist advection, moderate instability, very low LCLs, and strong, streamwise LLS but weak DLS. Together with the Candating, Arayat, and Camarines Norte cases, these findings add to the growing evidence that Philippine tornadic environments span multiple regimes, including the southwest monsoon, TCs, and easterly waves, and they emphasize the need for the continued systematic documentation of SCSs through Project SWAP and the national radar network.

\acknowledgments

The author is sincerely grateful to the residents of Iloilo City who reported and documented the tornado in their communities, and whose photographs and accounts made much of this analysis possible. Thanks are also due to Dr. Funing Li (Massachusetts Institute of Technology) and Dr. John T. Allen (Central Michigan University) for valuable guidance on the use of \textit{xcape}, and to the anonymous reviewers and the editor, whose careful and constructive comments substantially improved the manuscript. This work received no external funding and was carried out on the author's own time, dedicated to severe convective storm research in the Philippines. Finally, the author thanks family and loved ones for their unwavering support throughout this research. CGHA is supported by Planet Labs PBC (Planet) through the Education and Research Program.

\contribution

\textbf{Generich H. Capuli:} Writing -- original draft, Writing -- review \& editing, Visualization, Methodology, Investigation, Formal Analysis, Data curation, Conceptualization, Supervision.

\datastatement

The ERA5 reanalysis data on single and pressure levels \citep{Hersbach2020} are openly available from the Copernicus Climate Data Store (\url{https://cds.climate.copernicus.eu}). The individual full-disk Himawari-9 AHI bands were accessed through the THREDDS data server of the National Computational Infrastructure (NCI) Australia. PlanetScope imagery \citep{PlanetARPS2024} was obtained from Planet Labs PBC under its Education and Research Program and cannot be redistributed under the terms of that license. The 2023 administrative boundaries of the Philippines (levels 0--4) of the Philippine Statistics Authority (PSA) and the National Mapping and Resource Information Authority (NAMRIA) are available from the Humanitarian Data Exchange (\url{https://data.humdata.org/dataset/cod-ab-phl}), and the Copernicus GLO-30 Digital Elevation Model is distributed by OpenTopography (\url{https://doi.org/10.5069/G9028PQB}). Basemap drawn from OpenStreetMap (OSM) data licensed under the ODbL. The one-dimensional ERA5 model sounding profiles used in this study are archived in the fourth data release (DR4) of Project SWAP. The S-ILO radar volumes and the AWS observations are operational holdings of DOST-PAGASA and are not publicly redistributable. They may be requested from DOST-PAGASA through METTS, and the corresponding author will facilitate such requests upon reasonable inquiry. Users of these datasets should follow the attribution and licensing requirements of each provider.

This study made use of several open-source Python packages. The radar volumes were read, processed, and gridded with \textit{Py-ART} \citep{Helmus2016}. The sounding and ERA5 convective parameters were computed with \textit{MetPy}, \textit{SounderPy} v3.1.0 \citep{Gillett2025}, \textit{ecape-parcel}, and \textit{xcape} \citep{lepore2021}. The cloud and cloud-shadow masks of the PlanetScope imagery were generated with \textit{OmniCloudMask} \citep{Wright2025}, and the raster data were processed with \textit{Rasterio} and \textit{rioxarray}. The maps were produced with \textit{Cartopy}, \textit{GeoPandas}, \textit{Shapely}, and \textit{pyproj}, the reflectivity isosurfaces were extracted with \textit{scikit-image}, and all analyses and figures used \textit{NumPy}, \textit{SciPy}, \textit{pandas}, \textit{xarray}, \textit{netCDF4}, \textit{Matplotlib}, and \textit{cmweather}. The dual-PRF velocity errors were corrected with the code of \citet{Alford2022}, and the folded single-PRF tilts were dealiased with a reference-anchored, region-based procedure developed for this study, against which \textit{UNRAVEL} \citep{Louf2020} and the region-based dealiasing of \textit{Py-ART} were evaluated.

\interest

The author declares no competing interests.

\appendix
\appendixtitle{Supplementary Analysis}

Supplementary material is provided as a separate document. It contains (1) the complete damage survey inventory, listing the location, damage indicator, degree of damage, and IF rating of each of the 31 surveyed items; (2) the details of the radial velocity simulation, including the radar beam and gate model and the simulated Rankine and Burgers--Rott vortices; and (3) the supporting statistical analyses, namely the injection--recovery test and the Bayesian estimate of the vortex rotational velocity.

\bibliographystyle{ametsocV6}
\bibliography{references}

@STRING{AN        = "Astrophys.\ Norv."}

@STRING{JAM       = "J.\ Appl.\ Meteor."}

@STRING{JAOT      = "J.\ Atmos.\ Oceanic Technol."}

@STRING{MA        = "Meteor.\ Appl."}

@STRING{MAP       = "Meteor.\ Atmos.\ Phys."}

@STRING{TELLUS    = "Tellus"}

@STRING{WF        = "Wea.\ Forecasting"}

@article{Alford2022,
  author={Alford, A. A. and Biggerstaff, M. I. and Ziegler, C. L. and Jorgensen, D. P. and Carrie, G. D.},
  title={A Method for Correcting Staggered Pulse Repetition Time ({PRT}) and Dual Pulse Repetition Frequency ({PRF}) Processor Errors in Research Radar Datasets},
  journal=JAOT, year={2022}, volume={39}, number={11}, pages={1763--1780}, doi={10.1175/JTECH-D-21-0176.1}}

@article{Anderson2007,
      author = "Christopher J.  Anderson and Christopher K.  Wikle and Qin  Zhou and J. Andrew  Royle",
      title = {{Population Influences on Tornado Reports in the United States}},
      journal = "Weather and Forecasting",
      year = "2007",
      publisher = "American Meteorological Society",
      address = "Boston MA, USA",
      volume = "22",
      number = "3",
      doi = "10.1175/WAF997.1",
      pages=      "571 - 579",
      url = "https://journals.ametsoc.org/view/journals/wefo/22/3/waf997_1.xml"
}

@article {AndersonFrey2021,
      author = "Alexandra K. Anderson-Frey and Harold Brooks",
      title = {{Compared to What? Establishing Environmental Baselines for Tornado Warning Skill}},
      journal = "Bulletin of the American Meteorological Society",
      year = "2021",
      publisher = "American Meteorological Society",
      address = "Boston MA, USA",
      volume = "102",
      number = "4",
      doi = "10.1175/BAMS-D-19-0310.1",
      pages=      "E738 - E747",
      url = "https://journals.ametsoc.org/view/journals/bams/102/4/BAMS-D-19-0310.1.xml"
}

@article {Antonescu2017,
      author = "Bogdan Antonescu and David M. Schultz and Alois Holzer and Pieter Groenemeijer",
      title = {{Tornadoes in Europe: An Underestimated Threat}},
      journal = "Bulletin of the American Meteorological Society",
      year = "2017",
      publisher = "American Meteorological Society",
      address = "Boston MA, USA",
      volume = "98",
      number = "4",
      doi = "10.1175/BAMS-D-16-0171.1",
      pages=      "713 - 728",
      url = "https://journals.ametsoc.org/view/journals/bams/98/4/bams-d-16-0171.1.xml"
}

@article{Antonescu2019,
      author = "Bogdan Antonescu and Hugo M. A. M. Ricketts and David M. Schultz",
      title = {{100 Years Later: Reflecting on Alfred Wegener’s Contributions to Tornado Research in Europe}},
      journal = "Bulletin of the American Meteorological Society",
      year = "2019",
      publisher = "American Meteorological Society",
      address = "Boston MA, USA",
      volume = "100",
      number = "4",
      doi = "10.1175/BAMS-D-17-0316.1",
      pages=      "567 - 578"
}

@article{BAGAGLINI2022,
title = {{Synoptic patterns and mesoscale precursors of Italian tornadoes}},
journal = {Atmospheric Research},
volume = {253},
pages = {105503},
year = {2021},
issn = {0169-8095},
doi = {10.1016/j.atmosres.2021.105503},
url = {https://www.sciencedirect.com/science/article/pii/S0169809521000557},
author = {Leonardo Bagaglini and Roberto Ingrosso and Mario Marcello Miglietta},
}

@article{Bessho2016,
  title={{An Introduction to Himawari-8/9; Japan's New-Generation Geostationary Meteorological Satellites}},
  author="K. Bessho and K. Date and M. Hayashi and A. Ikeda and T.  Imai and H. Inoue and Y. Kumagai and T. Miyakawa and H. Murata and T. Ohno and A. Okuyama and R. Oyama and Y. Sasaki and  Y. Shimazu and K. Shimoji and Y. Sumida and M. Suzuki and H. Taniguchi and H. Tsuchiyama and D. Uesawa and H. Yokota and R. Yoshida",
  journal={Journal of the Meteorological Society of Japan. Ser. II},
  volume={94},
  number={2},
  pages={151-183},
  year={2016},
  doi={10.2151/jmsj.2016-009}
}

@article {Bodine2013,
      author = "David J.  Bodine and Matthew R.  Kumjian and Robert D.  Palmer and Pamela L.  Heinselman and Alexander V.  Ryzhkov",
      title = "Tornado Damage Estimation Using Polarimetric Radar",
      journal = "Weather and Forecasting",
      year = "2013",
      publisher = "American Meteorological Society",
      address = "Boston MA, USA",
      volume = "28",
      number = "1",
      doi = "10.1175/WAF-D-11-00158.1",
      pages=      "139 - 158",
      url = "https://journals.ametsoc.org/view/journals/wefo/28/1/waf-d-11-00158_1.xml"
}

@article{Boustead2013,
      author = "Joshua M.  Boustead and Barbara E.  Mayes and William  Gargan and Jared L.  Leighton and George  Phillips and Philip N.  Schumacher",
      title = {{Discriminating Environmental Conditions for Significant Warm Sector and Boundary Tornadoes in Parts of the Great Plains}},
      journal = "Weather and Forecasting",
      year = "2013",
      publisher = "American Meteorological Society",
      address = "Boston MA, USA",
      volume = "28",
      number = "6",
      doi = "10.1175/WAF-D-12-00102.1",
      pages=      "1498 - 1523",
      url = "https://journals.ametsoc.org/view/journals/wefo/28/6/waf-d-12-00102_1.xml"
}

@article {Brooks2003,
      author = "Harold E. Brooks and Charles A.  Doswell and Michael P. Kay",
      title = "Climatological Estimates of Local Daily Tornado Probability for the United States",
      journal = "Weather and Forecasting",
      year = "2003",
      publisher = "American Meteorological Society",
      address = "Boston MA, USA",
      volume = "18",
      number = "4",
      doi = "10.1175/1520-0434(2003)018<0626:CEOLDT>2.0.CO;2",
      pages=      "626 - 640"
}

@article{Browning1968,
  author={Browning, K. A. and Wexler, R.},
  title={The Determination of Kinematic Properties of a Wind Field Using {D}oppler Radar},
  journal=JAM, year={1968}, volume={7}, number={1}, pages={105--113}, doi={10.1175/1520-0450(1968)007<0105:TDOKPO>2.0.CO;2}}

@article{Bunkers2000,
      author = "Matthew J.  Bunkers and Brian A.  Klimowski and Jon W.  Zeitler and Richard L.  Thompson and Morris L.  Weisman",
      title = {{Predicting Supercell Motion Using a New Hodograph Technique}},
      journal = "Weather and Forecasting",
      year = "2000",
      publisher = "American Meteorological Society",
      address = "Boston MA, USA",
      volume = "15",
      number = "1",
      doi = "10.1175/1520-0434(2000)015<0061:PSMUAN>2.0.CO;2",
      pages= "61 - 79"
}

@article{Bunkers2006,
      author = "Matthew J.  Bunkers and Jeffrey S.  Johnson and Lee J.  Czepyha and Jason M.  Grzywacz and Brian A.  Klimowski and Mark R.  Hjelmfelt",
      title = {{An Observational Examination of Long-Lived Supercells. Part II: Environmental Conditions and Forecasting}},
      journal = "Weather and Forecasting",
      year = "2006",
      publisher = "American Meteorological Society",
      address = "Boston MA, USA",
      volume = "21",
      number = "5",
      doi = "10.1175/WAF952.1",
      pages=      "689 - 714",
      url = "https://journals.ametsoc.org/view/journals/wefo/21/5/waf952_1.xml"
}

@Article{Burow2020,
    AUTHOR = {Burow, Daniel and Herrero, Hannah V. and Ellis, Kelsey N.},
    TITLE = {{Damage Analysis of Three Long-Track Tornadoes Using High-Resolution Satellite Imagery}},
    JOURNAL = {Atmosphere},
    VOLUME = {11},
    YEAR = {2020},
    NUMBER = {6},
    ARTICLE-NUMBER = {613},
    ISSN = {2073-4433},
    DOI = {10.3390/atmos11060613}
}

@article {Byko2009,
      author = "Zack  Byko and Paul  Markowski and Yvette  Richardson and Josh  Wurman and Edwin  Adlerman",
      title = {{Descending Reflectivity Cores in Supercell Thunderstorms Observed by Mobile Radars and in a High-Resolution Numerical Simulation}},
      journal = "Weather and Forecasting",
      year = "2009",
      publisher = "American Meteorological Society",
      address = "Boston MA, USA",
      volume = "24",
      number = "1",
      doi = "10.1175/2008WAF2222116.1",
      pages=      "155 - 186",
      url = "https://journals.ametsoc.org/view/journals/wefo/24/1/2008waf2222116_1.xml"
}

@article{Capuli2024,
       author = {Generich H. Capuli},
        title = {{Project Severe Weather Archive of the Philippines (SWAP). Part 1: Establishing a Baseline Climatology for Severe Weather across the Philippine Archipelago}},
      journal = {Annals of Geophysics},
       volume = {67},
       number = {5},
         year = "2024",
        month = "12",
        pages = {GC554},
          doi = {10.4401/ag-9151}
}

@ARTICLE{Capuli2025,
  title    = {{Friday the 13th Hailstorm in the Province of Bulacan, Philippines (13 August 2021): A Case Study}},
  author   = "Capuli, Generich H.",
  journal  = "Asia-Pacific Journal of Atmospheric Sciences",
  volume   = "61",
  number   = "2",
  pages    = "13",
  month    = "apr",
  doi      = "10.1007/s13143-025-00396-6",
  year     = "2025"
}

@article{Capuli2026,
  author  = {Capuli, G. H. and Noveno, M. A. O. and Iba{\~n}ez, M. P. A.},
  title   = {{A Case Study of the Tornadic Supercell in the Province of Pampanga, Philippines (27 May 2024)}},
  journal = {Natural Hazards},
  volume  = {122},
  pages   = {89},
  year    = {2026},
  doi     = {10.1007/s11069-025-07915-8},
  url     = {https://doi.org/10.1007/s11069-025-07915-8}
}

@article{Choo2018,
  title     = {{A Case Study of Tsukuba Tornado in Japan on 6 May 2012}},
  author    = "Choo, Seonhee and Min, Ki-Hong and Kim, Kyung-Eak and Lee, Gyuwon",
  journal   = "J. Korean Earth Sci. Soc.",
  publisher = "The Korean Earth Science Society",
  volume    =  "39",
  number    =  "5",
  pages     = "403--418",
  month     =  "oct",
  year      =  "2018",
  doi       =  "10.5467/JKESS.2018.39.5.403"
}

@article{Clark2024,
    author = {Clark, Matthew R.},
    title = {{Polarimetric radar observations of the Jersey tornadic supercell on 1–2 November 2023}},
    journal = {Weather},
    volume = {79},
    number = {11},
    pages = {362-372},
    doi = {https://doi.org/10.1002/wea.7619},
    year = {2024}
}

@article{Coffer2017,
      author = "Brice E. Coffer and Matthew D. Parker",
      title = {{Simulated Supercells in Nontornadic and Tornadic VORTEX2 Environments}},
      journal = "Monthly Weather Review",
      year = "2017",
      publisher = "American Meteorological Society",
      address = "Boston MA, USA",
      volume = "145",
      number = "1",
      doi = "10.1175/MWR-D-16-0226.1",
      pages=      "149 - 180"
}

@article{Coffer2019,
      author = "Brice E. Coffer and Matthew D. Parker and Richard L. Thompson and Bryan T. Smith and Ryan E. Jewell",
      title = {{Using Near-Ground Storm Relative Helicity in Supercell Tornado Forecasting}},
      journal = "Weather and Forecasting",
      year = "2019",
      publisher = "American Meteorological Society",
      address = "Boston MA, USA",
      volume = "34",
      number = "5",
      doi = "10.1175/WAF-D-19-0115.1",
      pages=      "1417 - 1435"
}

@article {Coffer2020,
      author = "Brice E. Coffer and Mateusz Taszarek and Matthew D. Parker",
      title = {{Near-Ground Wind Profiles of Tornadic and Nontornadic Environments in the United States and Europe from ERA5 Reanalyses}},
      journal = "Weather and Forecasting",
      year = "2020",
      publisher = "American Meteorological Society",
      address = "Boston MA, USA",
      volume = "35",
      number = "6",
      doi = "10.1175/WAF-D-20-0153.1",
      pages=      "2621 - 2638"
}

@article{Coffer2023,
      author = "Brice E. Coffer and Matthew D. Parker and John M. Peters and Andrew R. Wade",
      title = {{Supercell Low-Level Mesocyclones: Origins of Inflow and Vorticity}},
      journal = "Monthly Weather Review",
      year = "2023",
      publisher = "American Meteorological Society",
      address = "Boston MA, USA",
      volume = "151",
      number = "9",
      doi = "10.1175/MWR-D-22-0269.1",
      pages = "2205 - 2232"
}

@article{Coniglio2020,
      author = "Michael C. Coniglio and Matthew D. Parker",
      title = {{Insights into Supercells and Their Environments from Three Decades of Targeted Radiosonde Observations}},
      journal = "Monthly Weather Review",
      year = "2020",
      publisher = "American Meteorological Society",
      address = "Boston MA, USA",
      volume = "148",
      number = "12",
      doi = "10.1175/MWR-D-20-0105.1",
      pages=      "4893 - 4915",
      url = "https://journals.ametsoc.org/view/journals/mwre/148/12/mwr-d-20-0105.1.xml"
}

@article{DAVIESJONES2015,
    title = {A review of supercell and tornado dynamics},
    journal = {Atmospheric Research},
    volume = {158-159},
    pages = {274-291},
    year = {2015},
    issn = {0169-8095},
    doi = {10.1016/j.atmosres.2014.04.007},
    author = {Robert Davies-Jones}
}

@article{DaviesJones1984,
      author = "Robert  Davies-Jones",
      title = {{Streamwise Vorticity: The Origin of Updraft Rotation in Supercell Storms}},
      journal = "Journal of Atmospheric Sciences",
      year = "1984",
      publisher = "American Meteorological Society",
      address = "Boston MA, USA",
      volume = "41",
      number = "20",
      doi = "10.1175/1520-0469(1984)041<2991:SVTOOU>2.0.CO;2",
      pages=      "2991 - 3006"
}

@book{Doviak1993,
  author={Doviak, R. J. and Zrni{\'c}, D. S.}, title={Doppler Radar and Weather Observations}, edition={2nd},
  publisher={Academic Press}, year={1993}, doi={10.1016/B978-0-12-221420-2.X5001-7}}

@book{Emanel1994,
  title={{Atmospheric Convection}},
  author={Emanuel, K.A.},
  isbn={9780195066302},
  lccn={lc93010414},
  year={1994},
  publisher={Oxford University Press}
}

@ARTICLE{Esterheld2008,
  title     = {{Discriminating between Tornadic and {Non-Tornadic} Supercells: A New Hodograph Technique}},
  author    = "Esterheld, John M and Giuliano, Donald J",
  journal   = "Electron. J. Sev. Storms Meteorol.",
  publisher = "E-Journals of Meteorology",
  volume    = "3",
  number    = "2",
  pages     = "1--50",
  month     = "sep",
  year      = "2008",
  doi       = "10.55599/ejssm.v3i2.15" 
}

@article{Evans2001,
      author = "Jeffry S. Evans and Charles A.  Doswell",
      title = {{Examination of Derecho Environments Using Proximity Soundings}},
      journal = "Weather and Forecasting",
      year = "2001",
      publisher = "American Meteorological Society",
      address = "Boston MA, USA",
      volume = "16",
      number = "3",
      doi = "10.1175/1520-0434(2001)016<0329:EODEUP>2.0.CO;2",
      pages=      "329 - 342",
      url = "https://journals.ametsoc.org/view/journals/wefo/16/3/1520-0434_2001_016_0329_eodeup_2_0_co_2.xml"
}

@article{Gillett2025,
    doi = {10.21105/joss.08087},
    url = {https://doi.org/10.21105/joss.08087},        year = {2025},
    publisher = {The Open Journal}, 
    volume = {10}, number = {112}, 
    pages = {8087}, 
    author = {Gillett, Kyle J.}, 
    title = {SounderPy: An atmospheric sounding visualization and analysis tool for Python}, 
    journal = {Journal of Open Source Software} 
}

@article {Groenemeijer2014,
      author = "Pieter Groenemeijer and Thilo Kühne",
      title = {{A Climatology of Tornadoes in Europe: Results from the European Severe Weather Database}},
      journal = "Monthly Weather Review",
      year = "2014",
      publisher = "American Meteorological Society",
      address = "Boston MA, USA",
      volume = "142",
      number = "12",
      doi = "10.1175/MWR-D-14-00107.1",
      pages=      "4775 - 4790",
      url = "https://journals.ametsoc.org/view/journals/mwre/142/12/mwr-d-14-00107.1.xml"
}

@techreport{Groenemeijer2023,
  author       = {Groenemeijer, Pieter and Holzer, Alois M. and Kühne, Thomas and Púčik, Tomáš},
  title        = {The International Fujita (IF) Scale for Tornado and Wind Damage Assessments},
  institution  = {European Severe Storms Laboratory (ESSL)},
  year         = {2023},
  pages        = {77},
  url          = {https://www.essl.org/cms/wp-content/uploads/IF-scale_v1.0e.pdf}
}

@article{HOQUE2017,
    title = {{Tropical cyclone disaster management using remote sensing and spatial analysis: A review}},
    journal = {International Journal of Disaster Risk Reduction},
    volume = {22},
    pages = {345-354},
    year = {2017},
    issn = {2212-4209},
    doi = {10.1016/j.ijdrr.2017.02.008},
    author = {Muhammad Al-Amin Hoque and Stuart Phinn and Chris Roelfsema and Iraphne Childs}
}

@article{Helmus2016,
 title = {{The Python ARM Radar Toolkit (Py-ART), a Library for Working with Weather Radar Data in the Python Programming Language}},
 author = {Helmus, Jonathan J and Collis, Scott M},
 doi = {10.5334/jors.119},
 journal = {Journal of Open Research Software},
 month = {Jul},
 volume = "4",
 number = "1",
 year = {2016}
}

@article{Hersbach2020,
    author = {Hersbach, Hans and Bell, Bill and Berrisford, Paul and Hirahara, Shoji and Horányi, András and Muñoz-Sabater, Joaquín and Nicolas, Julien and Peubey, Carole and Radu, Raluca and Schepers, Dinand and Simmons, Adrian and Soci, Cornel and Abdalla, Saleh and Abellan, Xavier and Balsamo, Gianpaolo and Bechtold, Peter and Biavati, Gionata and Bidlot, Jean and Bonavita, Massimo and De Chiara, Giovanna and Dahlgren, Per and Dee, Dick and Diamantakis, Michail and Dragani, Rossana and Flemming, Johannes and Forbes, Richard and Fuentes, Manuel and Geer, Alan and Haimberger, Leo and Healy, Sean and Hogan, Robin J. and Hólm, Elías and Janisková, Marta and Keeley, Sarah and Laloyaux, Patrick and Lopez, Philippe and Lupu, Cristina and Radnoti, Gabor and de Rosnay, Patricia and Rozum, Iryna and Vamborg, Freja and Villaume, Sebastien and Thépaut, Jean-Noël},
    title = {{The ERA5 global reanalysis}},
    year = {2020},
    journal = {Quarterly Journal of the Royal Meteorological Society},
    volume = {146},
    number = {730},
    pages = {1999-2049},
    doi = {10.1002/qj.3803}
}

@article{Hirose2019,
  title={{High Temporal Rainfall Estimations from Himawari-8 Multiband Observations Using the Random-Forest Machine-Learning Method}},
  author="H. Hirose and S. Shige and M. K. Yamamoto and A. Higuchi",
  journal={Journal of the Meteorological Society of Japan. Ser. II},
  volume={97},
  number={3},
  pages={689-710},
  year={2019},
  doi={10.2151/jmsj.2019-040}
}

@Article{Howard2020,
    AUTHOR = {Howard, Jeremy and Gugger, Sylvain},
    TITLE = {{Fastai: A Layered API for Deep Learning}},
    JOURNAL = {Information},
    VOLUME = {11},
    YEAR = {2020},
    NUMBER = {2},
    ARTICLE-NUMBER = {108},
    ISSN = {2078-2489},
    DOI = {10.3390/info11020108}
}

@article{Kilambi2018,
      author = "Alamelu Kilambi and Frédéric Fabry and Véronique Meunier",
      title = {{A Simple and Effective Method for Separating Meteorological from Nonmeteorological Targets Using Dual-Polarization Data}},
      journal = "Journal of Atmospheric and Oceanic Technology",
      year = "2018",
      publisher = "American Meteorological Society",
      address = "Boston MA, USA",
      volume = "35",
      number = "7",
      doi = "10.1175/JTECH-D-17-0175.1",
      pages=      "1415 - 1424"
}

@article{King2019,
      author = "Austin T. King and Aaron D. Kennedy",
      title = {{North American Supercell Environments in Atmospheric Reanalyses and RUC-2}},
      journal = "Journal of Applied Meteorology and Climatology",
      year = "2019",
      publisher = "American Meteorological Society",
      address = "Boston MA, USA",
      volume = "58",
      number = "1",
      doi = "10.1175/JAMC-D-18-0015.1",
      pages = "71 - 92"
}

@article{Kirshbaum2018,
    AUTHOR = {Kirshbaum, Daniel J. and Adler, Bianca and Kalthoff, Norbert and Barthlott, Christian and Serafin, Stefano},
    TITLE = {{Moist Orographic Convection: Physical Mechanisms and Links to Surface-Exchange Processes}},
    JOURNAL = {Atmosphere},
    VOLUME = {9},
    YEAR = {2018},
    NUMBER = {3},
    ARTICLE-NUMBER = {80},
    ISSN = {2073-4433},
    DOI = {10.3390/atmos9030080}
}

@article {Kumjian2008,
      author = "Matthew R. Kumjian and Alexander V. Ryzhkov",
      title = {{Polarimetric Signatures in Supercell Thunderstorms}},
      journal = "Journal of Applied Meteorology and Climatology",
      year = "2008",
      publisher = "American Meteorological Society",
      address = "Boston MA, USA",
      volume = "47",
      number = "7",
      doi = "10.1175/2007JAMC1874.1",
      pages=      "1940 - 1961"
}

@article {Kumjian2009,
      author = "Matthew R. Kumjian and Alexander V. Ryzhkov",
      title = {{Storm-Relative Helicity Revealed from Polarimetric Radar Measurements}},
      journal = "Journal of the Atmospheric Sciences",
      year = "2009",
      publisher = "American Meteorological Society",
      address = "Boston MA, USA",
      volume = "66",
      number = "3",
      doi = "10.1175/2008JAS2815.1",
      pages=      "667 - 685",
      url = "https://journals.ametsoc.org/view/journals/atsc/66/3/2008jas2815.1.xml"
}

@article{LEONCRUZ2025,
    title = {Tornadic environments in Mexico},
    journal = {Atmospheric Research},
    volume = {315},
    pages = {107916},
    year = {2025},
    issn = {0169-8095},
    doi = {10.1016/j.atmosres.2025.107916},
    url = {https://www.sciencedirect.com/science/article/pii/S0169809525000080},
    author = {José Francisco León-Cruz}
}

@article{Lee2013,
      author = "Sang-Ki Lee and Robert Atlas and David Enfield and Chunzai Wang and Hailong Liu",
      title = {{Is There an Optimal ENSO Pattern That Enhances Large-Scale Atmospheric Processes Conducive to Tornado Outbreaks in the United States?}},
      journal = "Journal of Climate",
      year = "2013",
      publisher = "American Meteorological Society",
      address = "Boston MA, USA",
      volume = "26",
      number = "5",
      doi = "10.1175/JCLI-D-12-00128.1",
      pages=      "1626 - 1642",
      url = "https://journals.ametsoc.org/view/journals/clim/26/5/jcli-d-12-00128.1.xml"
}

@article{LeonCruz2022,
  author  = {León-Cruz, J. F. and Pineda-Martínez, L. F. and Carbajal, N.},
  title   = {{Tornado climatology and potentially severe convective environments in Mexico}},
  journal = {Climate Research},
  volume  = {87},
  pages   = {147--165},
  year    = {2022},
  doi     = {10.3354/cr01692}
}

@ARTICLE{Lin2021,
  title = {{Removing interference echoes in Philippine radars using a fuzzy logic approach}},
  author    = "Lin, Pin-Fang and Lagrimas, Lester Kim M and Wang, Jung-Bin and Chang, Pao-Liang and Fang, Wei-Ting and Almario, John Grender C and Jou, Ben Jong-Dao and Cayanan, Esperanza O",
  journal   = "Terr. Atmos. Ocean. Sci.",
  publisher = "Springer Science and Business Media LLC",
  volume    = "32",
  number    = "5.1",
  pages     = "755-776",
  year      = "2021",
  doi  = "10.3319/TAO.2021.12.20.01"
}

@article {Lopes2025,
      author = "Murilo Machado Lopes and Maurício Ilha de Oliveira and Vitor Goede and Vanessa Ferreira and Ernani de Lima Nascimento",
      title = {{Damage Survey and Multisensor Analysis of a Cold Season Tornado Outbreak in Southern Brazil}},
      journal = "Journal of Applied Meteorology and Climatology",
      year = "2025",
      publisher = "American Meteorological Society",
      address = "Boston MA, USA",
      volume = "64",
      number = "7",
      doi = "10.1175/JAMC-D-24-0146.1",
      pages=      "831 - 855"
}

@article{Louf2020,
  author={Louf, V. and Protat, A. and Jackson, R. C. and Collis, S. M. and Helmus, J.},
  title={{UNRAVEL}: A Robust Modular Velocity Dealiasing Technique for {D}oppler Radar},
  journal=JAOT, year={2020}, volume={37}, number={5}, pages={741--758}, doi={10.1175/JTECH-D-19-0020.1}}

@article {Markowski2002,
      author = "Paul M. Markowski and Jerry M. Straka and Erik N. Rasmussen",
      title = "Direct Surface Thermodynamic Observations within the Rear-Flank Downdrafts of Nontornadic and Tornadic Supercells",
      journal = "Monthly Weather Review",
      year = "2002",
      publisher = "American Meteorological Society",
      address = "Boston MA, USA",
      volume = "130",
      number = "7",
      doi = "10.1175/1520-0493(2002)130<1692:DSTOWT>2.0.CO;2",
      pages=      "1692 - 1721",
      url = "https://journals.ametsoc.org/view/journals/mwre/130/7/1520-0493_2002_130_1692_dstowt_2.0.co_2.xml"
}

@article{Markowski2003,
      author = "Paul  Markowski and Christina  Hannon and Jeff  Frame and Elise  Lancaster and Albert  Pietrycha and Roger  Edwards and Richard L.  Thompson",
      title = {{Characteristics of Vertical Wind Profiles near Supercells Obtained from the Rapid Update Cycle}},
      journal = "Weather and Forecasting",
      year = "2003",
      publisher = "American Meteorological Society",
      address = "Boston MA, USA",
      volume = "18",
      number = "6",
      doi = "10.1175/1520-0434(2003)018<1262:COVWPN>2.0.CO;2",
      pages = "1262 - 1272"
}

@article {Mercer2012,
      author = "Andrew E. Mercer and Chad M. Shafer and Charles A. Doswell and Lance M. Leslie and Michael B. Richman",
      title = {{Synoptic Composites of Tornadic and Nontornadic Outbreaks}},
      journal = "Monthly Weather Review",
      year = "2012",
      publisher = "American Meteorological Society",
      address = "Boston MA, USA",
      volume = "140",
      number = "8",
      doi = "10.1175/MWR-D-12-00029.1",
      pages=      "2590 - 2608",
      url = "https://journals.ametsoc.org/view/journals/mwre/140/8/mwr-d-12-00029.1.xml"
}

@article {Nielsen2018,
      author = "Erik R. Nielsen and Russ S. Schumacher",
      title = "Dynamical Insights into Extreme Short-Term Precipitation Associated with Supercells and Mesovortices",
      journal = "Journal of the Atmospheric Sciences",
      year = "2018",
      publisher = "American Meteorological Society",
      address = "Boston MA, USA",
      volume = "75",
      number = "9",
      doi = "10.1175/JAS-D-17-0385.1",
      pages=      "2983 - 3009"
}

@article {Nielsen2020,
      author = "Erik R. Nielsen and Russ S. Schumacher",
      title = "Dynamical Mechanisms Supporting Extreme Rainfall Accumulations in the Houston “Tax Day” 2016 Flood",
      journal = "Monthly Weather Review",
      year = "2020",
      publisher = "American Meteorological Society",
      address = "Boston MA, USA",
      volume = "148",
      number = "1",
      doi = "10.1175/MWR-D-19-0206.1",
      pages=      "83 - 109"
}

@article{Nixon2021,
      author = "Cameron J. Nixon and John T. Allen",
      title = {{Anticipating Deviant Tornado Motion Using a Simple Hodograph Technique}},
      journal = "Weather and Forecasting",
      year = "2021",
      publisher = "American Meteorological Society",
      address = "Boston MA, USA",
      volume = "36",
      number = "1",
      doi = "10.1175/WAF-D-20-0056.1",
      pages = "219 - 235"
}

@article{Nixon2022,
      author = "Cameron J. Nixon and John T. Allen",
      title = {{Distinguishing between Hodographs of Severe Hail and Tornadoes}},
      journal = "Weather and Forecasting",
      year = "2022",
      publisher = "American Meteorological Society",
      address = "Boston MA, USA",
      volume = "37",
      number = "10",
      doi = "10.1175/WAF-D-21-0136.1",
      pages=      "1761 - 1782"
}

@article{Nixon2024,
      author = "Cameron J. Nixon and John T. Allen and Matthew B. Wilson and Matthew J. Bunkers and Mateusz Taszarek",
      title = {{Cell Mergers, Boundary Interactions, and Convective Systems in Cases of Significant Tornadoes and Hail}},
      journal = "Weather and Forecasting",
      year = "2024",
      publisher = "American Meteorological Society",
      address = "Boston MA, USA",
      volume = "39",
      number = "10",
      doi = "10.1175/WAF-D-23-0117.1",
      pages = "1435 - 1458"
}

@article{Nowotarski2018,
      author = "Christopher J. Nowotarski and Erin A. Jones",
      title = {{Multivariate Self-Organizing Map Approach to Classifying Supercell Tornado Environments Using Near-Storm, Low-Level Wind and Thermodynamic Profiles}},
      journal = "Weather and Forecasting",
      year = "2018",
      publisher = "American Meteorological Society",
      address = "Boston MA, USA",
      volume = "33",
      number = "3",
      doi = "10.1175/WAF-D-17-0189.1",
      pages = "661 - 670"
}

@article{Oliveira2022,
    author = {Maurício I. {Oliveira} and Franciano S.  {Puhales} and Ernani L. {Nascimento} and Vagner Anabor},
    title = {{Integrated damage, visual, remote sensing, and environmental analysis of a strong tornado in southern Brazil}},
    journal = {Atmospheric Research},
    volume = {274},
    pages = {106188},
    year = {2022},
    issn = {0169-8095},
    doi = {10.1016/j.atmosres.2022.106188}
}

@article{Overeem2020,
      author = "Aart Overeem and Remko Uijlenhoet and Hidde Leijnse",
      title = {{Full-Year Evaluation of Nonmeteorological Echo Removal with Dual-Polarization Fuzzy Logic for Two C-Band Radars in a Temperate Climate}},
      journal = "Journal of Atmospheric and Oceanic Technology",
      year = "2020",
      publisher = "American Meteorological Society",
      address = "Boston MA, USA",
      volume = "37",
      number = "9",
      doi = "10.1175/JTECH-D-19-0149.1",
      pages = "1643 - 1660"
}

@article{Peters2022,
      author = "John M. Peters and Jake P. Mulholland and Daniel R. Chavas",
      title = {{Generalized Lapse Rate Formulas for Use in Entraining CAPE Calculations}},
      journal = "Journal of the Atmospheric Sciences",
      year = "2022",
      publisher = "American Meteorological Society",
      address = "Boston MA, USA",
      volume = "79",
      number = "3",
      doi = "10.1175/JAS-D-21-0118.1",
      pages=      "815 - 836"
}

@article{Peters2023b,
      author = "John M. Peters and Daniel R. Chavas and Chun-Yian Su and Hugh Morrison and Brice E. Coffer",
      title = {{An Analytic Formula for Entraining CAPE in Midlatitude Storm Environments}},
      journal = "Journal of the Atmospheric Sciences",
      year = "2023",
      publisher = "American Meteorological Society",
      address = "Boston MA, USA",
      volume = "80",
      number = "9",
      doi = "10.1175/JAS-D-23-0003.1",
      pages=      "2165 - 2186"
}

@article{Pilguj2022,
author = {Pilguj, Natalia and Taszarek, Mateusz and Kryza, Maciej and Brooks, Harold E.},
title = {{Reconstruction of Violent Tornado Environments in Europe: High-Resolution Dynamical Downscaling of ERA5}},
journal = {Geophysical Research Letters},
volume = {49},
number = {11},
pages = {e2022GL098242},
doi = {https://doi.org/10.1029/2022GL098242},
url = {https://agupubs.onlinelibrary.wiley.com/doi/abs/10.1029/2022GL098242},
eprint = {https://agupubs.onlinelibrary.wiley.com/doi/pdf/10.1029/2022GL098242},
note = {e2022GL098242 2022GL098242},
year = {2022}
}

@article{Pilguj2022b,
      author = "Natalia Pilguj and Mateusz Taszarek and John T. Allen and Kimberly A. Hoogewind",
      title = {{Are Trends in Convective Parameters over the United States and Europe Consistent between Reanalyses and Observations?}},
      journal = "Journal of Climate",
      year = "2022",
      publisher = "American Meteorological Society",
      address = "Boston MA, USA",
      volume = "35",
      number = "12",
      doi = "10.1175/JCLI-D-21-0135.1",
      pages=      "3605 - 3626"
}

@misc{PlanetARPS2024,
  author       = {{Planet Team}},
  title        = {Technical Specification: Analysis-Ready PlanetScope (ARPS)},
  year         = {2024},
  howpublished = {\url{https://docs.planet.com/data/imagery/arps/techspec/}},
  note         = {Version 1.0.0, October 2024; accessed April 2026}
}

@article {Pucik2024,
      author = "Tomáš Púčik and David Rýva and Miloslav Staněk and Miroslav Šinger and Pieter Groenemeijer and Georg Pistotnik and Rainer Kaltenberger and Miloš Zich and Jan Koláček and Alois Holzer",
      title = {{The Violent Tornado on 24 June 2021 in Czechia: Damage Survey, Societal Impacts, and Lessons Learned}},
      journal = "Weather, Climate, and Society",
      year = "2024",
      publisher = "American Meteorological Society",
      address = "Boston MA, USA",
      volume = "16",
      number = "3",
      doi = "10.1175/WCAS-D-23-0080.1",
      pages=      "411 - 429",
      url = "https://journals.ametsoc.org/view/journals/wcas/16/3/WCAS-D-23-0080.1.xml"
}

@article {Rasmussen1998,
      author = "Erik N. Rasmussen and David O.  Blanchard",
      title = {{A Baseline Climatology of Sounding-Derived Supercell and Tornado Forecast Parameters}},
      journal = "Weather and Forecasting",
      year = "1998",
      publisher = "American Meteorological Society",
      address = "Boston MA, USA",
      volume = "13",
      number = "4",
      doi = "10.1175/1520-0434(1998)013<1148:ABCOSD>2A.0.CO;2",
      pages=      "1148 - 1164"
}

@article {Rasmussen2003,
      author = "Erik N. Rasmussen",
      title = "Refined Supercell and Tornado Forecast Parameters",
      journal = "Weather and Forecasting",
      year = "2003",
      publisher = "American Meteorological Society",
      address = "Boston MA, USA",
      volume = "18",
      number = "3",
      doi = "10.1175/1520-0434(2003)18<530:RSATFP>2.0.CO;2",
      pages=      "530 - 535"
}

@article {Rasmussen2006,
      author = "Erik N.  Rasmussen and Jerry M.  Straka and Matthew S.  Gilmore and Robert  Davies-Jones",
      title = {{A Preliminary Survey of Rear-Flank Descending Reflectivity Cores in Supercell Storms}},
      journal = "Weather and Forecasting",
      year = "2006",
      publisher = "American Meteorological Society",
      address = "Boston MA, USA",
      volume = "21",
      number = "6",
      doi = "10.1175/WAF962.1",
      pages=      "923 - 938",
}

@InProceedings{Ronnberger2015,
    author="Ronneberger, Olaf
and Fischer, Philipp
and Brox, Thomas",
    editor="Navab, Nassir
and Hornegger, Joachim
and Wells, William M.
and Frangi, Alejandro F.",
    title={{U-Net: Convolutional Networks for Biomedical Image Segmentation}},
    booktitle={{Medical Image Computing and Computer-Assisted Intervention -- MICCAI 2015}},
    year="2015",
    publisher="Springer International Publishing",
    address="Cham",
    pages="234--241",
    isbn="978-3-319-24574-4",
    doi="10.1007/978-3-319-24574-4_28"
}

@article {Ryzhkov2005,
      author = "Alexander V. Ryzhkov and Terry J. Schuur and Donald W. Burgess and Dusan S. Zrnic",
      title = {{Polarimetric Tornado Detection}},
      journal = "Journal of Applied Meteorology",
      year = "2005",
      publisher = "American Meteorological Society",
      address = "Boston MA, USA",
      volume = "44",
      number = "5",
      doi = "10.1175/JAM2235.1",
      pages=      "557 - 570"
}

@article {Sherburn2014,
      author = "Keith D.  Sherburn and Matthew D.  Parker",
      title = {{Climatology and Ingredients of Significant Severe Convection in High-Shear, Low-CAPE Environments}},
      journal = "Weather and Forecasting",
      year = "2014",
      publisher = "American Meteorological Society",
      address = "Boston MA, USA",
      volume = "29",
      number = "4",
      doi = "10.1175/WAF-D-13-00041.1",
      pages=      "854 - 877",
      url = "https://journals.ametsoc.org/view/journals/wefo/29/4/waf-d-13-00041_1.xml"
}

@article{SilvaDias2011,
      author = "Silva Dias, Maria A. F.",
      title = {{An Increase in the Number of Tornado Reports in Brazil}},
      journal = "Weather, Climate, and Society",
      year = "2011",
      publisher = "American Meteorological Society",
      address = "Boston MA, USA",
      volume = "3",
      number = "3",
      doi = "10.1175/2011WCAS1095.1",
      pages=      "209 - 217",
      url = "https://journals.ametsoc.org/view/journals/wcas/3/3/2011wcas1095_1.xml"
}

@article{Taszarek2020a,
      author = "Mateusz Taszarek and John T. Allen and Pieter Groenemeijer and Roger Edwards and Harold E. Brooks and Vanna Chmielewski and Sven-Erik Enno",
      title = {{Severe Convective Storms across Europe and the United States. Part I: Climatology of Lightning, Large Hail, Severe Wind, and Tornadoes}},
      journal = "Journal of Climate",
      year = "2020",
      publisher = "American Meteorological Society",
      address = "Boston MA, USA",
      volume = "33",
      number = "23",
      doi = "10.1175/JCLI-D-20-0345.1",
      pages=      "10239 - 10261"
}

@article{Taszarek2020b,
      author = "Mateusz Taszarek and John T. Allen and Tomáš Púčik and Kimberly A. Hoogewind and Harold E. Brooks",
      title = {{Severe Convective Storms across Europe and the United States. Part II: ERA5 Environments Associated with Lightning, Large Hail, Severe Wind, and Tornadoes}},
      journal = "Journal of Climate",
      year = "2020",
      publisher = "American Meteorological Society",
      address = "Boston MA, USA",
      volume = "33",
      number = "23",
      doi = "10.1175/JCLI-D-20-0346.1",
      pages=      "10263 - 10286",
      url = "https://journals.ametsoc.org/view/journals/clim/33/23/jcliD200346.xml"
}

@article{Taszarek2021a,
      author = "Mateusz Taszarek and Natalia Pilguj and John T. Allen and Victor Gensini and Harold E. Brooks and Piotr Szuster",
      title = {{Comparison of Convective Parameters Derived from ERA5 and MERRA-2 with Rawinsonde Data over Europe and North America}},
      journal = "Journal of Climate",
      year = "2021",
      publisher = "American Meteorological Society",
      address = "Boston MA, USA",
      volume = "34",
      number = "8",
      doi = "10.1175/JCLI-D-20-0484.1",
      pages = "3211 - 3237",
}

@ARTICLE{Taszarek2021b,
  title    = {{Global climatology and trends in convective environments from {ERA5} and rawinsonde data}},
  author   = "Taszarek, Mateusz and Allen, John T and Marchio, Mattia and Brooks, Harold E",
  journal  = "npj Climate and Atmospheric Science",
  doi      = "10.1038/s41612-021-00190-x",
  volume   =  4,
  number   =  1,
  pages    = "35",
  month    =  jun,
  year     =  2021
}

@article{Thompson2000,
      author = "Richard L.  Thompson and Roger  Edwards",
      title = {{An Overview of Environmental Conditions and Forecast Implications of the 3 May 1999 Tornado Outbreak}},
      journal = "Weather and Forecasting",
      year = "2000",
      publisher = "American Meteorological Society",
      address = "Boston MA, USA",
      volume = "15",
      number = "6",
      doi = "10.1175/1520-0434(2000)015<0682:AOOECA>2.0.CO;2",
      pages = "682 - 699"
}

@article {Thompson2003,
      author = "Richard L.  Thompson and Roger  Edwards and John A.  Hart and Kimberly L.  Elmore and Paul  Markowski",
      title = {{Close Proximity Soundings within Supercell Environments Obtained from the Rapid Update Cycle}},
      journal = "Weather and Forecasting",
      year = "2003",
      publisher = "American Meteorological Society",
      address = "Boston MA, USA",
      volume = "18",
      number = "6",
      doi = "10.1175/1520-0434(2003)018<1243:CPSWSE>2.0.CO;2",
      pages=      "1243 - 1261",
      url = "https://journals.ametsoc.org/view/journals/wefo/18/6/1520-0434_2003_018_1243_cpswse_2_0_co_2.xml"
}

@article{Thomspon2012,
      author = "Richard L.  Thompson and Bryan T.  Smith and Jeremy S.  Grams and Andrew R.  Dean and Chris  Broyles",
      title = {{Convective Modes for Significant Severe Thunderstorms in the Contiguous United States. Part II: Supercell and QLCS Tornado Environments}},
      journal = "Weather and Forecasting",
      year = "2012",
      publisher = "American Meteorological Society",
      address = "Boston MA, USA",
      volume = "27",
      number = "5",
      doi = "10.1175/WAF-D-11-00116.1",
      pages=      "1136 - 1154"
}

@article{Wagner2019,
      author = "Melissa Wagner and Robert K. Doe and Aaron Johnson and Zhiang Chen and Jnaneshwar Das and Randall S. Cerveny",
      title = {{Unpiloted Aerial Systems (UASs) Application for Tornado Damage Surveys: Benefits and Procedures}},
      journal = "Bulletin of the American Meteorological Society",
      year = "2019",
      publisher = "American Meteorological Society",
      address = "Boston MA, USA",
      volume = "100",
      number = "12",
      doi = "10.1175/BAMS-D-19-0124.1",
      pages=      "2405 - 2409",
      url = "https://journals.ametsoc.org/view/journals/bams/100/12/bams-d-19-0124.1.xml"
}

@article {Wakimoto2025,
      author = "Roger M. Wakimoto and David J. Bodine and Howard B. Bluestein and Trey Greenwood and Samuel Emmerson",
      title = {{Kinematic Structure of a Tornado Based on an Analysis of Lofted Debris Speeds, Debris Orientation, and Mobile Radar Data}},
      journal = "Monthly Weather Review",
      year = "2025",
      publisher = "American Meteorological Society",
      address = "Boston MA, USA",
      volume = "153",
      number = "8",
      doi = "10.1175/MWR-D-24-0176.1",
      pages=      "1365 - 1383",
      url = "https://journals.ametsoc.org/view/journals/mwre/153/8/MWR-D-24-0176.1.xml"
}

@article{Wang2022,
    author = "X. Wang and H. Iwabuchi and T. Yamashita",
    title = {{Cloud identification and property retrieval from Himawari-8 infrared measurements via a deep neural network}},
    journal = {Remote Sensing of Environment},
    volume = {275},
    pages = {113026},
    year = {2022},
    issn = {0034-4257},
    doi = {10.1016/j.rse.2022.113026}
}

@Inbook{Weisman1986,
    author="Weisman, Morris L. and Klemp, Joseph B.",
    title={{Characteristics of Isolated Convective Storms}},
    bookTitle={{Mesoscale Meteorology and Forecasting}},
    year="1986",
    publisher="American Meteorological Society",
    address="Boston, MA",
    pages="331--358",
    isbn="978-1-935704-20-1",
    doi="10.1007/978-1-935704-20-1_15"
}

@article{Wood1997,
  author={Wood, V. T. and Brown, R. A.},
  title={Effects of Radar Sampling on Single-{D}oppler Velocity Signatures of Mesocyclones and Tornadoes},
  journal=WF, year={1997}, volume={12}, number={4}, pages={928--938}, doi={10.1175/1520-0434(1997)012<0928:EORSOS>2.0.CO;2}}

@article{Wurman2021,
author = {Joshua Wurman  and Karen Kosiba  and Trevor White  and Paul Robinson },
title = {{Supercell tornadoes are much stronger and wider than damage-based ratings indicate}},
journal = {Proceedings of the National Academy of Sciences},
volume = {118},
number = {14},
pages = {e2021535118},
year = {2021},
doi = {10.1073/pnas.2021535118}
}

@article{Xie2016, 
    title={{Transfer Learning from Deep Features for Remote Sensing and Poverty Mapping}}, 
    volume={30}, 
    DOI={10.1609/aaai.v30i1.9906}, 
    number={1}, 
    journal={Proceedings of the AAAI Conference on Artificial Intelligence}, 
    author={Xie, Michael and Jean, Neal and Burke, Marshall and Lobell, David and Ermon, Stefano}, 
    year={2016}, 
    month={Mar.} 
}

@article{Xu1989,
      author = "Xu, Kuan-Man and Kerry A. Emanuel",
      title = {{Is the Tropical Atmosphere Conditionally Unstable?}},
      journal = "Monthly Weather Review",
      year = "1989",
      publisher = "American Meteorological Society",
      address = "Boston MA, USA",
      volume = "117",
      number = "7",
      doi = "10.1175/1520-0493(1989)117<1471:ITTACU>2.0.CO;2",
      pages = "1471 - 1479"
}

@ARTICLE{Yuan2024,
       author = {{Yuan}, Chao and {Bai}, Yang and {Sun}, Pengfei and {Bai}, Hua and {Xia}, Lu},
        title = "{Microphysical Insights into a Tornadic Supercell from Dual-Polarization Radar Observations in Jiangsu, China on 14 May 2021}",
      journal = {Journal of Meteorological Research},
         year = 2024,
        month = apr,
       volume = {38},
       number = {2},
        pages = {303-320},
          doi = {10.1007/s13351-024-3102-5}
}

@article{jiang2025,
  author  = {Jiang, Qin and Dawson II, Daniel T. and Li, Funing and Chavas, Daniel R.},
  title   = {Classifying synoptic patterns driving tornadic storms and associated spatial trends in the United States},
  journal = {npj Climate and Atmospheric Science},
  volume  = {8},
  pages   = {7},
  year    = {2025},
  doi     = {10.1038/s41612-025-00897-1}
}

@misc{lepore2021,
  author       = {Lepore, C. and Allen, J. and Abernathey, R.},
  title        = {xcape (v0.1.4)},
  year         = {2021},
  publisher    = {Zenodo},
  doi          = {10.5281/zenodo.5270332},
  url          = {https://doi.org/10.5281/zenodo.5270332}
}

@book{markowski2010,
  author    = {Markowski, Paul and Richardson, Yvette},
  title     = {Mesoscale Meteorology in Midlatitudes},
  year      = {2010},
  publisher = {John Wiley \& Sons},
  doi       = {10.1002/9780470682104}
}

@misc{rw2019timm,
  author = {Ross Wightman},
  title = {{PyTorch Image Models}},
  year = {2019},
  publisher = {GitHub},
  journal = {GitHub repository},
  doi = {10.5281/zenodo.4414861},
  howpublished = {\url{https://github.com/rwightman/pytorch-image-models}}
}

@article{schaefer2020,
  author  = {Schaefer, Martin and Teeuw, Richard and Day, Simon and Zekkos, Dimitrios and Weber, Paul and Meredith, Toby and van Westen, Cees J.},
  title   = {{Low-cost UAV surveys of hurricane damage in Dominica: automated processing with co-registration of pre-hurricane imagery for change analysis}},
  journal = {Natural Hazards},
  year    = {2020},
  volume  = {101},
  pages   = {755--784},
  doi     = {10.1007/s11069-020-03893-1}
}

@article{Aebischer1998,
  author  = {Aebischer, U. and Sch{\"a}r, C.},
  title   = {Low-Level Potential Vorticity and Cyclogenesis to the Lee of the {A}lps},
  journal = {Journal of the Atmospheric Sciences}, year = {1998}, volume = {55}, number = {2}, pages = {186--207}}

@article{Ashley2008,
  author  = {Ashley, W. S. and Krmenec, A. J. and Schwantes, R.},
  title   = {Vulnerability due to Nocturnal Tornadoes},
  journal = {Weather and Forecasting}, year = {2008}, volume = {23}, number = {5}, pages = {795--807}}

@article{Benjamin1968,
  author  = {Benjamin, T. B.},
  title   = {Gravity Currents and Related Phenomena},
  journal = {Journal of Fluid Mechanics}, year = {1968}, volume = {31}, number = {2}, pages = {209--248}}

@article{Besag1986,
  author  = {Besag, J.},
  title   = {On the Statistical Analysis of Dirty Pictures},
  journal = {Journal of the Royal Statistical Society: Series B}, year = {1986}, volume = {48}, number = {3}, pages = {259--302}}

@article{Blackadar1957,
  author  = {Blackadar, A. K.},
  title   = {Boundary Layer Wind Maxima and Their Significance for the Growth of Nocturnal Inversions},
  journal = {Bulletin of the American Meteorological Society}, year = {1957}, volume = {38}, number = {5}, pages = {283--290}}

@article{Brown1978,
  author  = {Brown, R. A. and Lemon, L. R. and Burgess, D. W.},
  title   = {Tornado Detection by Pulsed {D}oppler Radar},
  journal = {Monthly Weather Review}, year = {1978}, volume = {106}, number = {1}, pages = {29--38}}

@article{Brown2002,
  author  = {Brown, R. A. and Wood, V. T. and Sirmans, D.},
  title   = {Improved Tornado Detection Using Simulated and Actual {WSR-88D} Data with Enhanced Resolution},
  journal = {Journal of Atmospheric and Oceanic Technology}, year = {2002}, volume = {19}, number = {11}, pages = {1759--1771}}

@article{Bunkers2022,
  author  = {Bunkers, M. J. and Wilson, M. B. and Van Den Broeke, M. S. and Healey, D. J.},
  title   = {Scan-by-Scan Storm-Motion Deviations for Concurrent Tornadic and Nontornadic Supercells},
  journal = {Weather and Forecasting}, year = {2022}, volume = {37}, number = {5}, pages = {749--770}}

@article{CapuliCN2026,
  author  = {Capuli, G. H.},
  title   = {Localized Tornado Outbreak at the Upstream of a Tropical Easterly Wave in {C}amarines {N}orte, {P}hilippines (13 {S}eptember 2025)},
  journal = {Earth Systems and Environment}, year = {2026}, note = {Accepted, arXiv:2604.20222}}

@article{Firdaus2025,
  author  = {Firdaus, I. M. and Yamazaki, T. and Abdillah, M. R. and Riawan, E.},
  title   = {{I}ndonesia Tornado Database: Tornado Climatology of {I}ndonesia},
  journal = {Natural Hazards and Earth System Sciences}, year = {2025}, volume = {25}, number = {11}, pages = {4331--4341},
  doi     = {10.5194/nhess-25-4331-2025}}

@misc{Grams2026,
  author       = {Grams, J. and Bunting, B.},
  title        = {{SPC} Convective Outlooks},
  year         = {2026},
  howpublished = {NOAA/NWS Storm Prediction Center},
  note         = {Accessed 15 November 2026, \url{https://www.spc.noaa.gov/misc/SPC_probotlk_info.html}}}

@article{Gropp2018,
  author  = {Gropp, M. E. and Davenport, C. E.},
  title   = {The Impact of the Nocturnal Transition on the Lifetime and Evolution of Supercell Thunderstorms in the {G}reat {P}lains},
  journal = {Weather and Forecasting}, year = {2018}, volume = {33}, number = {4}, pages = {1045--1061},
  doi     = {10.1175/WAF-D-17-0150.1}}

@article{Holleman2003,
  author  = {Holleman, I. and Beekhuis, H.},
  title   = {Analysis and Correction of Dual {PRF} Velocity Data},
  journal = {Journal of Atmospheric and Oceanic Technology}, year = {2003}, volume = {20}, number = {4}, pages = {443--453}}

@article{Holton1967,
  author  = {Holton, J. R.},
  title   = {The Diurnal Boundary Layer Wind Oscillation above Sloping Terrain},
  journal = {Tellus}, year = {1967}, volume = {19}, number = {2}, pages = {200--205}}

@article{Houser2015,
  author  = {Houser, J. L. and Bluestein, H. B. and Snyder, J. C.},
  title   = {Rapid-Scan, Polarimetric, {D}oppler Radar Observations of Tornadogenesis and Tornado Dissipation in a Tornadic Supercell: The ``{E}l {R}eno, {O}klahoma'' Storm of 24 {M}ay 2011},
  journal = {Monthly Weather Review}, year = {2015}, volume = {143}, number = {7}, pages = {2685--2710}}

@article{James2001,
  author  = {James, C. N. and Houze, Jr., R. A.},
  title   = {A Real-Time Four-Dimensional {D}oppler Dealiasing Scheme},
  journal = {Journal of Atmospheric and Oceanic Technology}, year = {2001}, volume = {18}, number = {10}, pages = {1674--1683}}

@article{Joe2003,
  author  = {Joe, P. and May, P. T.},
  title   = {Correction of Dual {PRF} Velocity Errors for Operational {D}oppler Weather Radars},
  journal = {Journal of Atmospheric and Oceanic Technology}, year = {2003}, volume = {20}, number = {4}, pages = {429--442}}

@article{Kennedy2007,
  author  = {Kennedy, A. and Straka, J. M. and Rasmussen, E. N.},
  title   = {A Statistical Study of the Association of {DRCs} with Supercells and Tornadoes},
  journal = {Weather and Forecasting}, year = {2007}, volume = {22}, number = {6}, pages = {1191--1199},
  doi     = {10.1175/2007WAF2006095.1}}

@article{Kis2010,
  author  = {Kis, A. K. and Straka, J. M.},
  title   = {Nocturnal Tornado Climatology},
  journal = {Weather and Forecasting}, year = {2010}, volume = {25}, number = {2}, pages = {545--561}}

@article{McCaul1991,
  author  = {McCaul, Jr., E. W.},
  title   = {Buoyancy and Shear Characteristics of Hurricane-Tornado Environments},
  journal = {Monthly Weather Review}, year = {1991}, volume = {119}, number = {8}, pages = {1954--1978}}

@article{McLaughlin2009,
  author  = {McLaughlin, D. and others},
  title   = {Short-Wavelength Technology and the Potential for Distributed Networks of Small Radar Systems},
  journal = {Bulletin of the American Meteorological Society}, year = {2009}, volume = {90}, number = {12}, pages = {1797--1817},
  doi     = {10.1175/2009BAMS2507.1}}

@article{Mead2011,
  author  = {Mead, C. M. and Thompson, R. L.},
  title   = {Environmental Characteristics Associated with Nocturnal Significant-Tornado Events in the {G}reat {P}lains},
  journal = {Electronic Journal of Severe Storms Meteorology}, year = {2011}, volume = {6}, number = {6}}

@article{Meng2018,
  author  = {Meng, Z. and others},
  title   = {The Deadliest Tornado ({EF4}) in the Past 40 Years in {C}hina},
  journal = {Weather and Forecasting}, year = {2018}, volume = {33}, number = {3}, pages = {693--713},
  doi     = {10.1175/WAF-D-17-0085.1}}

@article{Pazmany2013,
  author  = {Pazmany, A. L. and Mead, J. B. and Bluestein, H. B. and Snyder, J. C. and Houser, J. B.},
  title   = {A Mobile Rapid-Scanning {X}-band Polarimetric ({RaXPol}) {D}oppler Radar System},
  journal = {Journal of Atmospheric and Oceanic Technology}, year = {2013}, volume = {30}, number = {7}, pages = {1398--1413},
  doi     = {10.1175/JTECH-D-12-00166.1}}

@article{Shapiro2016,
  author  = {Shapiro, A. and Fedorovich, E. and Rahimi, S.},
  title   = {A Unified Theory for the {G}reat {P}lains Nocturnal Low-Level Jet},
  journal = {Journal of the Atmospheric Sciences}, year = {2016}, volume = {73}, number = {8}, pages = {3037--3057}}

@article{Smith2020,
  author  = {Smith, B. T. and Thompson, R. L. and Speheger, D. A. and Dean, A. R. and Karstens, C. D. and Anderson-Frey, A. K.},
  title   = {{WSR-88D} Tornado Intensity Estimates. {P}art {II}: Real-Time Applications to Tornado Warning Time Scales},
  journal = {Weather and Forecasting}, year = {2020}, volume = {35}, number = {6}, pages = {2493--2506},
  doi     = {10.1175/WAF-D-20-0011.1}}

@article{Snyder2015,
  author  = {Snyder, J. C. and Ryzhkov, A. V.},
  title   = {Automated Detection of Polarimetric Tornadic Debris Signatures Using a Hydrometeor Classification Algorithm},
  journal = {Journal of Applied Meteorology and Climatology}, year = {2015}, volume = {54}, number = {9}, pages = {1861--1870},
  doi     = {10.1175/JAMC-D-15-0138.1}}

@article{Takahashi2019,
  author  = {Takahashi, N. and others},
  title   = {Development of Multi-Parameter Phased Array Weather Radar ({MP-PAWR}) and Early Detection of Torrential Rainfall and Tornado Risk},
  journal = {Journal of Disaster Research}, year = {2019}, volume = {14}, number = {2}, pages = {235--247},
  doi     = {10.20965/jdr.2019.p0235}}

@inproceedings{Thompson2004,
  author    = {Thompson, R. L. and Edwards, R. and Mead, C. M.},
  title     = {An Update to the Supercell Composite and Significant Tornado Parameters},
  booktitle = {Preprints, 22nd Conf. on Severe Local Storms}, address = {Hyannis, MA},
  publisher = {Amer. Meteor. Soc.}, year = {2004}, pages = {P8.1}}

@article{Waldteufel1979,
  author  = {Waldteufel, P. and Corbin, H.},
  title   = {On the Analysis of Single-{D}oppler Radar Data},
  journal = {Journal of Applied Meteorology}, year = {1979}, volume = {18}, number = {4}, pages = {532--542}}

@article{Wright2025,
  author  = {Wright, N. and Duncan, J. M. A. and Callow, J. N. and Thompson, S. E. and George, R. J.},
  title   = {Training Sensor-Agnostic Deep Learning Models for Remote Sensing: Achieving State-of-the-Art Cloud and Cloud Shadow Identification with {OmniCloudMask}},
  journal = {Remote Sensing of Environment}, year = {2025}, volume = {322}, pages = {114694},
  doi     = {10.1016/j.rse.2025.114694}}

\end{document}